\documentclass[preprint,3p]{elsarticle}
\usepackage{amsmath,amssymb}
\usepackage{xspace}
\usepackage{longtable}
\usepackage{ulem}
\usepackage{color}
\usepackage[mathscr]{eucal}
\usepackage{pifont}
\usepackage{listings}
\usepackage{slashed}
\usepackage{float}
\usepackage{subfig}
\usepackage{hyperref}
\newcommand{\SARAH}{{\tt SARAH}\xspace}
\newcommand{\Bs}{B^0_s}
\newcommand{\Bd}{B^0_d}
\newcommand{\SPheno}{{\tt SPheno}\xspace}
\newcommand{\FeynArts}{{\tt FeynArts}\xspace}
\newcommand{\FormCalc}{{\tt FormCalc}\xspace}

\newcommand{\abs}[1]{\left|{#1}\right|}

\newcommand{\ampM}[0]{\mathcal{M}}

\newcommand{\kl}[1]{\left(#1\right)}

\newcommand{\nn}{\nonumber}
\newcommand{\PLR}{P_{L/R}}

\newcommand{\mli}{m_{i}}
\newcommand{\mlj}{m_{j}}
\newcommand{\mlisq}{m_{i}^2}
\newcommand{\mljsq}{m_{j}^2}
\newcommand{\scn}[2]{\ensuremath{#1\cdot 10^{#2}}}

\newcommand{\Bsll}{\ensuremath{B^0_s\to \ell\bar{\ell}}\xspace}
\newcommand{\Bdll}{\ensuremath{B^0_d\to \ell\bar{\ell}}\xspace}

\newcommand{\Bsdll}{\ensuremath{B^0_{s,d}\to \ell \bar{\ell} \xspace}}

\newcommand{\BRx}[1]{\ensuremath{\text{BR}(#1)\,}}
\newcommand{\BR}{\ensuremath{\text{BR}\,}}

\newenvironment{eqnsub}{\begin{subequations}\begin{eqnarray}}{\end{eqnarray}\end{subequations}}

\newcommand{\qref}[1]{eq.~\eqref{#1}}

\newcommand{\MeV}{\text{Me}\hspace{-0.05cm}\text{V}}
\newcommand{\GeV}{\text{Ge}\hspace{-0.05cm}\text{V}}

\begin{document}
\begin{flushright}
Bonn-TH-2012-34
\end{flushright}

\title{Full 1-loop calculation of BR($\Bsdll$) \\ in models beyond the MSSM with \SARAH and \SPheno}

\author[bonn]{H.~Dreiner}
\ead{dreiner@uni-bonn.de}

\author[bonn]{K.~Nickel}
\ead{nickel@th.physik.uni-bonn.de}

\author[wu]{W.~Porod}
\ead{porod@physik.uni-wuerzburg.de}

\author[bonn]{F.~Staub}
\ead{fnstaub@th.physik.uni-bonn.de}

\address[bonn]{Bethe Center for Theoretical Physics \& Physikalisches Institut der
Universit\"at Bonn, Nu{\ss}allee 12, 53115 Bonn, Germany}
\address[wu]{Institut f\"ur Theoretische Physik und Astrophysik,
 Universit\"at  W\"urzburg, 97074  W\"urzburg, Germany}

\begin{abstract}
  We present the possibility of calculating the quark
  flavor changing neutral current decays $\Bsll$ and $\Bdll$
  for a large variety of supersymmetric models. For this purpose, the
  complete one--loop calculation has been implemented in a generic form
  in the {\tt Mathematica} package \SARAH. This information is used by
  \SARAH to generate {\tt Fortran} source code for \SPheno for a
  numerical evaluation of these processes in a given model. We comment
  also on the possibility to use this setup for non--supersymmetric
  models.
\end{abstract}

\maketitle

\section{Introduction}
With the recent discovery of a bosonic resonance
\cite{Atlas:2012gk,CMS:2012gu} showing all the characteristics of the
SM Higgs boson a long search might soon come to a successful end.  In
contrast there are no hints for a signal of supersymmetric (SUSY)
particles or particles predicted by any other extension of the
standard model (SM)
\cite{:2012rz,Chatrchyan:2012jx,CMS-PAS-SUS-11-022,CMS-PAS-SUS-12-005,:2012mfa}. 
Therefore, large areas of the parameter space of the simplest SUSY
models are excluded. The allowed mass spectra as well as the best fit
mass values to the data are pushed to higher and higher values
\cite{pMSSM}.  This has lead to an increasing interest in the study of
SUSY models which provide new features. For instance, models with
broken $R$-parity \cite{rpv,rpvsearches} or compressed spectra
\cite{compressed} might be able to hide much better at the LHC, while
for other models high mass spectra are a much more natural feature
than this is the case in the minimal-supersymmetric standard model
(MSSM) \cite{finetuning}.

However, bounds on the masses and couplings of beyond the SM
(BSM) models follow not only from direct searches at colliders. New
particles also have an impact on SM processes 
via virtual quantum corrections, leading in many
instances to sizable deviations from the SM expectations. This holds
in particular for the anomalous magnetic moment of the muon
\cite{Stockinger:2006zn} and 
processes which are highly suppressed in the SM. 
The latter are mainly lepton flavor violating (LFV) or decays involving
quark flavor changing neutral currents (qFCNC). While the prediction
of LFV decays in the SM is many orders of magnitude below the
experimental sensitivity \cite{Cheng:1985bj}, qFCNC is
experimentally well established. For instance, the observed rate of $b
\to s\gamma$ is in good agreement with the SM expectation and this
observable has put for several years strong
constraints on qFCNCs beyond the SM \cite{bsgamma}. 

The experiments at the LHC have reached now a sensitivity to test also
the SM prediction for BR($B^0_s\to\mu\bar\mu$) as well as BR($B^0_d\to
\mu \bar{\mu}$) \cite{Buras:2012ru}
\begin{eqnarray}
  \BRx{B^0_s \to \mu \bar{\mu}}_{\text{SM}} &=& \scn{(3.23\pm 0.27)}{-9} \label{eq:Bsintro1}, \\
  \BRx{B^0_d \to \mu \bar{\mu}}_{\text{SM}} &=& \scn{(1.07\pm 0.10)}{-10}. \label{eq:Bsintro2}
\end{eqnarray}
Using the measured finite width difference of the $B$ mesons the time integrated branching ratio
which should be compared to experiment is \cite{Buras:2013uqa}
\begin{eqnarray}
  \BRx{B^0_s \to \mu \bar{\mu}}_{\text{theo}} &=& \scn{(3.56\pm 0.18)}{-9} \, .
\end{eqnarray}
Recently, LHCb reported the first evidence for $B^0_s \to \mu
\bar{\mu}$. The observed rate  \cite{LHCb:2012ct}
\begin{equation}
\BRx{B^0_s \to \mu \bar{\mu}} = (3.2^{+1.5}_{-1.2})\times 10^{-9}  \\
\end{equation}
fits nicely to the SM prediction. For $\BRx{B^0_d \to \mu \bar{\mu}}$
the current upper 
bound: $9.4 \cdot 10^{-10}$
is already of the same order as the SM expectation.

This leads to new constraints for BSM models and each model has to be
confronted with these measurements. So far, there exist several public
tools which can calculate 
$\BRx{B^0_{s,d}\to \ell\bar{\ell}}$ as well as other observables in the context of the MSSM
or partially also for the next-to-minimal supersymmetric standard
model (NMSSM) \cite{NMSSM}: {\tt superiso} \cite{superiso}, {\tt
SUSY\_Flavor} \cite{susyflavor}, {\tt NMSSM-Tools} \cite{NMSSMTools},
{\tt MicrOmegas} \cite{MicrOmegas} or {\tt SPheno} \cite{spheno}. 
However, for more complicated SUSY models none of the available tools
provides the possibility to calculate 
these decays easily. This gap is now closed by the interplay of the {\tt
Mathematica} package \SARAH \cite{sarah} and the spectrum generator
\SPheno. \SARAH 
already has many SUSY models incorporated but allows also an easy and
efficient implementation of new models. For all of these models \SARAH
can generate new modules for \SPheno for a comprehensive numerical
evaluation. This functionality is extended, as described in this paper, by a
full 1-loop calculation of $B^0_{s,d}\to\ell\bar{\ell}$.

The rest of the paper is organized as follows: in
sec.~\ref{sec:analytical} we recall briefly the analytical calculation
for BR($B^0_{s,d}\to\ell\bar{\ell}$).  In
sec.~\ref{sec:implementation} we discuss the implementation of this
calculation in \SARAH and \SPheno before we conclude in
sec.~\ref{sec:conclusion}. The appendix contains more information
about the calculation and generic results for the amplitudes.

\section{Calculation of BR($B^0_{s,d}\to\ell\bar{\ell}$)}
\label{sec:analytical}
In the SM this decay was first calculated in ref~\cite{Inami:1980fz},
in the analogous context of kaons. The higher order corrections were
first presented in \cite{Buchalla:1993bv}; see also
\cite{Misiak:1999yg}. In the context of supersymmetry this was 
considered in \cite{bsmumu-susy}. See also the interesting correlation
between BR$(B_s^0\to\mu\bar\mu)$ and $(g-2)_\mu$ \cite{Dedes:2001fv}.

We present briefly the main steps of the calculation of BR($B^0_{q}\to
\ell_k\bar{\ell}_l$) with $q=s,d$. We follow
closely the notation of ref.~\cite{Dedes:2008iw}. The effective
Hamiltonian can be parametrized by
\begin{eqnarray}
  \label{eq:effectiveH} \mathcal{H}&=&\frac{1}{16\pi^2}
  \sum_{X,Y=L,R}\kl{C_{SXY}\mathcal{O}_{SXY}+C_{VXY}\mathcal{O}_{VXY}+C_{TX}\mathcal{O}_{TX}}
  \, ,
\end{eqnarray}
with the Wilson coefficients $C_{SXY},C_{VXY},C_{TX}$ corresponding to
the scalar, vector and tensor operators
\begin{equation}
  \mathcal{O}_{SXY} = (\bar q_j P_X q_i)(\bar \ell_l P_Y \ell_k) \,, \hspace{0.5cm}
  \mathcal{O}_{VXY} = (\bar q_j \gamma^\mu P_X q_i)(\bar \ell_l \gamma_\mu P_Y
\ell_k) \,, \hspace{0.5cm}
  \mathcal{O}_{TX} = (\bar q_j \sigma^{\mu\nu} P_X q_i)(\bar \ell_l \sigma_{\mu\nu} \ell_k) \, . 
\end{equation}
$P_L$ and $P_R$ are the projection operators on left respectively
right handed states.  The expectation value of the axial vector matrix
element is defined as
\begin{eqnarray}
  \label{eq:fBs}
  \langle 0\left| \bar b \gamma^\mu \gamma^5 q \right| B^0_q(p)\rangle 
&\equiv&  ip^\mu f_{B^0_q} \, .
\end{eqnarray}
Here, we introduced the meson decay constants $f_{B^0_q}$ which can be
obtained from lattice QCD simulations \cite{Laiho:2009eu}. The current
values for $B^0_s$ and $B^0_d$ are given by \cite{Davies:2012qf}
\begin{equation}
  \label{eq:fBsValue}
  f_{B^0_s} = (227\pm 8)~{\MeV} \,,\hspace{1cm} f_{B^0_d} = (190\pm 8)~{\MeV} \, . 
\end{equation}
Since the momentum $p$ of the meson is the only four-vector available,
the matrix element in \qref{eq:fBs} can only depend on
$p^\mu$. The incoming momenta of the $b$ antiquark and the $s$ (or $d$) quark are $p_1,p_2$  respectively, where $p=p_1+p_2$. Contracting \qref{eq:fBs} with $p_\mu$ and using the
equations of motion $\bar b \slashed p_1=-\bar b m_b$ and $\slashed
p_2 q=m_q q$ leads to an
expression for the pseudoscalar current
\begin{eqnarray}
  \label{eq:fBspseudo}
   \langle 0 \left| \bar b \gamma^5 q\right| B^0_q(p) \rangle &=& - i \frac{M_{B_q^0}^2 f_{B^0_q}}{m_b+m_q} \, .
\end{eqnarray}
The vector and scalar currents vanish 
\begin{equation}
  \label{eq:vanish}
  \langle 0\left| \bar b \gamma^\mu  q \right| B^0_q(p)\rangle=
\langle 0 \left| \bar b q\right| B^0_q(p)
\rangle =0 \, .
\end{equation}
From eqs.~(\ref{eq:fBs}) and (\ref{eq:fBspseudo}) we obtain
\begin{equation} 
\langle 0 \left| \bar b \gamma^\mu\PLR q \right| B^0_q(p)\rangle = 
\mp \frac i2 p^\mu f_{B^0_q} \label{eq:fBsPLR1} \, , \hspace{1cm} 
\langle 0 \left| \bar b \PLR q
\right| B^0_q(p)\rangle = \pm \frac i2 \frac {M_{B^0_q}^2
f_{B^0_q}}{m_b+m_q} \, .  
\end{equation}
In general, the matrix element $\mathcal M$ is a function of the form
factors $F_S,F_P,F_V,F_A$ of the scalar, pseudoscalar, vector and
axial-vector current and can be expressed by
\begin{equation}
  \label{eq:matrixelementBs}
    (4\pi)^2\mathcal M = F_S \bar \ell \ell + F_P \bar \ell \gamma^5 \ell 
+ F_V p_\mu \bar \ell
  \gamma^\mu \ell + F_A p_\mu \bar \ell \gamma^\mu \gamma^5 \ell \, . 
\end{equation}
Note that there is no way of building an antisymmetric 2-tensor out of
just one vector $p^\mu$. The matrix element of the tensor operator
$\mathcal{O}_{TX}$ must therefore vanish. The form factors can be
expressed by linear combinations of the Wilson coefficients of
eq.~(\ref{eq:effectiveH})
\begin{eqnarray}
    F_S &=& \frac i4 \frac{M_{B^0_q}^2 f_{B^0_q}}{m_b+m_q}  \kl{ C_{SLL} +
    C_{SLR} - C_{SRR}-C_{SRL}},\label{eq:formfactorsBs1}\\
   F_P &=& \frac i4 \frac{M_{B^0_q}^2 f_{B^0_q}}{m_b+m_q}  \kl{ -C_{SLL} +
    C_{SLR} - C_{SRR}+C_{SRL}}, \label{eq:formfactorsBs2}\\
  F_V &=& -\frac i4 f_{B^0_q} \kl{ C_{VLL} + C_{VLR} - C_{VRR}-C_{VRL}}, \label{eq:formfactorsBs3}\\
  F_A &=& -\frac i4 f_{B^0_q} \kl{ -C_{VLL} + C_{VLR} - C_{VRR}+C_{VRL}}.\label{eq:formfactorsBs4} 
  \end{eqnarray}

The main task is to calculate the different Wilson coefficients for a
given model.  These Wilson coefficients receive at the 1-loop level
contributions from 
various wave, penguin
and box diagrams, see Figures~\ref{fig:wave}-\ref{fig:box2} in
\ref{app:amplitudes}. Furthermore, in some models these decays could 
also happen already at tree-level \cite{Dreiner:2006gu}.  The
amplitudes for all possible, generic diagrams which can contribute to
the Wilson coefficients have been calculated with \FeynArts/\FormCalc
\cite{feynarts} and the results are listed in
\ref{app:amplitudes}. This calculation has been performed in the
$\overline{\text{DR}}$ scheme and 't Hooft gauge. How these results
are used together with \SARAH and \SPheno to get numerical results
will be discussed in the next section.

After the calculation of the form factors, the squared amplitude is
\begin{align}
  \label{eq:squaredMBsllp}
  (4\pi)^4 \abs{\ampM}^2&=2\abs{F_S}^2\kl{M_{B^0_q}^2-(m_\ell+m_k)^2}
+2\abs{F_P}^2\kl{M_{B^0_q}^2-(m_\ell-m_k)^2}
  \\
\nn &+ 2\abs{F_V}^2\kl{M_{B^0_q}^2(m_k-m_\ell)^2-(m_k^2-m_\ell^2)^2} \\
\nn &+ 2\abs{F_A}^2\kl{M_{B^0_q}^2(m_k+m_\ell)^2-(m_k^2-m_\ell^2)^2} \\
\nn &+ 4\Re (F_s F_V^*) (m_\ell-m_k)\kl{M_{B^0_q}^2+(m_k+m_\ell)^2} \\
\nn &+ 4\Re (F_P F_A^*) (m_\ell+m_k)\kl{M_{B^0_q}^2-(m_k-m_\ell)^2} \, .
\end{align}
Here, $m_\ell$ and $m_k$ are the lepton masses. In the case $k=\ell$, 
this expression simplifies to
\begin{eqnarray}
\label{eq:ampSquaredsimp}
\abs{\ampM}^2&=&\frac{2}{(16\pi^2)^2} \kl{(M_{B^0_q}^2-4 m_\ell^2)\abs{F_S}^2+
M_{B^0_q}^2 \abs{F_P+2m_\ell F_A}^2}. 
\end{eqnarray}
Note, the result is independent of the form factor $F_V$ in this
limit. In the SM the leading 1-loop contributions proceed via
the exchange of virtual gauge bosons. They are thus helicity
suppressed. Furthermore, since these are flavor changing neutral
currents, they are GIM suppressed. The diagrams involving virtual
Higgs bosons are suppressed due to small Yukawa couplings. In BSM
scenarios these suppressions can be absent.

The branching ratio is then given by
\begin{equation}
  \label{eq:Bsllpbranching}
  \BR(B_q^0\to \ell_k\bar{\ell}_l)=\frac{\tau_{B^0_q}}{16\pi}
  \frac{\abs{\mathcal{M}}^2}{M_{B_q^0}}\sqrt{1-\kl{\frac{m_k+m_l}{M_{B_q^0}}}^2}\sqrt{1-\kl{\frac{m_k-m_l}{M_{B_q^0}}}^2} 
\end{equation}
with $\tau_{B^0_q}$ as the life time of the mesons.

\section{Automatized calculation of $\Bsdll$}
\label{sec:implementation}
\subsection{Implementation in \SARAH and \SPheno}
\SARAH is the first 'spectrum-generator-generator' on the market which
means that it can generate Fortran source for \SPheno to obtain a
full-fledged spectrum generator for models beyond the MSSM. The main
features of a \SPheno module written by \SARAH are a precise mass
spectrum calculation based on 2-loop renormalization group equations
(RGEs) and a full 1-loop calculation of the mass spectrum. Two-loop
results known for the MSSM can be included. Furthermore, also the
decays of SUSY and Higgs particles are calculated as well as
observables like $\ell_i \to \ell_j \gamma$, $\ell_i \to 3 \ell_j$, $b\to
s\gamma$, $\delta\rho$, $(g-2)$, or electric dipole moments. For more
information about the interplay between \SARAH and \SPheno we refer
the interested reader to Ref.~\cite{Staub:2011dp}.

Here we extend the list of observables  by BR($B^0_s\to\ell\bar
{\ell}$) and BR($B^0_d\to \ell\bar{\ell}$). For this purpose, the
generic tree--level and 1--loop amplitudes calculated with
\FeynArts/\FormCalc given in \ref{app:amplitudes} have been implemented
in \SARAH. When \SARAH generates the output for \SPheno it checks for
all possible field combinations which can populate the generic
diagrams in the given model. This information is then used to generate
Fortran code for a numerical evaluation of all of these diagrams. The
amplitudes are then combined to the Wilson coefficients which again
are used to calculate the form factors
eqs.~(\ref{eq:formfactorsBs1})-(\ref{eq:formfactorsBs4}).  The
branching ratio is finally calculated by using
eq.~(\ref{eq:Bsllpbranching}). Note, the known 2--loop QCD corrections
of Refs.~\cite{Buchalla:1993bv,Misiak:1999yg,Bobeth:2001jm} are 
not included in this calculation.

The Wilson coefficients for $\Bsdll$ are
calculated at a scale $Q=160$~GeV by all modules generated by \SARAH,
as this is done by default by \SPheno in the MSSM. Hence, as input
parameters for the calculation the running SUSY masses and couplings at
this scale obtained by a 2-loop RGE evaluation down from the SUSY scale are used. 
In the standard model gauge sector we use the running value of $\alpha_{em}$, the on-shell 
Weinberg angle $\sin \Theta_W^2 = 1 -\frac{m_W^2}{m_Z^2}$ with $m_W$ calculated 
from $\alpha_{em}(M_Z)$, $G_F$ and the $Z$ mass. In addition, the CKM matrix calculated 
from the Wolfenstein parameters ($\lambda$, $A$,$\rho$,$\eta$) as well as 
the running quark masses enter the calculation. To obtain the running SM parameters at $Q=160$~GeV 
 we use 2-loop standard model RGEs of Ref.~\cite{Arason:1991ic}. 
The default SM values as well as the derived parameters are given in Tab.~\ref{tab:sm}. 
Note, even if CP violation is not switched on the calculation of the SUSY spectrum, always the phase
of the CKM matrix is taken into account in these calculations. This is especially important for $B_d^0$
decays.
\begin{table}[bt]
\centering
\small
\begin{tabular}{|l|l|l|l|l|}
\hline \hline
\multicolumn{5}{|c|}{default SM input parameters} \\
\hline
$\alpha^{-1}_{em}(M_Z) = 127.93  $ & $\alpha_s(M_Z) = 0.1190  $ & $G_F = 1.16639\cdot 10^{-5}~\text{GeV}^{-2}$
 & $\rho = 0.135$ & $\eta = 0.349$  \\
$m_t^{pole} = 172.90~\text{GeV}$ & $M_Z^{pole} = 91.1876~\text{GeV}  $ & $m_b(m_b) = 4.2~\text{GeV} $ & $\lambda = 0.2257 $ & $A = 0.814$ \\
\hline \hline
\multicolumn{5}{|c|}{derived parameters} \\
\hline
 $m_t^{\overline{DR}} = 166.4~\text{GeV}$ &
$| V_{tb}^* V_{ts}| = 4.06*10^{-2}  $ & $| V_{tb}^* V_{td}| = 8.12*10^{-3} $ & $m_W = 80.3893 $ & $\sin^2 \Theta_W = 0.2228 $ \\
\hline
\end{tabular}
\caption{SM input values and derived parameters by default used for the numerical evaluation of $B^0_{s,d} \to l\bar{l}$ in \SPheno.} 
  \label{tab:sm}
\end{table}
All standard model parameters can be adjusted by using the 
corresponding standard blocks
of the SUSY LesHouches Accord 2 (SLHA2) \cite{SLHA}. 
Furthermore, the default input values for the
hadronic parameters given in Table~\ref{tab:input} are used. These 
can be changed in the Les Houches input accordingly to the Flavor Les Houches
Accord (FLHA) \cite{Mahmoudi:2010iz} using the following blocks:
\begin{verbatim}
      Block FLIFE          #
      511   1.525E-12      # tau_Bd
      531   1.472E-12      # tau_Bs
      Block FMASS          #
      511   5.27950   0    0    # M_Bd
      531   5.3663    0    0    # M_Bs
      Block FCONST    #
      511   1   0.190    0    0   # f_Bd
      531   1   0.227    0    0   # f_Bs
\end{verbatim}
While \SPheno includes the
chiral resummation for the MSSM, this is not taken into account in the
routines generated by \SARAH because of its large model dependence.
\begin{table}
\centering
\begin{tabular}{|l|l|l|}
\hline
\multicolumn{3}{|c|}{Default hadronic parameters} \\
\hline
$m_{\Bs} = 5.36677$ GeV & $f_{\Bs} = 227(8)$ MeV & $\tau_{\Bs} = 1.466(31)$ ps \\
$m_{\Bd} = 5.27958$ GeV  & $f_{\Bd} = 190(8)$ MeV & $\tau_{\Bd} = 1.519(7)$ ps  \\
\hline
 \end{tabular}
 \caption{Hadronic input values by default used for the numerical evaluation of $B^0_{s,d} \to l\bar{l}$ in \SPheno.} 
  \label{tab:input}
\end{table}
\subsection{Generating and running the source code}
We describe briefly the main steps necessary to generate and run the
\SPheno code for a given model: after starting {\tt Mathematica} and
loading \SARAH it is just necessary to evaluate the demanded model and
call the function to generate the \SPheno source code. For instance,
to get a \SPheno module for the B-L-SSM \cite{Khalil:2007dr,FileviezPerez:2010ek,O'Leary:2011yq}, use
\begin{verbatim}
<<[$SARAH-Directory]/SARAH.m;
Start["BLSSM"];
MakeSPheno[];
\end{verbatim}
{\tt MakeSPheno[]} calculates first all necessary information
(\textit{i.e.} vertices, mass matrices, tadpole equations, RGEs,
self-energies) and then exports this
information to Fortran code and writes all necessary auxiliary
functions needed to compile the code together with \SPheno. The entire
output is saved in the directory
\begin{verbatim}
[$SARAH-Directory]/Output/BLSSM/EWSB/SPheno/ 
\end{verbatim}
The content of this directory has to be copied into a new
subdirectory of \SPheno called {\tt BLSSM} and afterwards the code can
be compiled:
\begin{verbatim}
cp [$SARAH-Directory]/Output/BLSSM/EWSB/SPheno/*  [$SPheno-Directory]/BLSSM/
cd [$SPheno-Directory]
make Model=BLSSM
\end{verbatim}
This creates a new binary {\tt SPhenoBLSSM} in the directory {\tt bin}
of \SPheno. To run the spectrum calculation a file called {\tt
LesHouches\!.in.BLSSM} containing all input parameters in the Les
Houches format has to be provided.  \SARAH writes also a template for
such a file which has been copied with the other files to {\tt
/BLSSM}. This example can be evaluated via
\begin{verbatim}
./bin/SPhenoBLSSM BLSSM/LesHouches.in.BLSSM 
\end{verbatim}
and the output is written to {\tt SPheno.spc.BLSSM}. This file
contains all information like the masses, mass matrices, decay widths
and branching ratios, and observables. For the $B^0_{s,d}\to \ell \bar{\ell}$ 
decays 
the results are given twice for easier comparison: once for the full 
calculation and once including only the SM contributions. All results
are written to the block  {\tt SPhenoLowEnergy} in the spectrum file
using the following numbers:
\begin{center}
\begin{tabular}{|cl|cl|}
\hline 
{\tt 4110}   & $\text{BR}^{SM}(B^0_d\to e^+e^-)$ & {\tt 4111}  & $\text{BR}^{full}(B^0_d\to e^+e^-)$ \\
 {\tt 4220}   & $\text{BR}^{SM}(B^0_d\to \mu^+\mu^-)$ & {\tt 4221}  & $\text{BR}^{full}(B^0_d\to \mu^+\mu^-)$ \\
 {\tt 4330}   & $\text{BR}^{SM}(B^0_d\to \tau^+\tau^-)$ & {\tt 4331}  & $\text{BR}^{full}(B^0_d\to \tau^+\tau^-)$ \\
 {\tt 5110}   & $\text{BR}^{SM}(B^0_s\to e^+e^-)$ & {\tt 5111}  & $\text{BR}^{full}(B^0_s\to e^+e^-)$ \\
 {\tt 5210}   & $\text{BR}^{SM}(B^0_s\to \mu^+e^-)$ & {\tt 5211}  & $\text{BR}^{full}(B^0_s\to \mu^+e^-)$ \\
 {\tt 5220}   & $\text{BR}^{SM}(B^0_s\to \mu^+\mu^-)$ & {\tt 5221}  & $\text{BR}^{full}(B^0_s\to \mu^+\mu^-)$ \\
 {\tt 5330}   & $\text{BR}^{SM}(B^0_s\to \tau^+\tau^-)$ & {\tt 5331}  & $\text{BR}^{full}(B^0_s\to \tau^+\tau^-)$ \\
\hline
\end{tabular}
\end{center}
Note, we kept for completeness and as cross-check $\text{BR}^{SM}(B^0_s\to \mu^+e^-)$ which has to vanish.
The same steps can be repeated for any other model implemented in
\SARAH, or the {\tt SUSY-Toolbox} scripts \cite{Staub:2011dp} can be
used for an automatic implementation of new models in \SPheno as well
as in other tools based on the \SARAH output.

\subsection{Checks}
\begin{figure}[hbtp]
 \centering
 \includegraphics[width=0.5\linewidth]{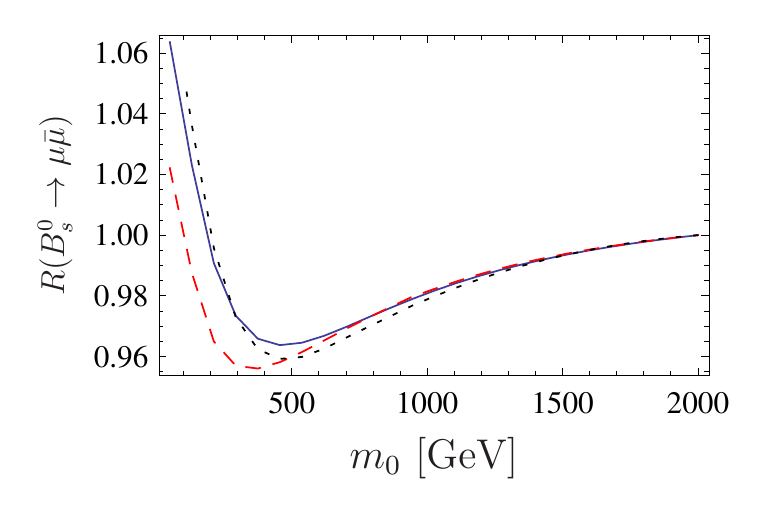}  \\
 \includegraphics[width=0.5\linewidth]{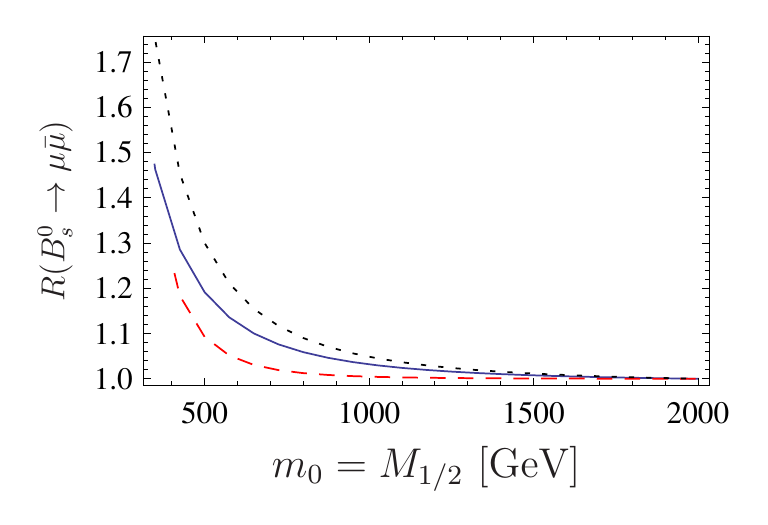}  \\
 \includegraphics[width=0.5\linewidth]{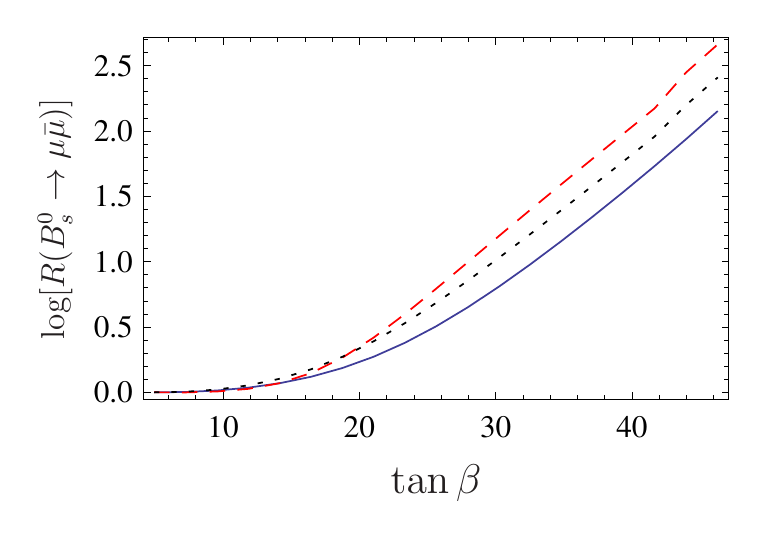} 
 \caption{The top figure: $R=\text{BR}(B_s^0\to \mu^+
   \mu^-)/\text{BR}(B_s^0\to \mu^+\mu^-)_{SM}$ for the constrained 
   MSSM and as function of $m_0$. The other parameters were set to
   $M_{1/2}=140$~GeV, $\tan\beta=10$, $\mu>0$. In the middle $m_0$
   and $M_{1/2}$ were varied simultaneously, while $\tan\beta=30$ 
   was fixed. In the bottom figure we show $\log(R)$ as a function
   of $\tan\beta$, while $m_0=M_{1/2} = 150$~GeV were kept fix. In
   all figures $A_0=0$ and $\mu>0$ was used. The color code is as 
   follows: {\tt superiso 3.3} (dotted black), {\tt SPheno 3.2.1}
   (dashed red) and \SPheno by \SARAH (solid blue).}
 \label{fig:comparison}
\end{figure}

We have performed several cross checks of the code generated by
\SARAH: the first, trivial check has been that we reproduce the known 
SM results and that those agree with the full calculation in the limit of 
heavy SUSY spectra. For the input parameters 
of Tab.~\ref{tab:sm} we obtain $\text{BR}(B^0_s\to \mu^+ \mu^-)^{SM} = 3.28\cdot 10^{-9}$ 
and $\text{BR}(B^0_d\to \mu^+ \mu^-)^{SM} = 1.08\cdot 10^{-10}$ which are in good agreement 
with eqs.~(\ref{eq:Bsintro1})-(\ref{eq:Bsintro2}).
Secondly, as mentioned in
the introduction there are several codes which calculate these decays
for the MSSM or NMSSM. A detailed comparison of all of these codes is
beyond the scope of the presentation here and will 
be presented elsewhere \cite{comparison}. 
However, a few comments are in order:
the code generated by \SARAH as well as most other codes usually show
the same behavior. There are differences in the numerical
values calculated by the programs because of different values
for the SM inputs. For instance, there is an especially strong
dependence on the value of the 
electroweak mixing angle and, of course, of the hadronic parameters used in the
calculation \cite{Buras:2012ru}.  In addition, these processes are
implemented with different accuracy in different tools: the treatment
of NLO QCD corrections \cite{Bobeth:2001jm}, chiral resummation
\cite{Crivellin:2011jt}, or SUSY box diagrams is not the
same. Therefore, we depict in Fig.~\ref{fig:comparison} a comparison
between {\tt SPheno 3.2.1}, {\tt Superiso 3.3} and {\tt SPheno by
SARAH} using the results normalized to the SM limit of each program.

\begin{figure}[hbt]
 \centering
 \includegraphics[width=0.6\linewidth]{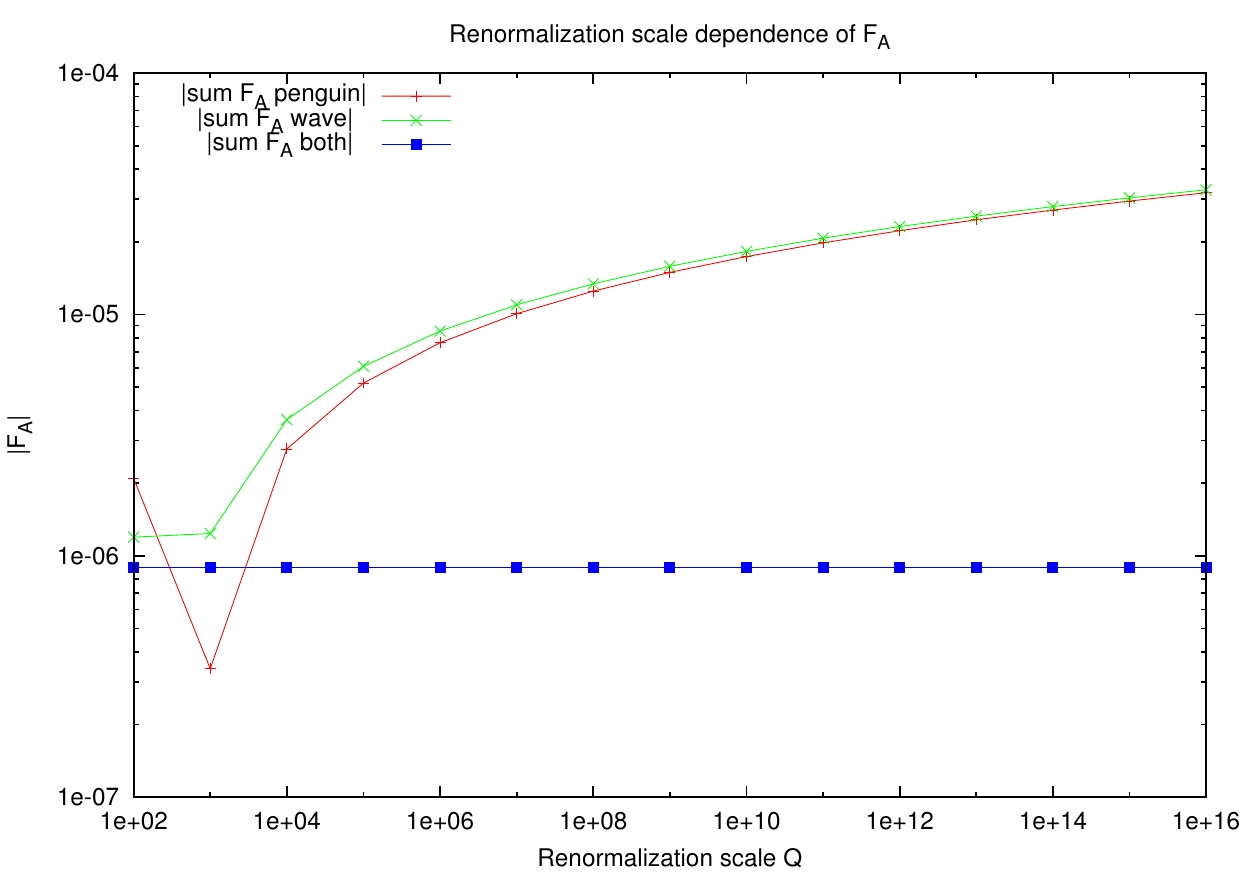}
 \caption{The figure shows $|\sum F_A|_{\text{penguin}}$ and $|\sum F_A|_{\text{wave}}$ as well as the sum of both $|\sum F_A|$. Penguin and wave contributions have opposite signs that interchange between $Q=10^2{\GeV}$ to $Q=10^3{\GeV}$.}
 \label{fig:scaledependence}
\end{figure}

It is also possible to perform a check of self-consistency: the
leading-order contribution has to be finite which leads to non-trivial
relations among the amplitudes for all 
wave and penguin diagrams are given in \ref{sec:waveBapp} and \ref{sec:penguinB}.
Therefore, we can check these relations numerically by
varying the renormalization scale used in all loop integrals. The
dependence on this scale should cancel and the branching ratios should
stay constant. This is shown in Figure~\ref{fig:scaledependence}:
while single contributions can change by several orders the sum of all
is numerically very stable.

\subsection{Non-supersymmetric models}
We have focused our discussion so far on SUSY models. However, even if
\SARAH is optimized for the study of SUSY models it is also able to
handle non-SUSY models to some extent. The main drawback at the moment
for non-SUSY models is that the RGEs can not be calculated
because the results of Refs.~\cite{Martin:1993zk,Fonseca:2011vn} which
are used by \SARAH are not valid in this case. However, all other
calculations like the ones for the vertices, mass matrices and
self-energies don't use SUSY properties and therefore apply to
any model. Hence, it is also possible to generate \SPheno code for
these models which calculates $B^0_{s,d}\to\ell\bar{\ell}$. The main
difference in the calculation comes from the missing possibility to
calculate the RGEs: the user has to provide numerical values for all
parameters at the considered scale which then enter the
calculation. We note that in order to fully support non-supersymmetric
models with \SARAH the calculation of the corresponding RGEs at 2-loop
level will be included in \SARAH in the future \cite{nonsusyrges}.

\section{Conclusion}
\label{sec:conclusion}
We have presented a model independent implementation of the flavor
violating decays $B^0_{s,d} \to \ell\bar{\ell}$ in \SARAH and \SPheno. Our
approach provides the possibility to generate source code which
performs a full 1-loop calculation of these observables for any
model which can be implemented in \SARAH.  Therefore, it takes care of
the necessity to confront many BSM models in the future with the
increasing constraints coming from the measurements of $B^0_{s,d} \to
\ell\bar{\ell}$ at the LHC.

\section*{Acknowledgements}
W.P.\ thanks L.~Hofer for discussions.  This work has been supported
by the Helmholtz alliance `Physics at the Terascale' and W.P.\ in part
by the DFG, project No. PO-1337/2-1. HKD acknowledges support from
BMBF grant 00160200.

\begin{appendix}
\section{Conventions}
\label{conventions}
\subsection{Passarino-Veltman integrals}

We use in the following the conventions of \cite{Pierce:1996zz}
for the Passarino-Veltman integrals. All Wilson coefficients appearing
in the following can be expressed by the integrals
\begin{eqnarray}
  \label{eq:B0withPsq0}
  B_0(0,x,y)&=& \Delta + 1 + \ln \kl{\frac{Q^2}{y}} +
  \frac{x}{x-y}\ln \kl{\frac{y}{x}} \\
\Delta &=& \frac 2{4-D} - \gamma_E + \log 4\pi \\
  B_1(x,y,z) &=& \frac 12 (z-y)\frac{B_0(x,y,z)-B_0(0,y,z)}{x}-\frac 12  B_0(x,y,z)  \\
  C_0(x,y,z) &=& \frac{1}{y-z} \left[ \frac{y}{x-y} \log \frac yx - \frac{z}{x-z} \log \frac zx \right] \label{eq:C0} \\
 \nn C_{00}(x,y,z) 
  &=& \frac 14 \kl{1-\frac 1{y-z}\kl{ \frac {x^2\log x - y^2\log y}{x-y}-\frac{x^2\log x-z^2\log z}{x-z}}} \\
  D_0(x,y,z,t) &=& \frac{C_0(x,y,z)-C_0(x,y,t)}{z-t} \label{eq:D0reduce} \\
\nn &=& -\left[ \frac{y\log \frac yx}{(y-x)(y-z)(y-t)} +
    \frac{z\log \frac zx}{(z-x)(z-y)(z-t)}+\frac{t\log \frac
      tx}{(t-x)(t-y)(t-z)} \right] \\
  D_{00}(x,y,z,t)  & = &  -\frac 14\left[ \frac{y^2\log \frac yx}{(y-x)(y-z)(y-t)} +
    \frac{z^2log \frac zx}{(z-x)(z-y)(z-t)}+\frac{t^2\log \frac
      tx}{(t-x)(t-y)(t-z)} \right].
\end{eqnarray}
Note, the conventions of ref.~\cite{Pierce:1996zz} (Pierce, Bagger
[PB]) are different than those presented in ref.~\cite{Dedes:2008iw}
(Dedes, Rosiek, Tanedo [DRT]). The box integrals are related by
\begin{eqnsub}
  \label{eq:Dloopsall} 
D_{0} &=& D_0^{(\text{PB})}=- D_0^{(DRT)}, \\
\label{eq:Dloopsall2}
D_{00}&=& D_{27}^{(\text{PB})}=- \frac 14 D_{2}^{(DRT)} \,.
\end{eqnsub}

\subsection{Massless limit of loop integrals}
In some amplitudes (i.e. penguin diagrams $(a-b)$, box diagram $(v)$) the
following combinations of loop integrals appear:
\begin{align}
  I_1 &= B_0(s,M_{F1}^2,M_{F2}^2)+M_S^2 C_0(s,0,0,M_{F2}^2,M_{F1}^2,M_S^2), \\
  I_2 &= C_0(0,0,0,M_{F2}^2,M_{F1}^2,M_{V2}^2) + M_{V1}^2 D_0(M_{F2}^2,M_{F1}^2,M_{V1}^2,M_{V2}^2).
\end{align}
The loop functions $B_0,C_0,D_0$ diverge for massless fermions (e.g. neutrinos in the MSSM) but
 the expressions $I_1,I_2$ are finite. However, this limit must be taken analytically in order to avoid
numerical instabilities. 
In a generalized  form and in the limit of zero external momenta, $I_i$ can be expressed by
\begin{align}
  I_1(a,b,c) &= B_0(0,a,b)+c C_0(0,0,0,a,b,c) \equiv B_0(0,a,b)+c C_0(a,b,c), \\
  I_2(a,b,c,d) &= C_0(0,0,0,a,b,d) + c D_0(a,b,c,d) \equiv C_0(a,b,d)+c D_0(a,b,c,d).
\end{align}
Using eqs.~{\ref{eq:B0withPsq0},\ref{eq:C0},\ref{eq:D0reduce}} we obtain in the limit $a\to 0$ 
\begin{align}
  I_1(0,b,c) &= B_0(0,0,b) + c C_0(0,b,c) \\
 &= \Delta + 1 - \log \frac b{Q^2} + c \frac 1{b-c}\log \frac cb \\
 &= \Delta + 1 + \log Q^2  + \frac {c}{b-c} \log c + \kl{-1-\frac c{b-c}}\log
 b 
\end{align}
The term proportional to $\log b$ vanishes in the limit $b\to 0$ 
\begin{equation}
 I_1(0,0,c) = \Delta + 1 - \log \frac c{Q^2}
\end{equation}
The same strategy works for $I_2$:
\begin{align}
  I_2(0,b,c,d) &= C_0(0,b,d)+c D_0(0,b,c,d) \\
&= \frac 1{b-d} \log \frac{d}{b} + c  \frac{C_0(0,b,c)-C_0(0,b,d)}{c-d} \\
&= \frac 1{b-d} \log \frac{d}{b} +  \frac c{c-d} \frac 1{b-c} \log \frac{c}{b} -
\frac{c}{c-d}\frac 1{b-d}\log \frac {d}{b} \\
&= \frac{(c-d)(b-c)\log \frac db + c (b-d)\log \frac cb - c(b-c) \log \frac
  db}{(b-d)(c-d)(b-c)} \label{eq:I2last}
\end{align}
The denominator of \qref{eq:I2last} is finite for $b\to 0$ and in the
numerator, the $\log b$ terms cancel each other:
\begin{equation}
  ( (c-d)c + cd -c^2 ) \log b =0.
\end{equation}
Hence, we end up with
\begin{align}
I_2(0,0,c,d) &= \frac{-c(c-d)\log d - cd \log c + c^2 \log
  d}{cd(c-d)} = \frac{\log \frac dc}{c-d}.
\end{align}
\subsection{Parametrization of vertices}
We are going to express the amplitude in the following in terms of
generic vertices. For this purpose, we parametrize a vertex between 
two fermions and one vector or scalar respectively as
\begin{eqnarray}
\label{eq:chiralvertices1}
 & G_A \gamma_\mu P_L + G_B  \gamma_\mu P_R\, , &  \\
\label{eq:chiralvertices2}
 & G_A P_L + G_B   P_R \, . & 
\end{eqnarray}
$P_{L,R}=\frac 12 (1\mp \gamma^5)$ are the polarization operators. In
addition, for the vertex between three vector bosons and the one between one
vector boson and two scalars the conventions are as follows
\begin{eqnarray}
& G_{VVV} \cdot \kl{g_{\mu\nu}(k_2-k_1)_\rho+g_{\nu\rho}(k_3-k_2)_\mu
+g_{\rho\mu}(k_1-k_3)_\nu}  \, , \\
& G_{SSV}\cdot (k_1-k_2)_\mu\, . &  
\end{eqnarray}
Here, $k_i$ are the (ingoing) momenta of the external particles.

\section{Generic amplitudes}
\label{app:amplitudes}
We present in the following the expressions for the generic amplitudes obtained with \FeynArts and \FormCalc. All coefficients that are not explicitly listed are zero. Furthermore, the Wilson coefficients are left--right symmetric, i.e. 
\begin{equation}
 C_{XRR} = C_{XLL} (L \leftrightarrow R) \,, \hspace{1cm} C_{XRL}
   = C_{XLR}  (L \leftrightarrow R),
\end{equation}
with $X=S,V$ and where $(L \leftrightarrow R)$ means that 
the coefficients of the left and right polarization part of each vertex have to be interchanged.

\allowdisplaybreaks
\subsection{Tree Level Contributions}
\label{sec:treelevel}
\begin{figure}[htbp]
  \centering
  \begin{tabular}{cc}
  \includegraphics[width=4cm]{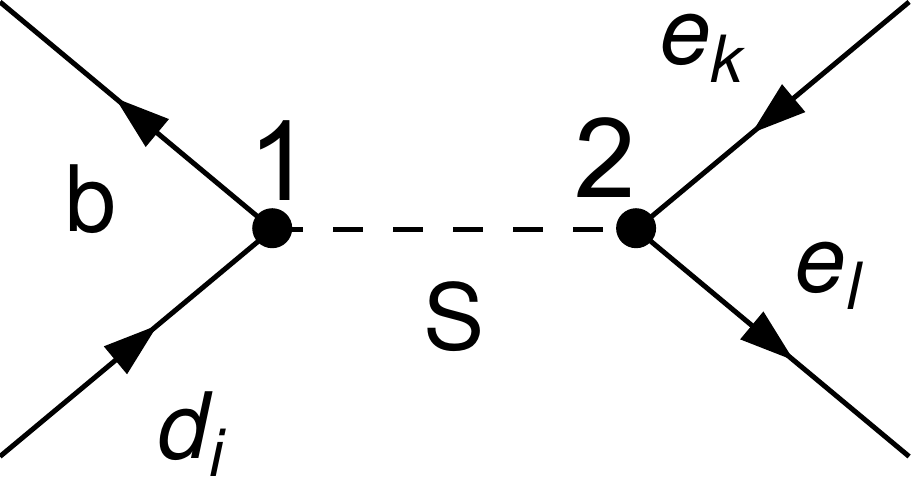} & \includegraphics[width=4cm]{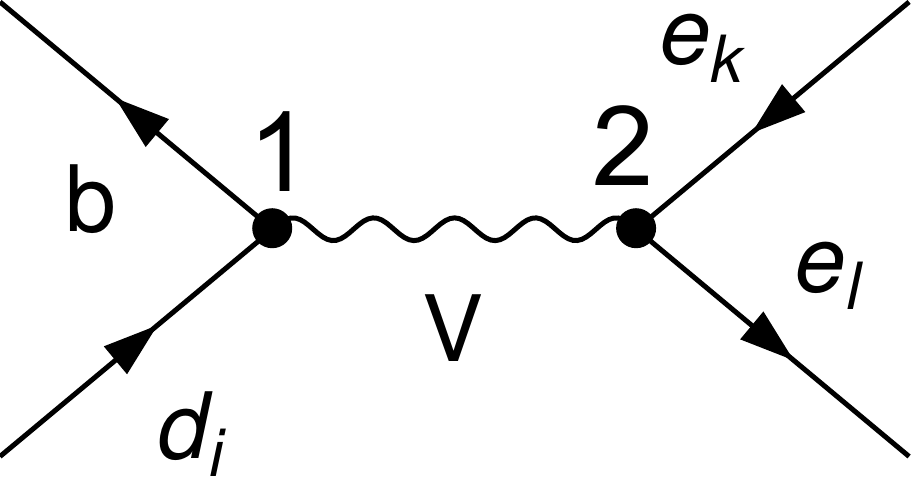} \\
   (a) & (b) \\
  \includegraphics[width=4cm]{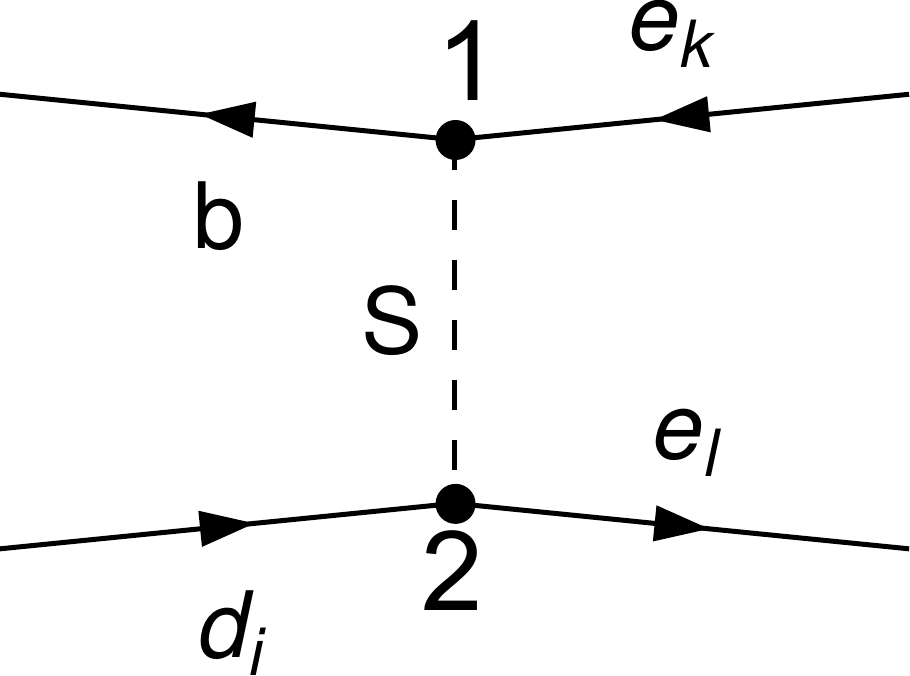} & \includegraphics[width=4cm]{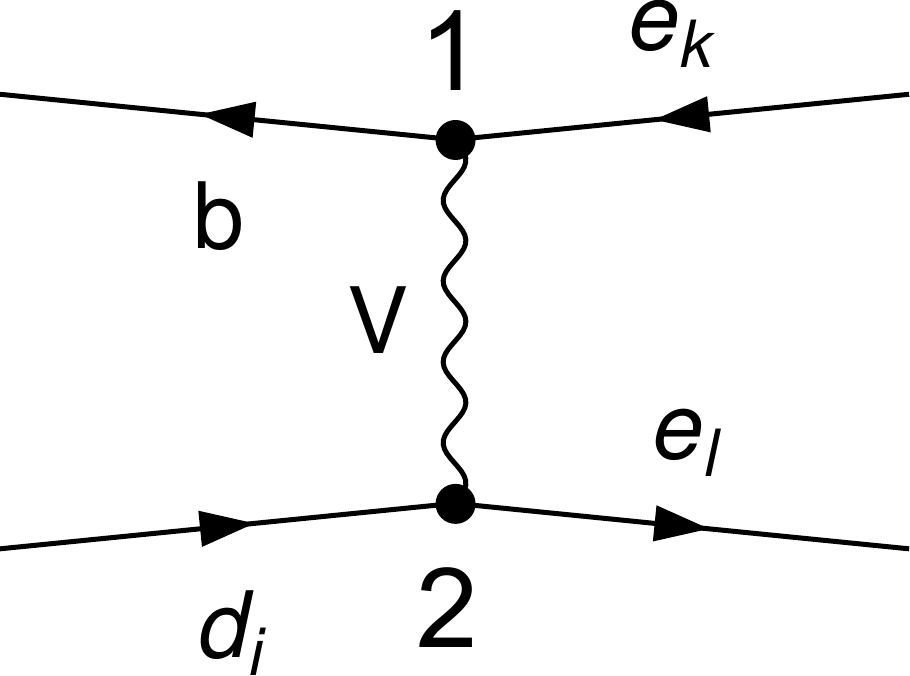} \\
  (c) & (d) \\
    \includegraphics[width=4cm]{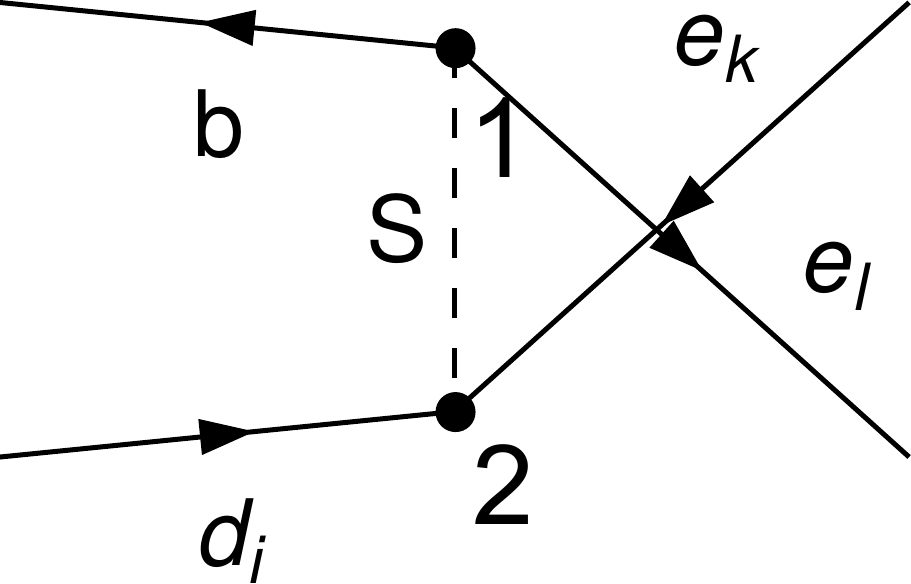} &    \includegraphics[width=4cm]{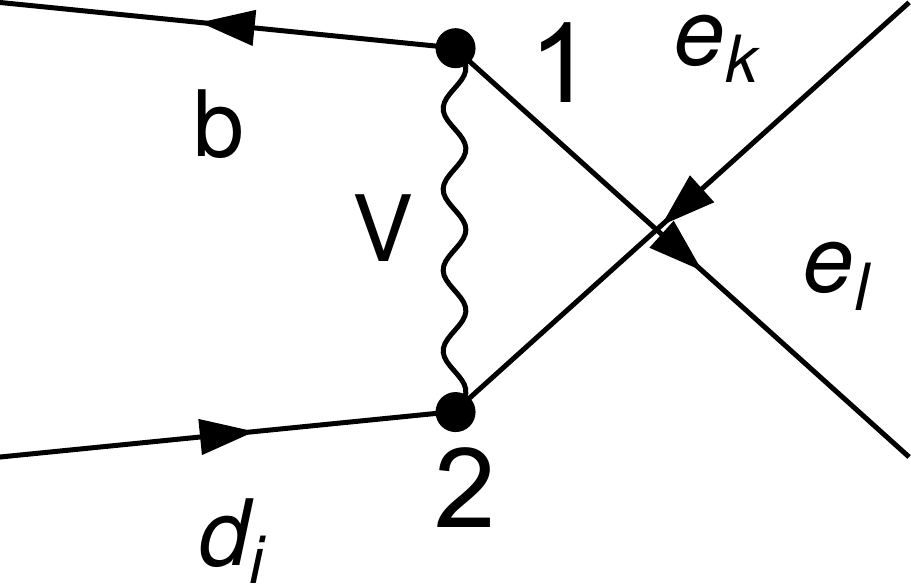} \\
   (e) & (f) 
   \end{tabular}
  \caption{Tree level diagrams with vertex numbering}
  \label{fig:treeleveldiagrams}
\end{figure}
Since in models beyond the MSSM it might be possible that $\Bsll$ is already possible at tree--level. This is for instance the case for trilinear $R$-parity violation \cite{Dreiner:2006gu}. The generic diagrams are given in Figure~\ref{fig:treeleveldiagrams}. The chiral vertices are parametrized as in eqs.~(\ref{eq:chiralvertices1})-(\ref{eq:chiralvertices2}) with $A=1,B=2$ for vertex 1 and $A=3,B=4$ for vertex 2. Using these conventions, the corresponding contributions to the Wilson coefficients read
\begin{eqnarray}
  \label{eq:treeSchannel}
  C^{(a)}_{SLL}&=&16\pi^2 \frac{G_1G_3}{M_S^2-s} \,, \hspace{1cm}  C^{(a)}_{SLR} = 16\pi^2 \frac{G_1G_4}{M_S^2-s}\\
  C^{(b)}_{VLL}&=&16\pi^2 \frac{-G_1G_3}{M_V^2-s}\,, \hspace{1cm}  C^{(b)}_{VLR} = 16\pi^2 \frac{-G_1G_4}{M_V^2-s}\\ 
  C^{(c)}_{SLL}&=& 16\pi^2\frac{-G_1G_3}{2(M_S^2-t)}\,, \hspace{1cm} C^{(c)}_{VLR} =  16\pi^2\frac{-G_2G_3}{2(M_S^2-t)} \\ 
  C^{(d)}_{SLR}&=&16\pi^2\frac{2G_2G_3}{M_V^2-t} \,, \hspace{1cm} C^{(d)}_{VLL} = 16\pi^2\frac{-G_1G_3}{M_V^2-t} \\
  C^{(e)}_{SLL} &=& 16\pi^2\frac{-G_1G_3}{2(M_S^2-u)} \,, \hspace{1cm} C^{(e)}_{VLL} = 16\pi^2\frac{G_2G_3}{2(M_S^2-u)} \\
  C^{(f)}_{SLR} &=& 16\pi^2\frac{-2G_2G_4}{M_V^2-u}  \,, \hspace{1cm} C^{(f)}_{VLR} =  16\pi^2\frac{-G_1G_4}{M_V^2-u} 
 \end{eqnarray}
Here, $s$, $t$ and $u$ are the usual Mandelstam variables. 

\subsection{Wave Contributions}
\label{sec:waveBapp}
\begin{figure}[htpb]
\centering
\begin{tabular}{cc}
\includegraphics[width=3cm]{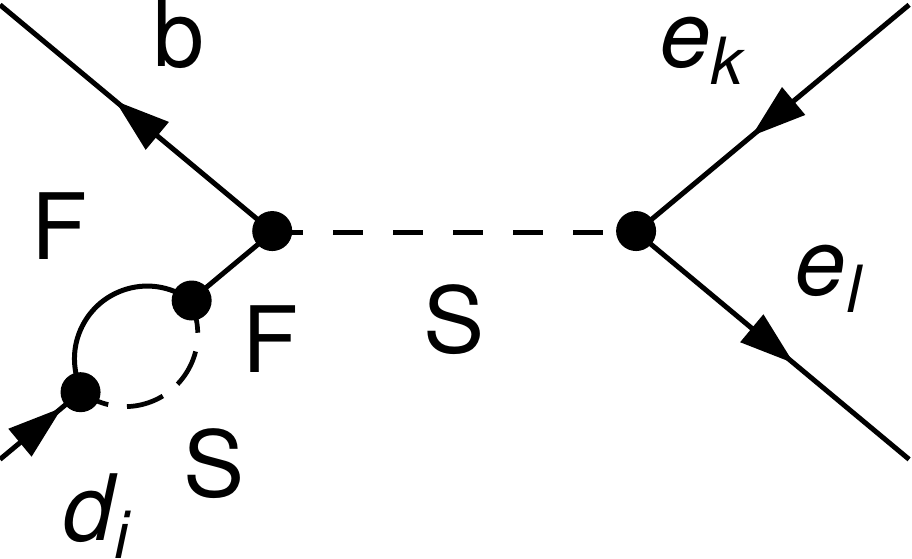} 
& \includegraphics[width=3cm]{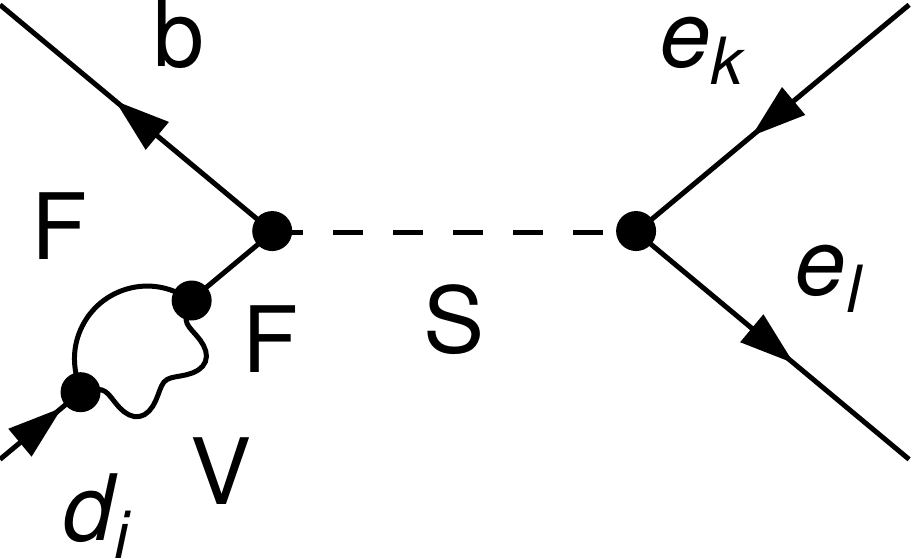}  \\
(a) & (b) \\
\includegraphics[width=3cm]{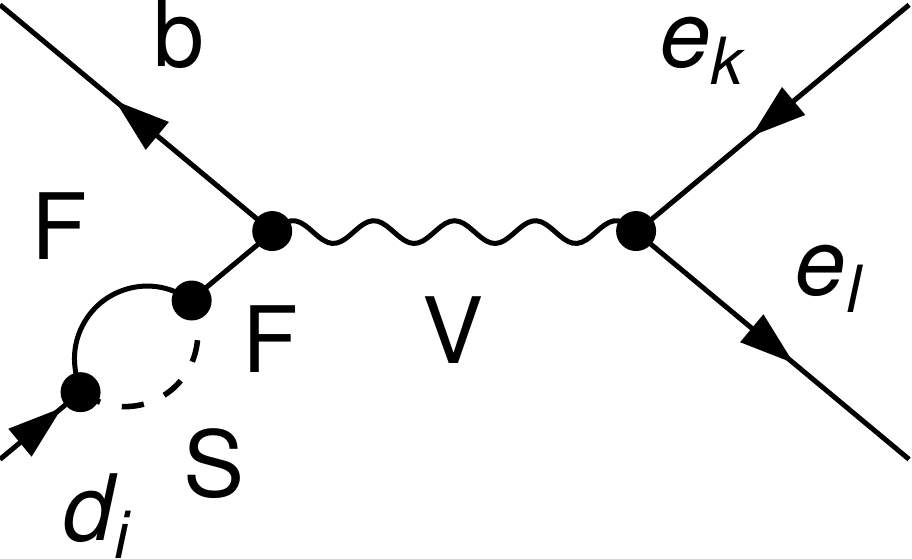} &    \includegraphics[width=3cm]{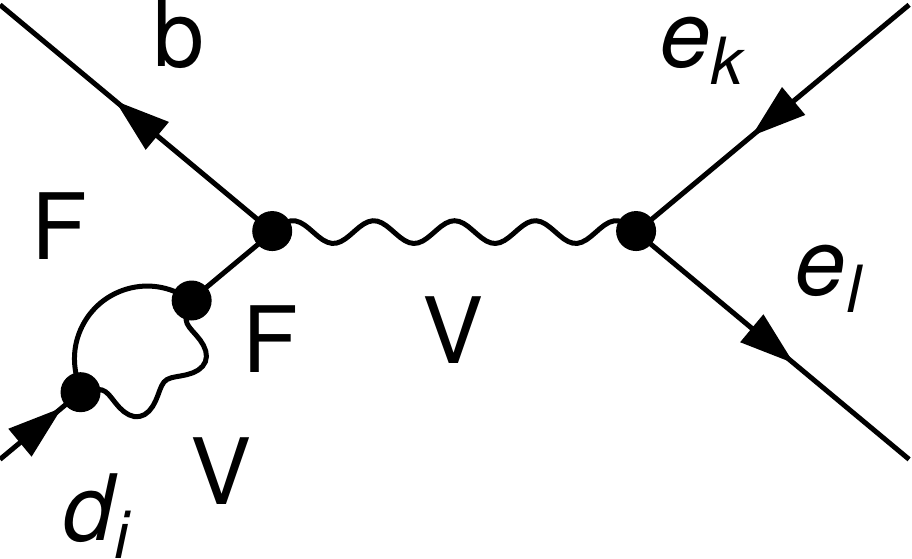}\\
(c) & (d) \\
\includegraphics[width=3cm]{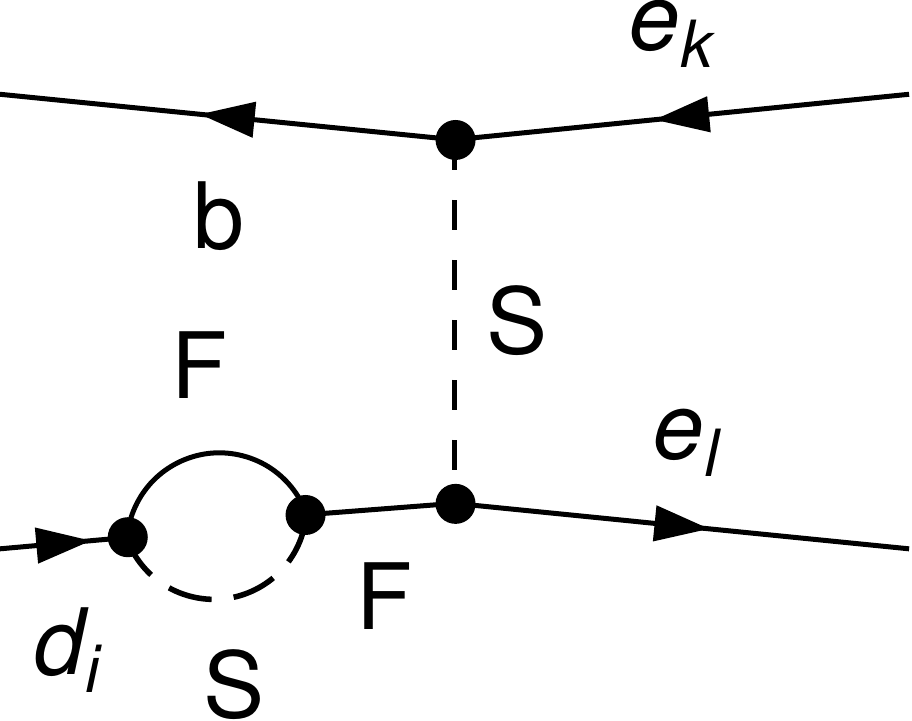} 
 & \includegraphics[width=3cm]{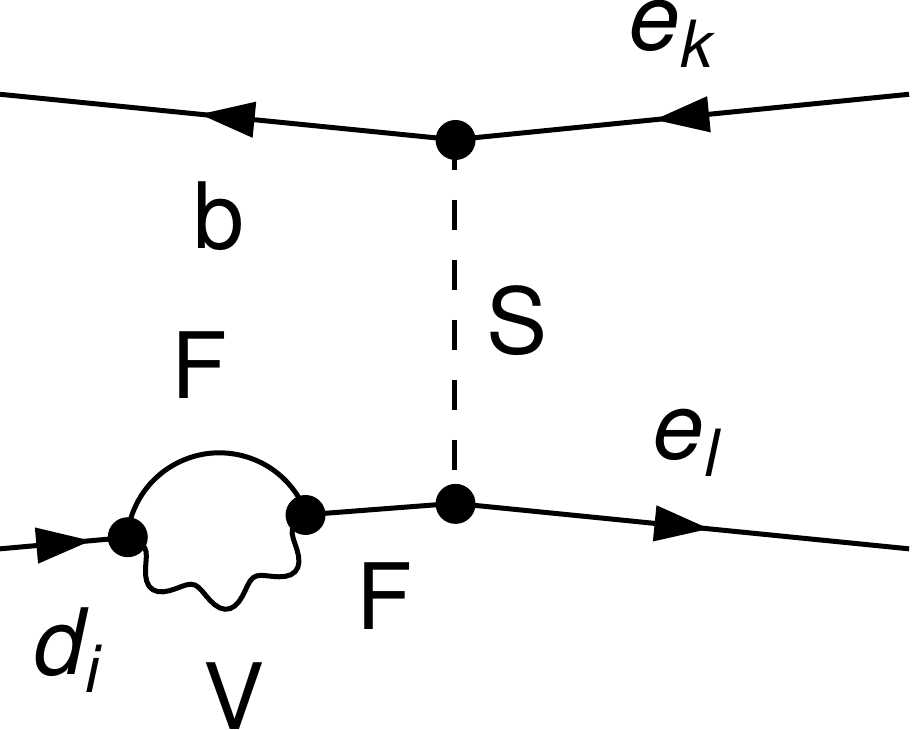} 
\\
(e) & (f) \\
\includegraphics[width=3cm]{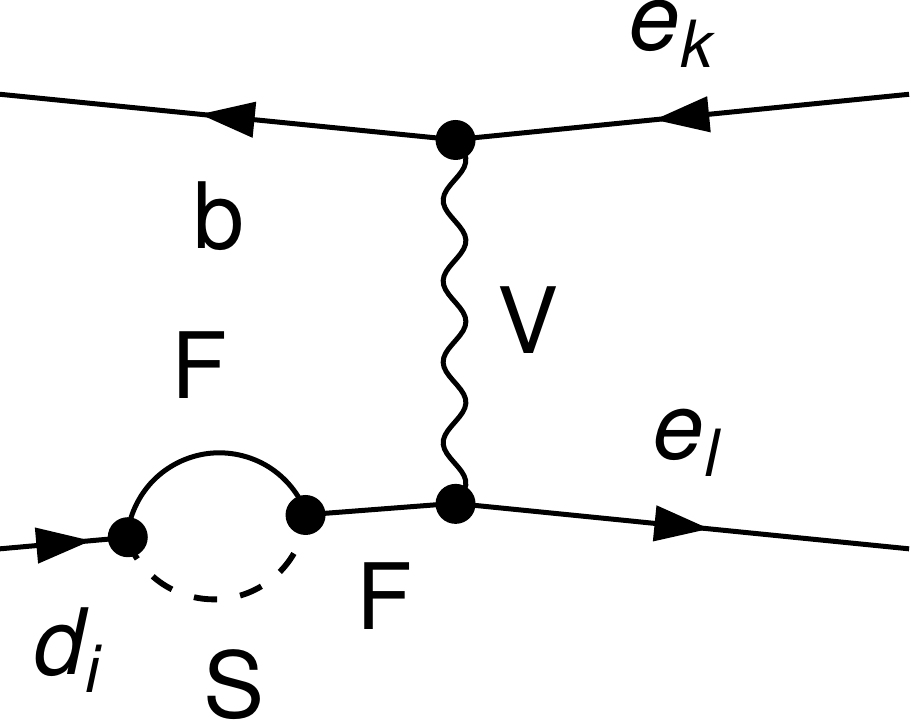}
 &   \includegraphics[width=3cm]{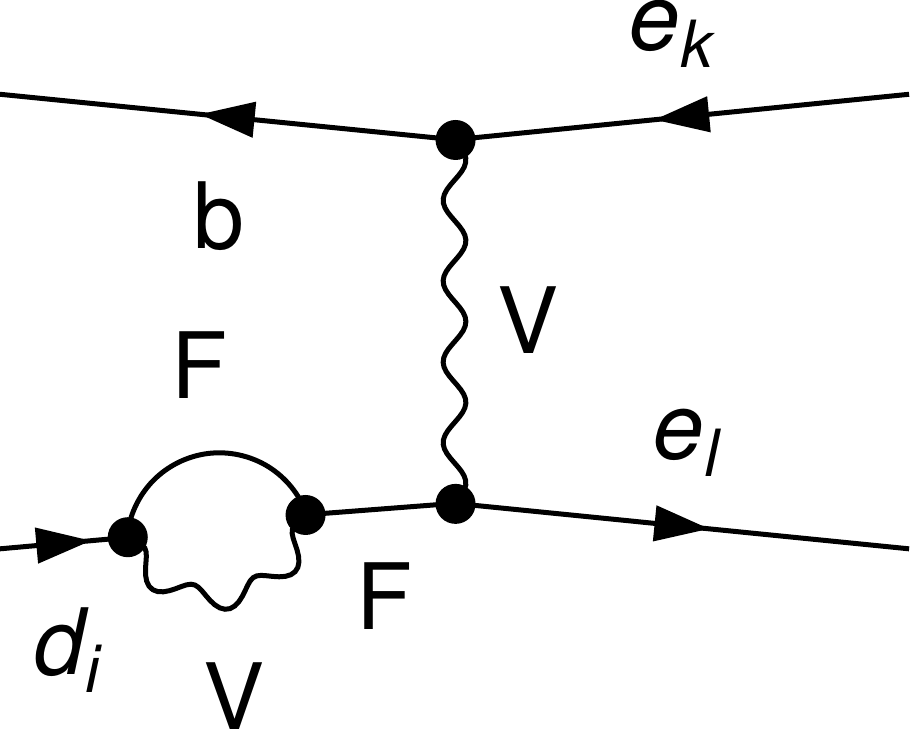}  
 \\
(g) & (h) \\
\includegraphics[width=3cm]{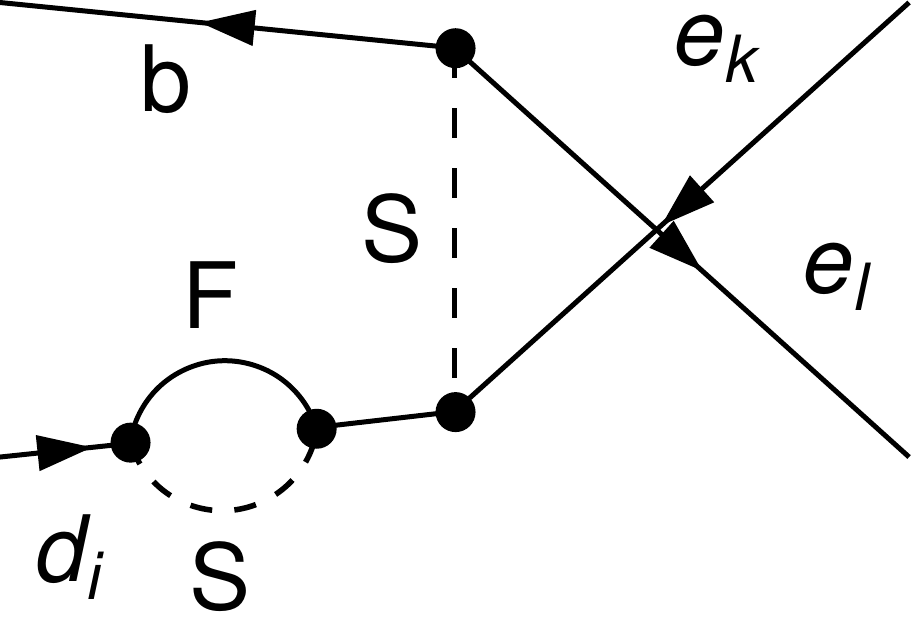}
 & \includegraphics[width=3cm]{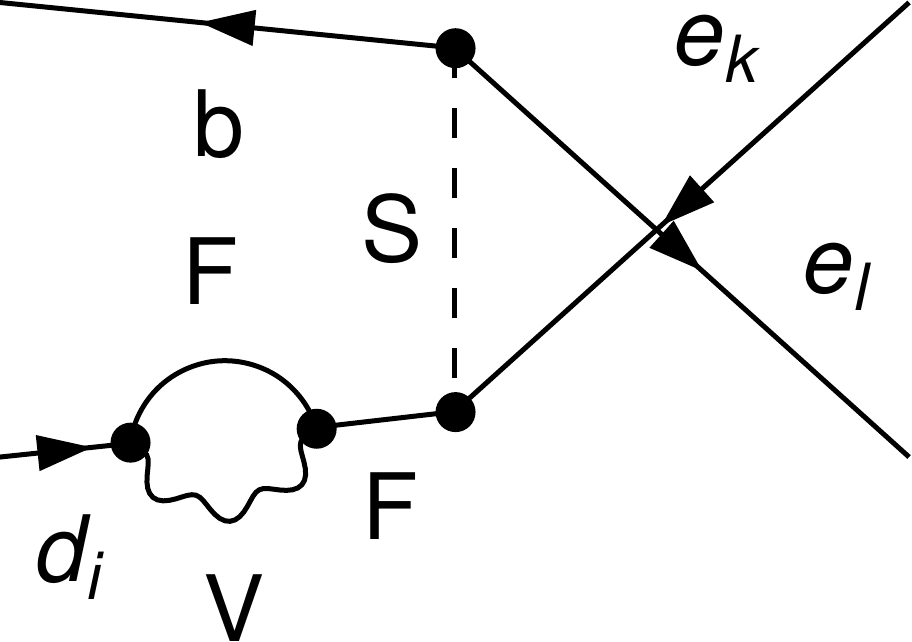} 
\\
(i) & (j) \\
\includegraphics[width=3cm]{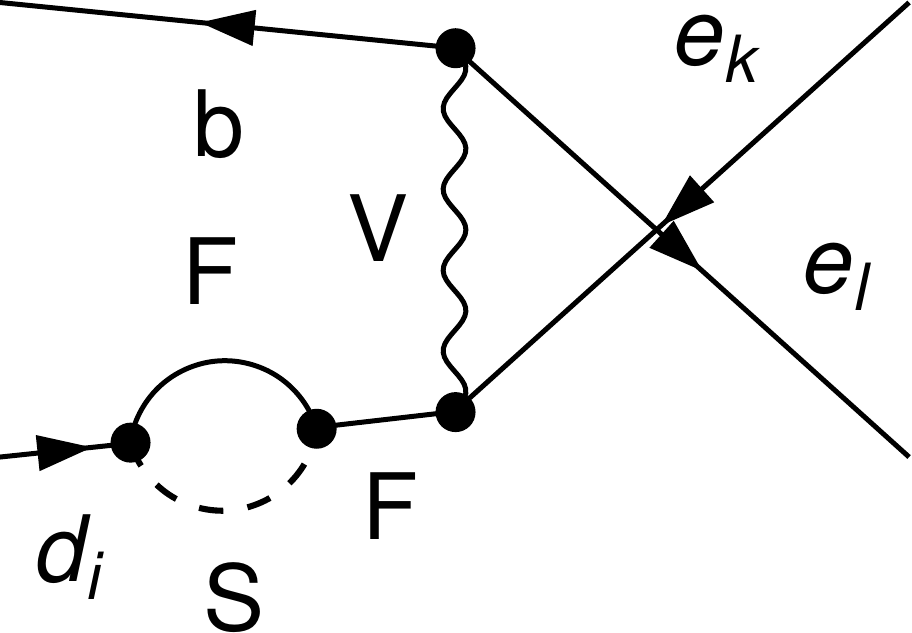}
& \includegraphics[width=3cm]{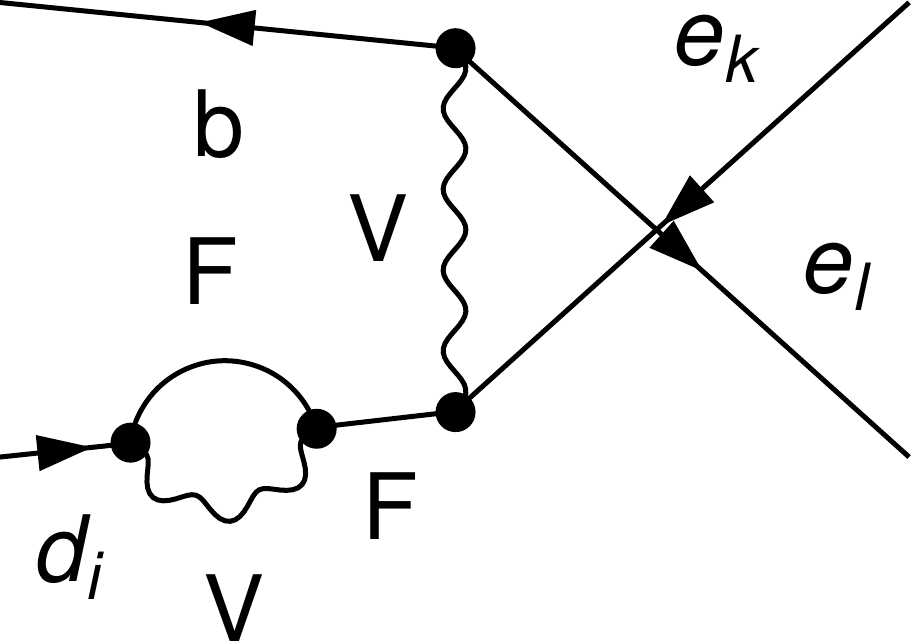} 
\\
(k) & (l) \\
\end{tabular}
\caption[Generic wave diagrams]{Generic wave diagrams. For every diagram there is a crossed version, where the loop attaches to the other external quark.}
\label{fig:wave}
\end{figure}
\begin{figure}[htpb]
  \centering
  \includegraphics[width=4cm]{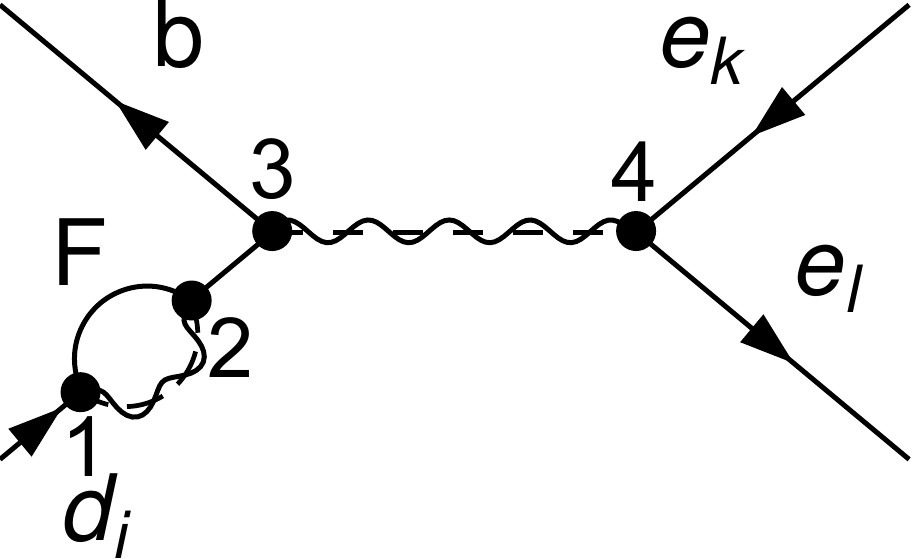}
  \includegraphics[width=4cm]{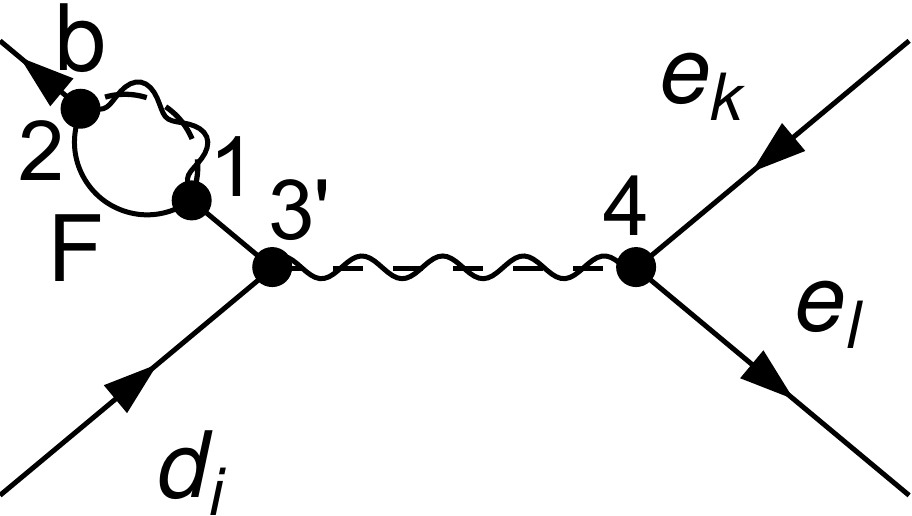} \\
  \includegraphics[width=4cm]{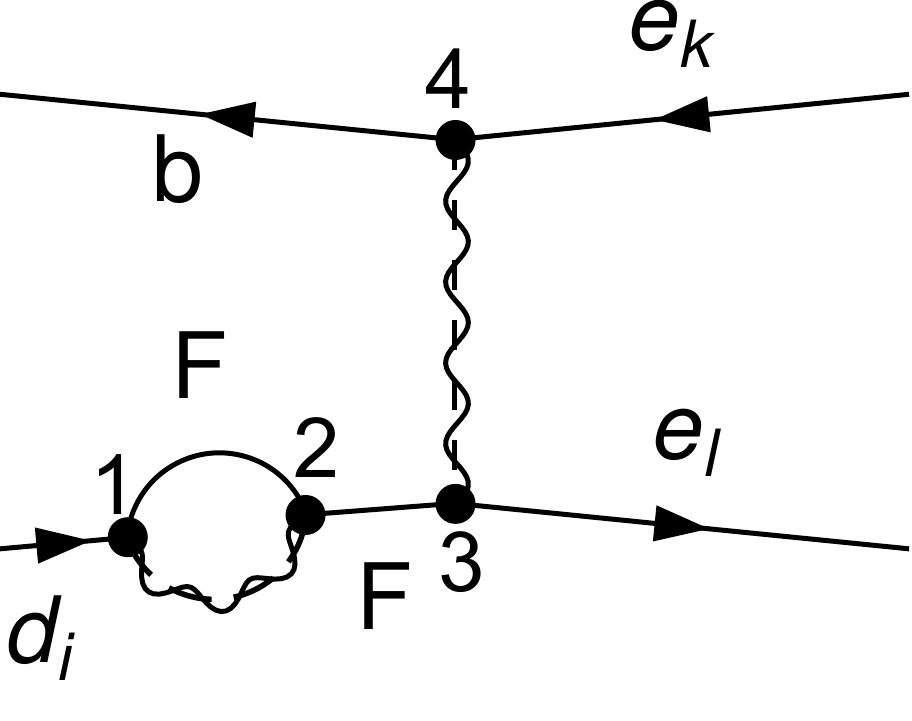}
  \includegraphics[width=4cm]{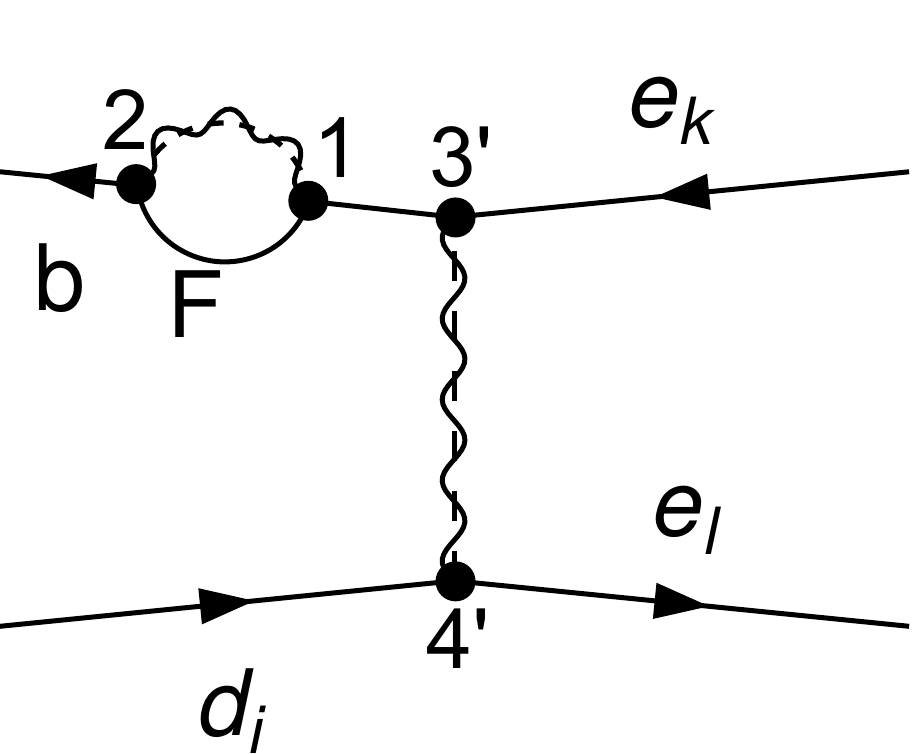} \\
  \includegraphics[width=4cm]{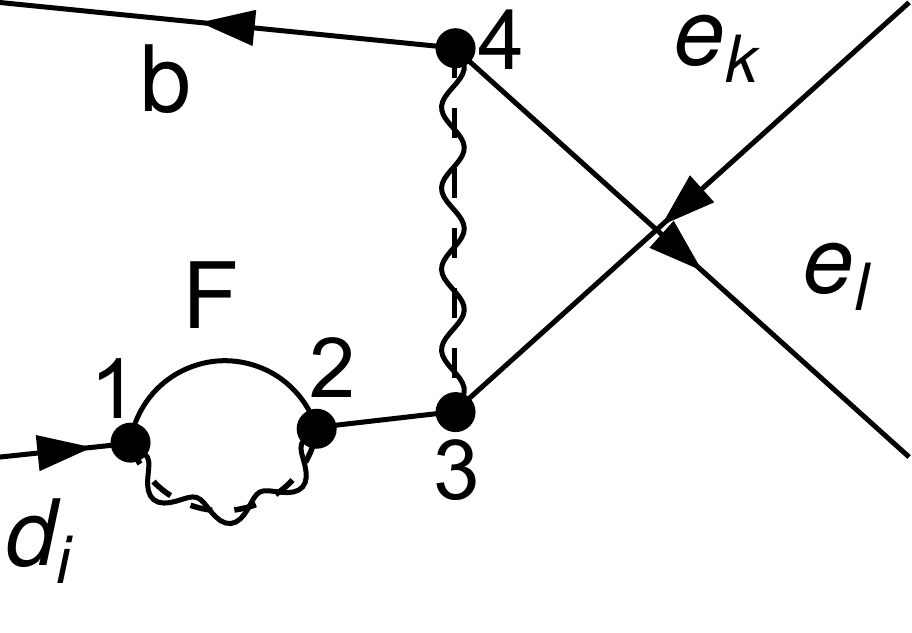}
  \includegraphics[width=4cm]{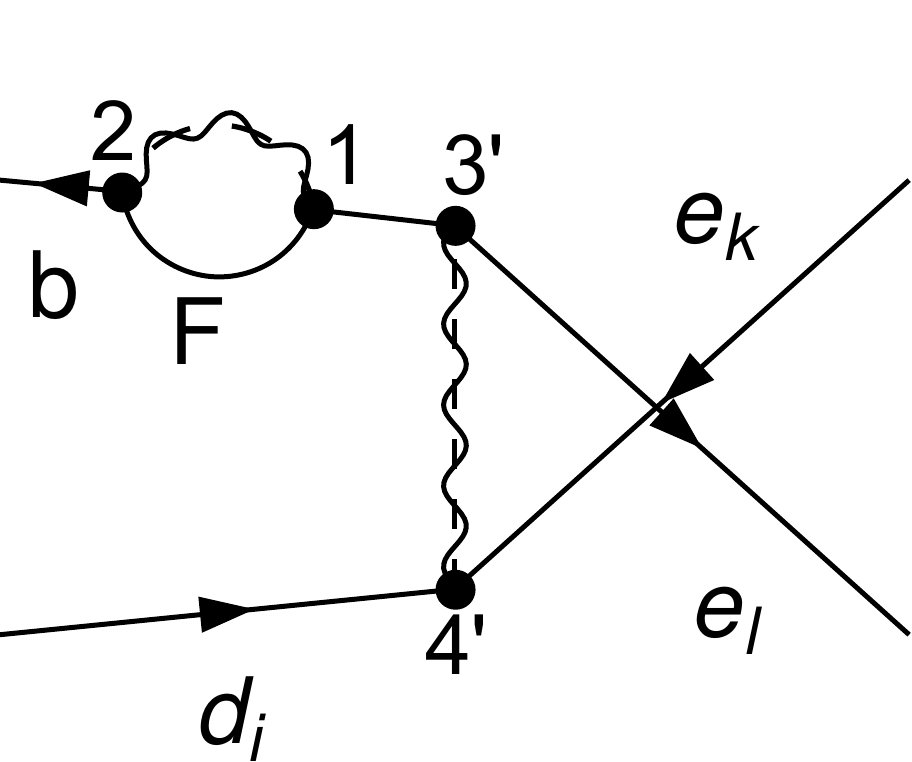}
  \caption{Generic wave diagram vertex numbering}
  \label{fig:wavenumbering}
\end{figure}
The generic wave diagrams are given in Figure~\ref{fig:wave}. The internal
quark which attaches to the vector or scalar propagator has generation index
$n$. Couplings that depend on $n$ carry it as an additional index. The chiral
vertices are parametrized as in
eqs.~(\ref{eq:chiralvertices1})-(\ref{eq:chiralvertices2}) with $A=1,B=2$ for
vertex 1, $A=3,B=4$ for vertex 2, $A=5,B=6$ for vertex 3 and $A=7,B=8$ for
vertex 4, see also Figure~\ref{fig:wavenumbering} for the numbering of the
vertices. If a vertex is labelled $3^\prime$ for instance, the corresponding couplings
are $G_5^\prime, G_6^\prime$. Furthermore, we define the following abbreviations:
\begin{align}
  \label{eq:wavebubbleSapp}
  f_{S1} &= \frac{1}{m_n^2-\mlisq} \kl{-M_F(G_1G_{3n}m_n+G_2G_{4n}\mli)B_0^{(i)}+(G_2G_{3n}m_n\mli+G_1G_{4n}\mlisq)B_1^{(i)}},
  \\
  f_{S2} &= \frac{1}{\mljsq-m_n^2}\kl{M_F(G_{2n}G_4\mlj+G_{1n}G_3m_n)B_0^{(j)}-(G_{2n}G_3\mljsq+G_{1n}G_4\mlj
    m_n)B_1^{(j)}}, \\
  \tilde f_{S2} &= \frac{1}{\mljsq-m_n^2}\kl{M_F(G_{1n}G_3\mlj+G_{2n}G_4 m_n)B_0^{(j)}-(G_{1n}G_4\mljsq+G_{2n}G_3\mlj
    m_n)B_1^{(j)}}, \\
\label{eq:wavebubbleVapp}
  f_{V1} &= \frac{1}{m_n^2-\mlisq}\kl{2M_F(G_1G_{4n} m_n+G_2G_{3n}\mli)B_0^{(i)}+(G_2G_{4n}m_n\mli+G_1G_{3n}\mlisq)B_1^{(i)}}, \\
  f_{V2} &= \frac{1}{\mljsq-m_n^2}\kl{2M_F(G_{2n}G_3m_j+G_{1n}G_4 m_n)B_0^{(j)}+(G_{2n}G_4\mljsq+G_{1n}G_3\mlj m_n)B_1^{(j)}},  \\
\tilde f_{V2} &= \frac{1}{\mljsq-m_n^2}\kl{2M_F(G_{1n}G_4m_j+G_{2n}G_3 m_n)B_0^{(j)}+(G_{1n}G_3\mljsq+G_{2n}G_4\mlj m_n)B_1^{(j)}}.
\end{align}
The $m_i,m_j$ are the quark masses and $B_{0,1}^{(i)} = B_{0,1}(m_i^2,M_F^2,M_S^2)$ (or $M_V^2$ instead of $M_S^2$). $m_n$ is the mass of the internal quark with generation index $n$. Couplings that involve the internal quark are also labelled with $n$ (e.g. $G_{3n}$). Using these conventions, the contributions to the Wilson coefficients are
\begin{eqnarray}
C^{(a)}_{SLL}&=& \frac{G_7}{M_{S0}^2-s} \kl{ G_{5n} f_{S1}+ G_{5n}^\prime f_{S2} } \\
C^{(c)}_{VLL}&=& \frac{G_7}{M_{V0}^2-s} \kl{ -G_{5n} f_{S1} - G_{5n}^\prime \tilde f_{S2} }  \to \frac{G_7}{M_{V0}^2-s} G_5 G_1G_4B_1(0,M_F,M_S)  \\
C^{(b)}_{SLL}&=&\frac{2G_7}{M_{S0}^2-s} \kl{ G_{5n} f_{V1} - G_{5n}^\prime f_{V2} } \\
C^{(d)}_{VLL}&=&\frac{2G_7}{M_{V0}^2-s} \kl{-G_{5n} f_{V1} + G_{5n}^\prime \tilde f_{V2}} \to \frac{2G_7}{M_{V0}^2-s} G_5 G_1G_3B_1(0,M_F,M_V)  \\
 C^{(e)}_{SLL} &=& \frac{-1}{2(M_{S0}^2-t)} \kl{ G_{5n} G_7 f_{S1} + G_{5n}^\prime G_7^\prime f_{S2} } \\
 C^{(e)}_{VLR} &=& \frac{ -1}{2(M_{S0}^2-t)} \kl{ G_{5n} G_8 f_{S1} + G_{6n}^\prime G_7^\prime \tilde f_{S2} } \\
C^{(g)}_{SLR}&=& \frac{+2}{M_{V0}^2-t} \kl{ G_{5n} G_8 f_{S1} + G_{6n}^\prime  G_7^\prime f_{S2} } \\
C^{(g)}_{VLL}&=&\frac{-1}{M_{V0}^2-t} \kl{ G_{5n}G_7 f_{S1} + G_{5n}^\prime  G_7^\prime \tilde f_{S2} } \\ 
    C^{(f)}_{SLL}&=& \frac{-1}{M_{S0}^2-t} \kl{ G_{5n}G_7 f_{V1} - G_{5n}^\prime G_7^\prime f_{V2} }  \\
    C^{(f)}_{VLR} &=& \frac{-1}{M_{S0}^2-t} \kl{ G_{5n}G_8 f_{V1} - G_{6n}^\prime   G_7^\prime \tilde f_{V2} } \\
    C^{(h)}_{SLR} &=& \frac{+4}{M_{V0}^2-t} \kl{ G_{5n}G_8 f_{V1} - G_{6n}^\prime G_7^\prime f_{V2} } \\
    C^{(h)}_{VLL} &=& \frac{+2}{M_{V0}^2-t} \kl{ - G_{5n}G_7 f_{V1} + G_{5n}^\prime     G_7^\prime \tilde f_{V2} }\\
    C^{(i)}_{SLL} &=& \frac{-1}{2(M_{S0}^2-u)} \kl{ G_{5n}G_7 f_{S1} + G_{5n}^\prime  G_7^\prime f_{S2} } \\
C^{(i)}_{VLL} &=& \frac{+1}{2(M_{S0}^2-u)} \kl{ G_{5n}G_8 f_{S1} + G_{6n}^\prime  G_{7}^\prime \tilde f_{S2} }\\
C^{(k)}_{SLR}&=& \frac{-2}{M_{V0}^2-u}\kl{G_{6n}G_8 f_{S1}+G_{6n}^\prime G_8^\prime f_{S2}} \\
C^{(k)}_{VLR} &=& \frac{-1}{M_{V0}^2-u} \kl{G_{6n}G_7 f_{S1}+G_{5n}^\prime  G_8^\prime \tilde f_{S2}} \\
    C^{(j)}_{SLL}&=& \frac{-1}{M_{S0}^2-u} \kl{ G_{5n} G_7 \tilde f_{V1} -   G_{5n}^\prime G_7^\prime f_{V2}  } \\
    C^{(j)}_{VLL} &=& \frac{-1}{M_{S0}^2-u} \kl{ - G_{5n}G_8 \tilde f_{V1} +   G_{6n}^\prime G_7^\prime \tilde f_{V2}   } \\
    C^{(l)}_{SLR} &=& \frac{-4}{M_{V0}^2-u} \kl{ G_{6n}G_8 \tilde f_{V1} -    G_{6n}^\prime G_8^\prime f_{V2} }\\
    C^{(l)}_{VLR} &=& \frac{-2}{M_{V0}^2-u} \kl{ G_{6n}G_7 \tilde f_{V1} -    G_{5n}^\prime G_8^\prime \tilde f_{V2} }
\end{eqnarray}

\subsection{Penguin Contributions}
\label{sec:penguinB}
\begin{figure}[htpb]
  \centering
  \includegraphics[width=4cm]{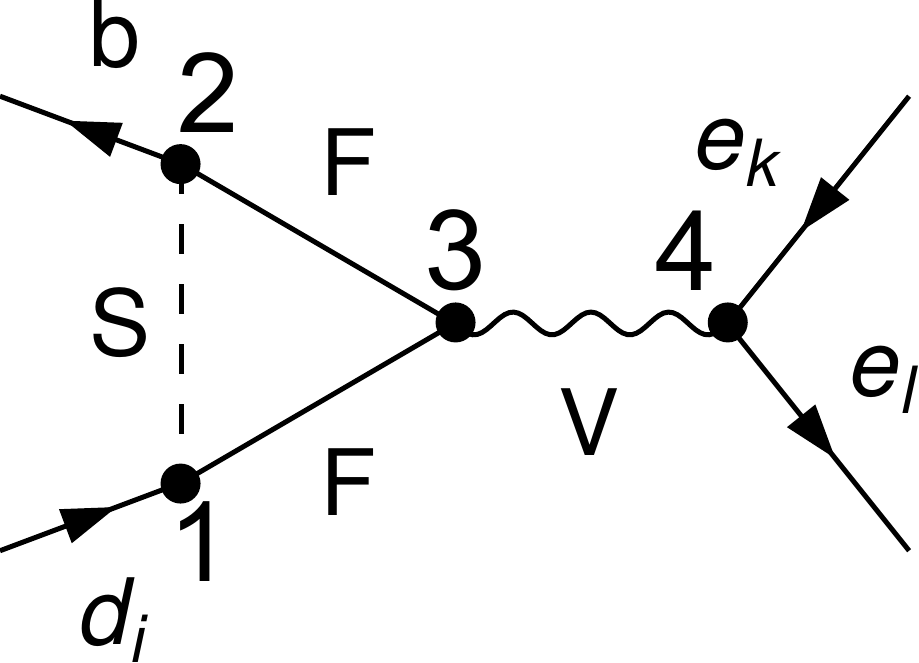}
  \caption{Vertex number conventions for a representative penguin diagram}
  \label{fig:penguinvertex}
\end{figure}
\begin{figure}[hbt]
 \centering
 \begin{tabular}{cccc}
  \includegraphics[width=3cm]{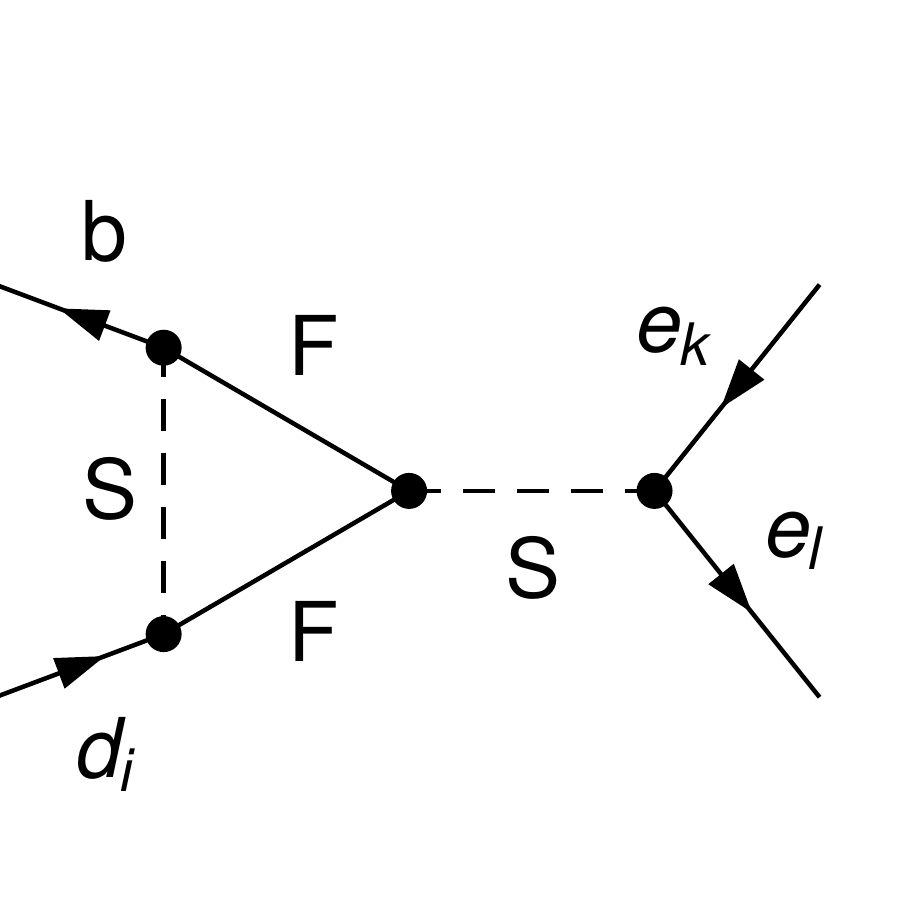} &  \includegraphics[width=3cm]{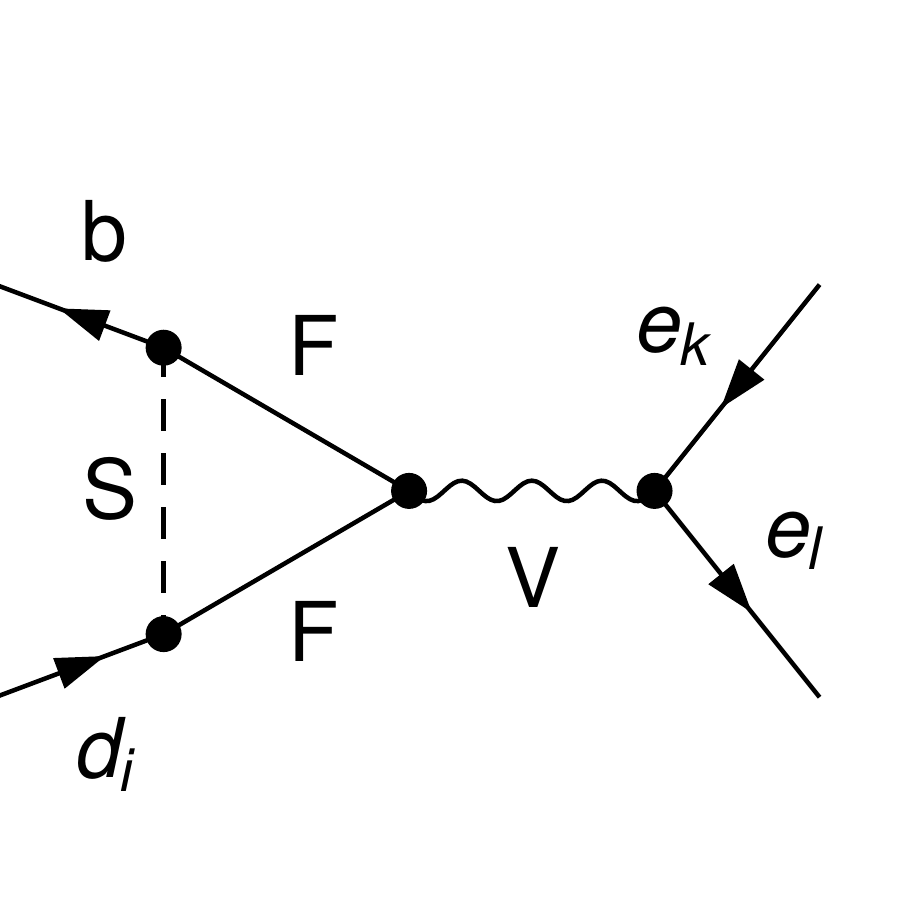} & \includegraphics[width=3cm]{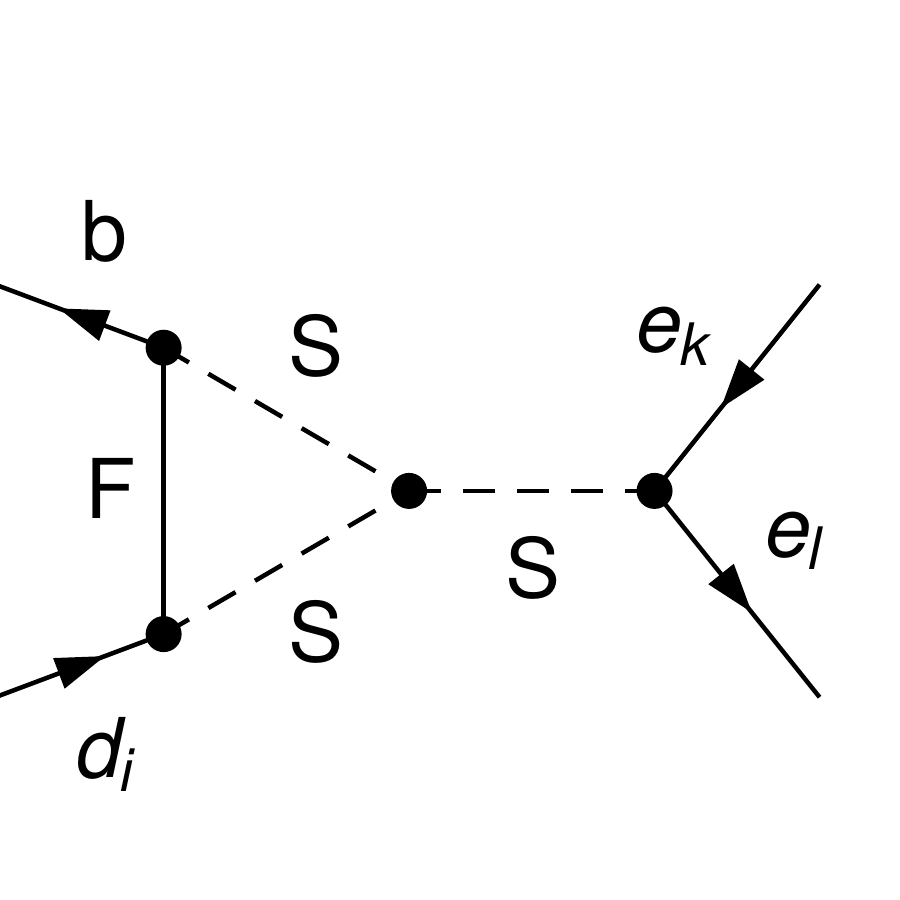} &  \includegraphics[width=3cm]{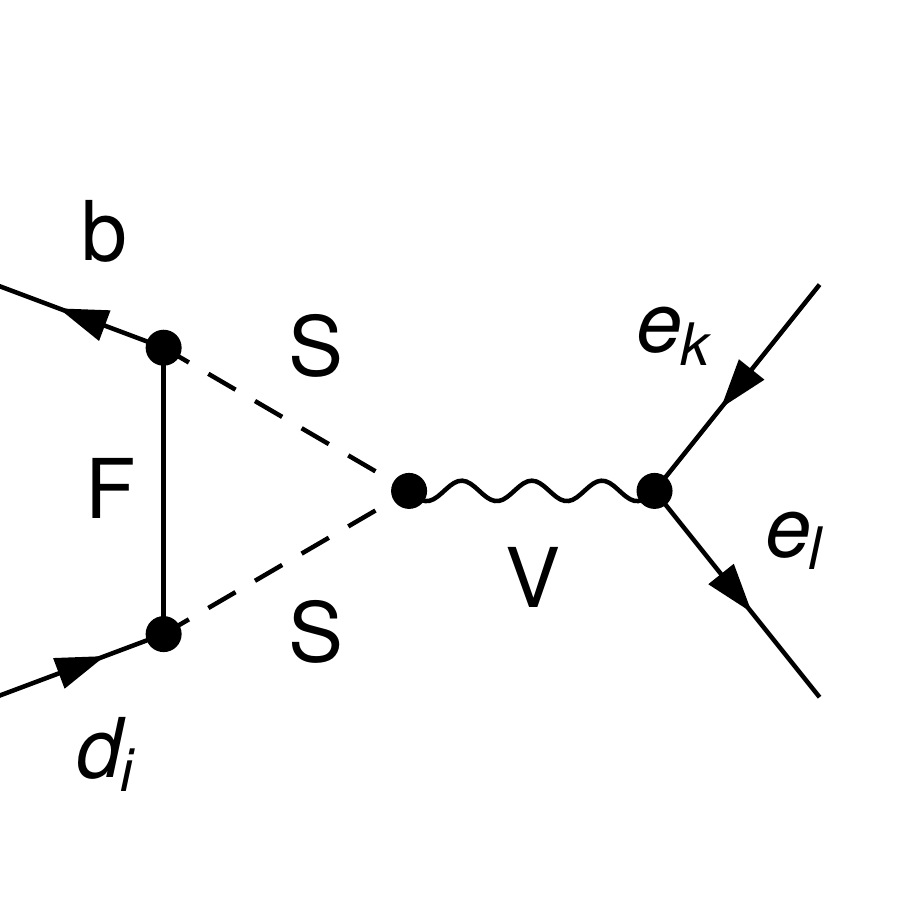} \\
  (a) & (b) &  (c) & (d) \\
  \includegraphics[width=3cm]{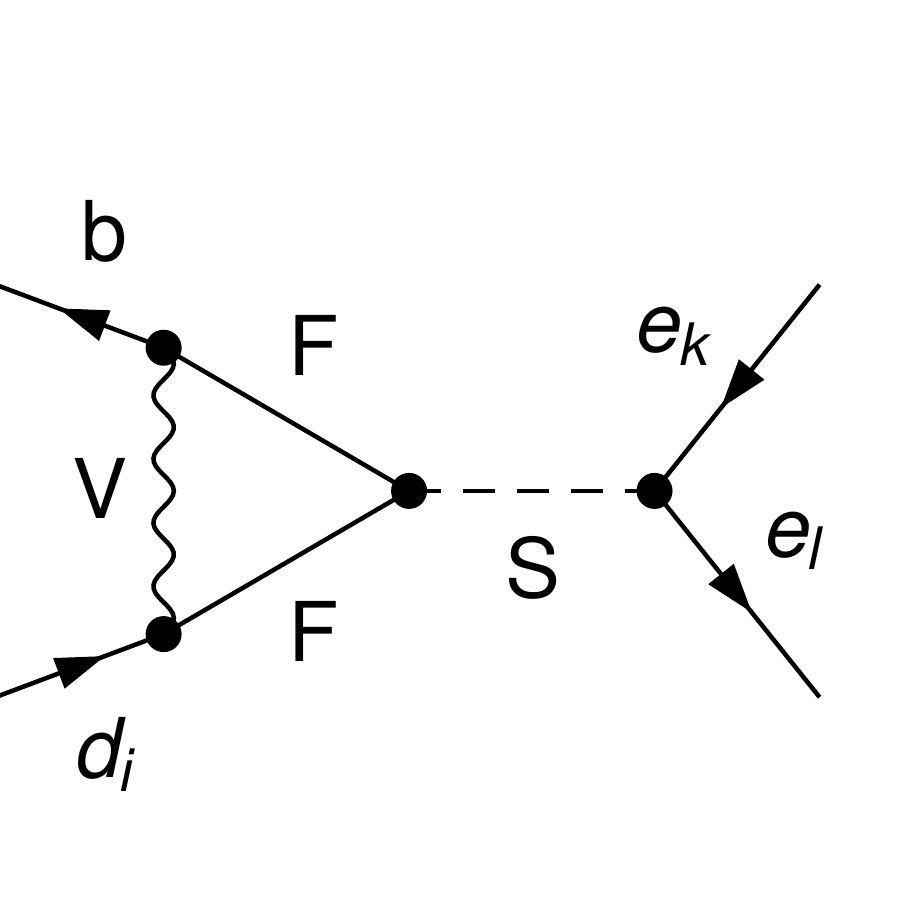} &  \includegraphics[width=3cm]{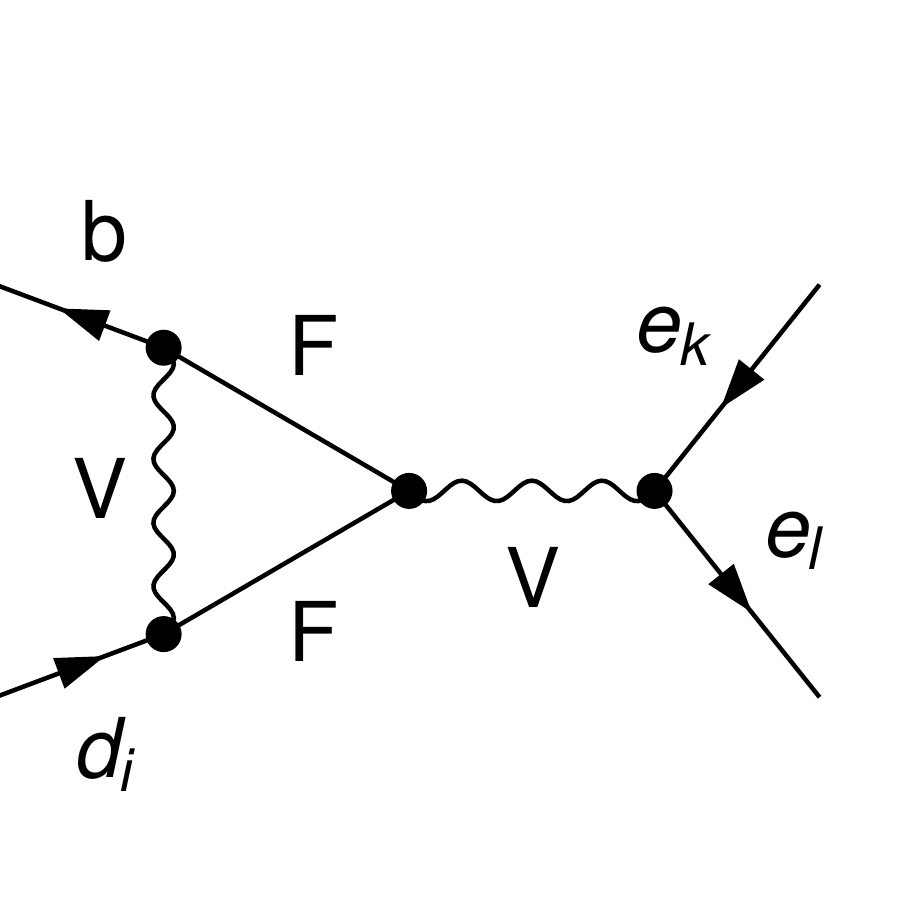} & 
  \includegraphics[width=3cm]{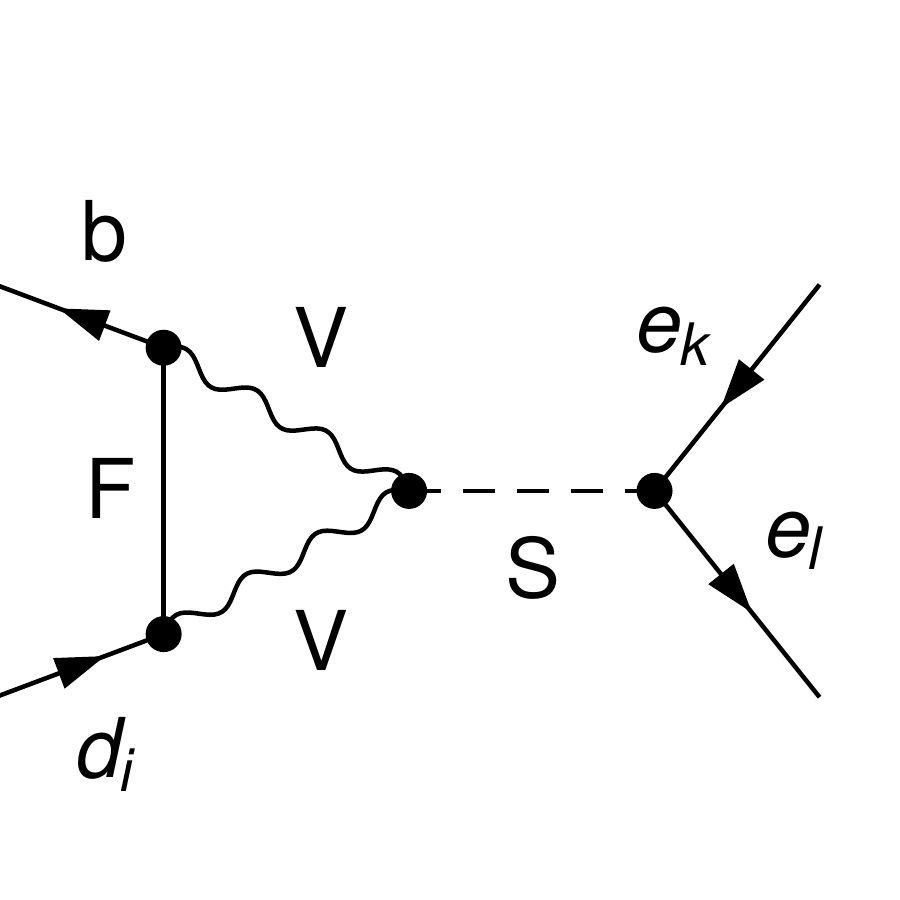} &  \includegraphics[width=3cm]{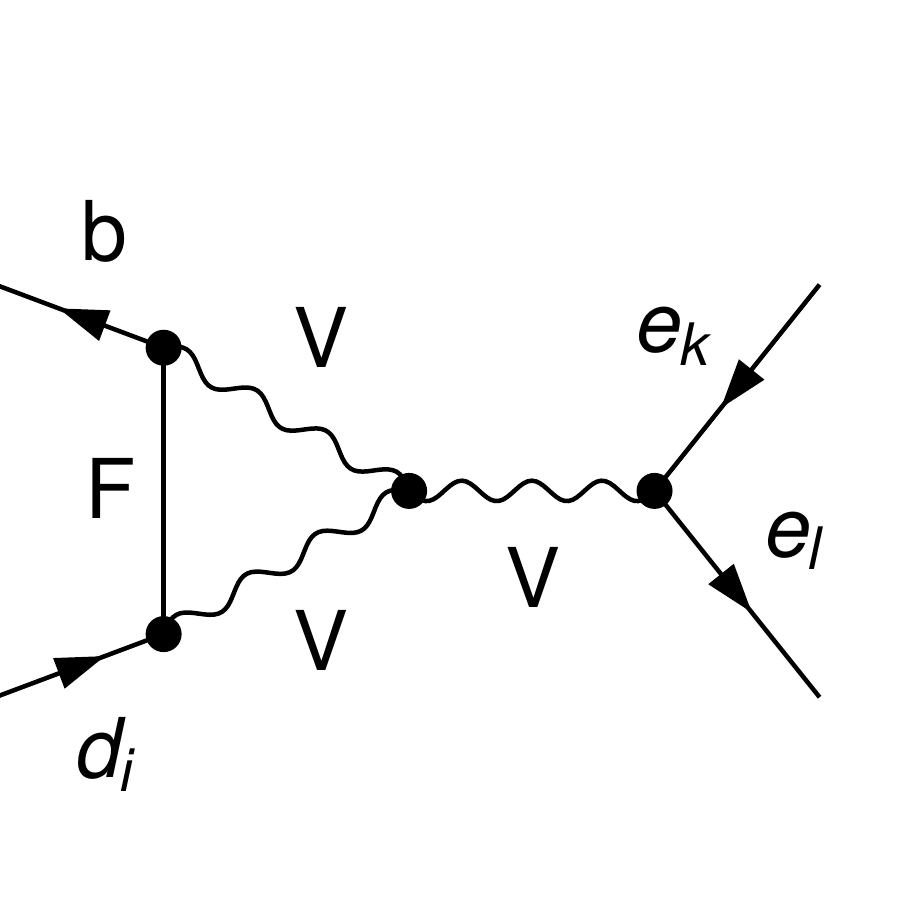} \\
  (e) & (f) &  (g) & (h) \\
  \includegraphics[width=3cm]{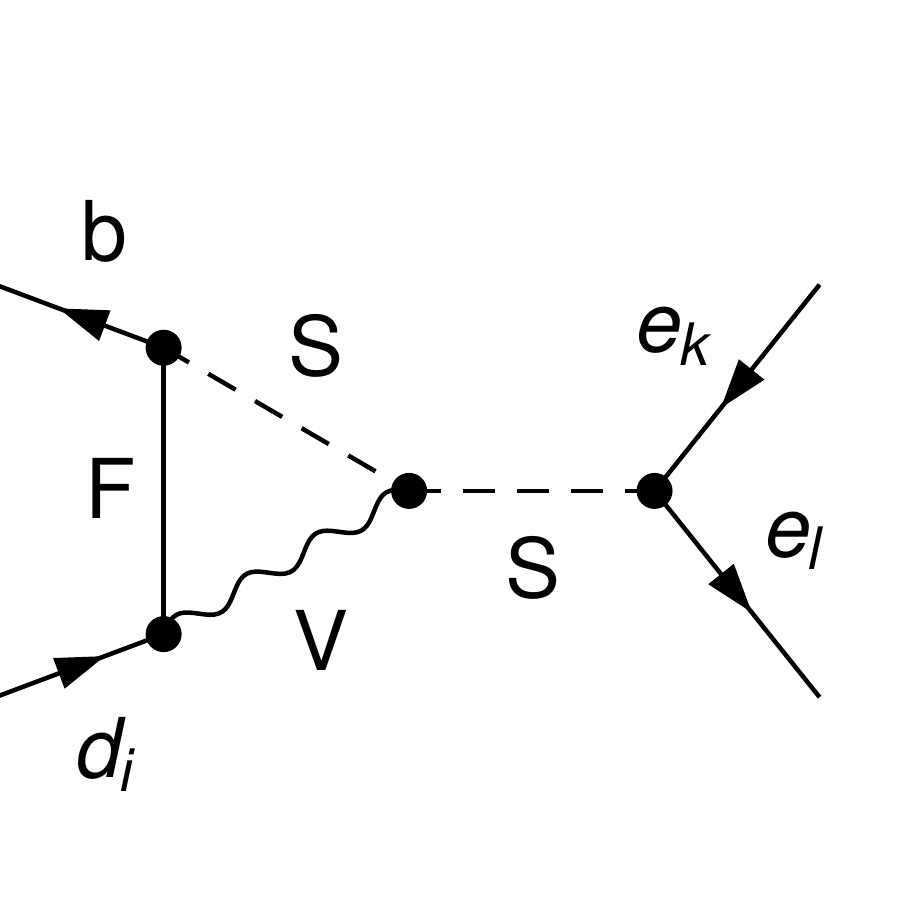} & \includegraphics[width=3cm]{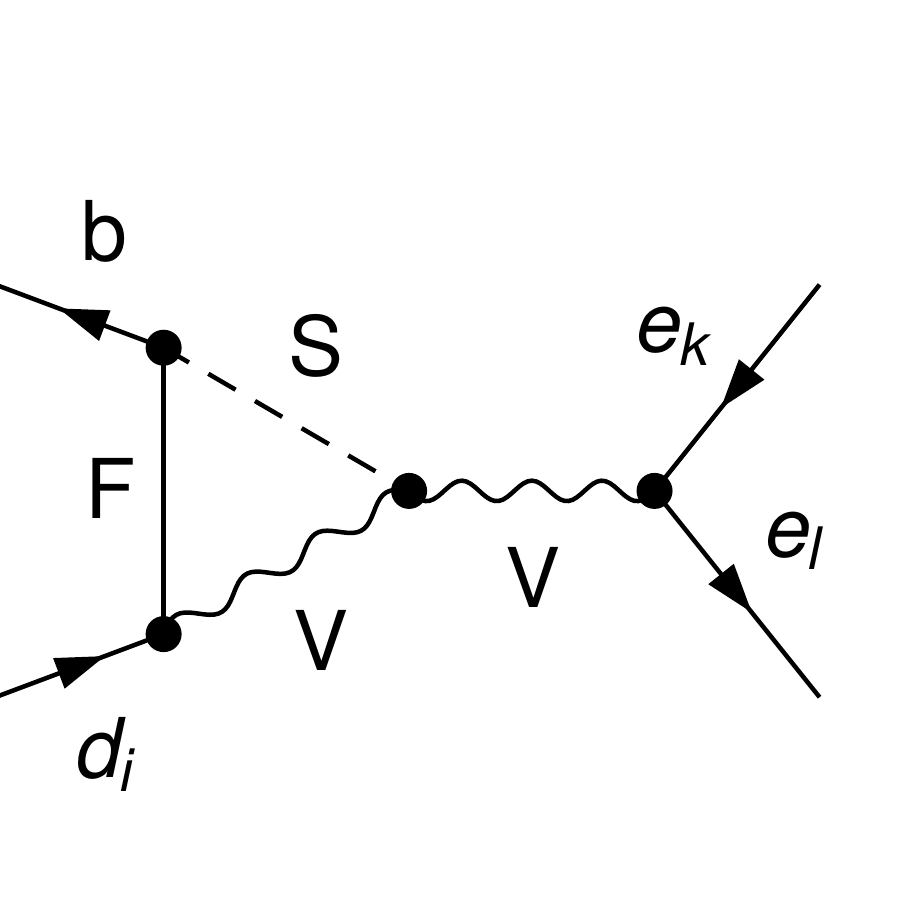} &
  \includegraphics[width=3cm]{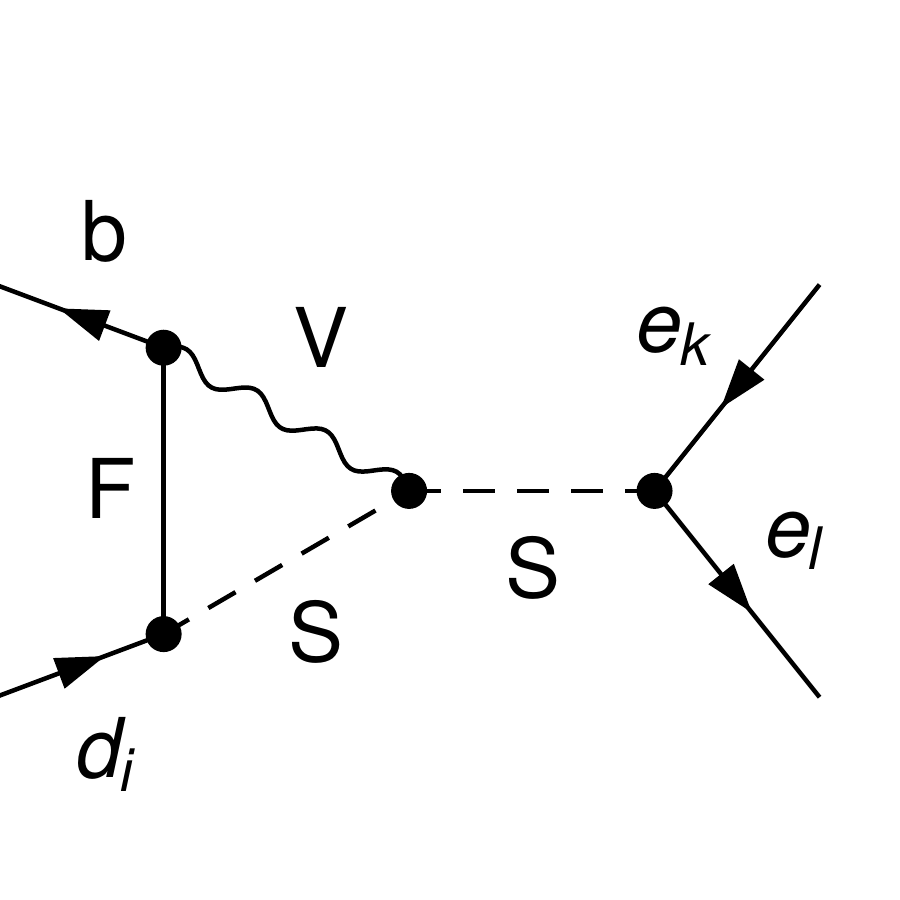} &  \includegraphics[width=3cm]{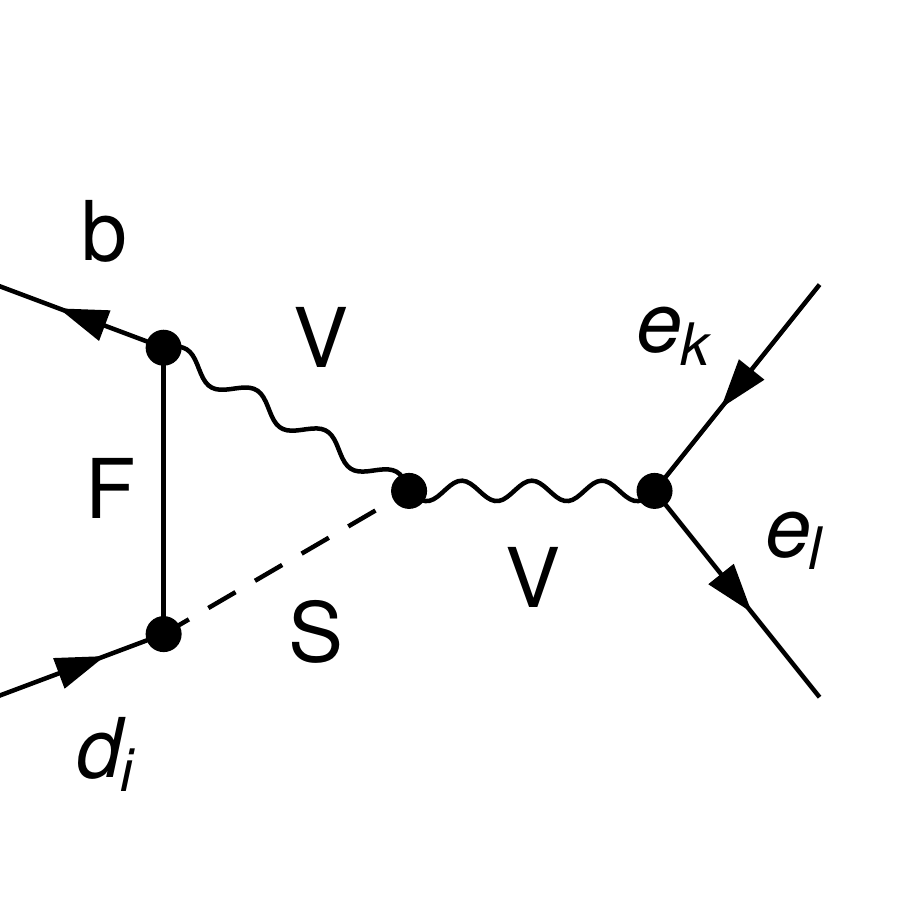} \\
    (i) & (j) &  (k) & (l) 
 \end{tabular}
\caption{Generic penguin diagrams}
\label{fig:penguin}
\end{figure}
Diagrams with scalar propagators have $C_{VXY}=0$ and those with vector propagators have \mbox{$C_{SXY}=0$}. The vertex number conventions are given in fig.~\ref{fig:penguinvertex} and all possible diagrams are depicted in Figure~\ref{fig:penguin}. The chiral vertices are parametrized as in eqs.~(\ref{eq:chiralvertices1})-(\ref{eq:chiralvertices2}) with $A=1,B=2$ for vertex 1, $A=3,B=4$ for vertex 2 and $A=7,B=8$ for vertex 4. Vertex 3 can be a chiral vertex, in this case $A=5,B=6$ is used. Otherwise, we will denote it with a index $5$ and give as additional subscript the kind of vertex. The contributions to the Wilson coefficients from these diagrams read
\begin{align}
  C^{(a)}_{SLL}&= \frac 1{M_{S0}^2-s}G_1G_3G_7\kl{G_6 B^{(a,b)}_0+(G_5M_{F1}M_{F2}+G_6M_S^2)C^{(a,b)}_0 } \\
  C^{(a)}_{SLR}&= \frac 1{M_{S0}^2-s}G_1G_3G_8\kl{G_6 B^{(a,b)}_0+(G_5M_{F1}M_{F2}+G_6M_S^2)C^{(a,b)}_0 } \\
  C^{(b)}_{VLL}&= \frac 1{M_{V0}^2-s} G_1G_4G_7 \kl{G_6B^{(a,b)}_0+(-G_5M_{F1}M_{F2}+G_6M_S^2)C^{(a,b)}_0-2G_6C^{(a,b)}_{00} } \\
  C^{(b)}_{VLR}&= \frac 1{M_{V0}^2-s} G_1G_4G_8 \kl{G_6B^{(a,b)}_0+(-G_5M_{F1}M_{F2}+G_6M_S^2)C^{(a,b)}_0-2G_6C^{(a,b)}_{00} } \\
  C^{(c)}_{SLL}&= \frac 1{M_{S0}^2-s}G_1G_3G_{5,SSS}G_7 M_F C^{(c,d)}_0 \\
  C^{(c)}_{SLR}&= \frac 1{M_{S0}^2-s}G_1G_3G_{5,SSS}G_8M_F C^{(c,d)}_0 \\
  C^{(d)}_{VLL}&= - \frac 2{M_{V0}^2-s} G_1G_4G_{5,SSV} G_7C^{(c,d)}_{00}\\ 
  C^{(d)}_{VLR}&= - \frac 2{M_{V0}^2-s} G_1G_4G_{5,SSV} G_8 C^{(c,d)}_{00} \\
  C^{(e)}_{SLL}&= -\frac 4{M_{S0}^2-s} G_1G_4G_7 \kl{G_5B^{(e,f)}_0+(G_6M_{F1}M_{F2}+G_5M_V^2)C^{(e,f)}_0} \\
  C^{(e)}_{SLR}&= -\frac 4{M_{S0}^2-s} G_1G_4G_8 \kl{G_5B^{(e,f)}_0+(G_6M_{F1}M_{F2}+G_5M_V^2)C^{(e,f)}_0} \\
  C^{(f)}_{VLL}&= \frac{2}{M_{V0}^2-s} G_1 G_3 G_7 \kl{G_5B^{(e,f)}_0+(-G_6M_{F1}M_{F2}+G_5M_V^2)C^{(e,f)}_0-2 G_5 C^{(e,f)}_{00}} \\
  C^{(f)}_{VLR}&= \frac{2}{M_{V0}^2-s} G_1 G_3 G_8  \kl{G_5B^{(e,f)}_0+(-G_6M_{F1}M_{F2}+G_5M_V^2)C^{(e,f)}_0-2 G_5 C^{(e,f)}_{00}} \\
    C^{(g)}_{SLL}&= \frac 4{M_{S0}^2-s} G_1G_4G_{5,SVV} G_7 M_{F} C^{(g,h)}_0\\
  C^{(g)}_{SLR}&= \frac 4{M_{S0}^2-s} G_1G_4G_{5,SVV} G_8 M_{F} C^{(g,h)}_0\\
  C^{(h)}_{VLL}&= -\frac 2{M_{V0}^2-s} G_1G_3G_{5,VVV}G_7  \kl{B^{(g,h)}_0+M_{F}^2C^{(g,h)}_0+2C^{(g,h)}_{00}} \\
  C^{(h)}_{VLR}&= -\frac 2{M_{V0}^2-s} G_1G_3G_{5,VVV}G_8  \kl{B^{(g,h)}_0+M_{F}^2C^{(g,h)}_0+2C^{(g,h)}_{00}} \\
  C^{(i)}_{SLL}&= \frac 1{M_{S0}^2-s} G_1G_3G_{5,SSV}  G_7 \kl{B^{(i-l)}_0+M_{F}^2C^{(i-l)}_0} \\
  C^{(i)}_{SLR}&= \frac 1{M_{S0}^2-s} G_1G_3G_{5,SSV}  G_8 \kl{B^{(i-l)}_0+M_{F}^2C^{(i-l)}_0} \\
 C^{(k)}_{SLL}&= -\frac 1{M_{S0}^2-s} G_1G_4G_{5,SSV}  G_7 \kl{B^{(i-l)}_0+M_{F}^2C^{(i-l)}_0} \\
  C^{(k)}_{SLR}&= -\frac 1{M_{S0}^2-s} G_1G_4G_{5,SSV}  G_8 \kl{B^{(i-l)}_0+M_{F}^2C^{(i-l)}_0} \\
  C^{(j)}_{VLL}&= \frac{1}{M_{V0}^2-s} G_1G_4G_{5,SVV}G_7 M_F C^{(i-l)}_0 \\
  C^{(j)}_{VLR}&= \frac{1}{M_{V0}^2-s} G_1G_4G_{5,SVV}G_8 M_F C^{(i-l)}_0 \\
  C^{(l)}_{VLL}&= \frac{1}{M_{V0}^2-s} G_1G_3G_{5,SVV}G_7 M_F C^{(i-l)}_0 \\
  C^{(l)}_{VLR}&= \frac{1}{M_{V0}^2-s} G_1G_3G_{5,SVV}G_8 M_F C^{(i-l)}_0 
\end{align}
Here, the arguments of the Passarino-Veltman integrals are as follows, with $s=M_{B^0_q}^2$:
\begin{align}
C_X^{(a,b)} &=C_X (s,0,0,M_{F2}^2,M_{F1}^2,M_{S}^2) \,\hspace{1cm}
B_X^{(a,b)}  = B_X(s,M_{F1}^2,M_{F2}^2)  \\
C_X^{(c,d)} & = C_X(0,s,0,M_{F}^2,M_{S1}^2,M_{S2}^2) \\
C_X^{(e,f)} &=C_X(s,0,0,M_{F2}^2,M_{F1}^2,M_{V}^2) \,\hspace{1cm}
B_X^{(e,f)} = B_X(s,M_{F1}^2,M_{F2}^2)\\
C_X^{(g,h)}&=C_X(0,s,0,M_{F}^2,M_{V1}^2,M_{V2}^2) \,\hspace{1cm}
B_X^{(g,h)}=B_X(s,M_{V1}^2,M_{V2}^2) \\
C_X^{(i-l)} &= C_X(0,s,0,M_{F}^2,M_{S}^2,M_{V}^2) \,\hspace{1cm}
B_X^{(i-l)} = B_X(s,M_{S}^2,M_{V}^2)
\end{align}

\subsection{Box Contributions}
\begin{figure}[htpb]
  \centering
  \begin{tabular}{ccc}
  \includegraphics[width=4cm]{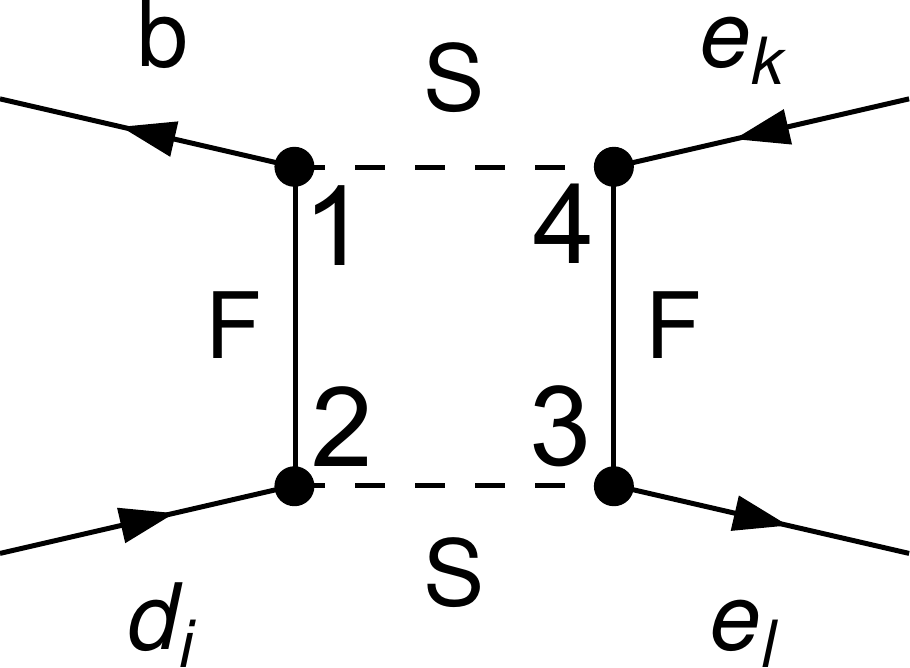} & \includegraphics[width=4cm]{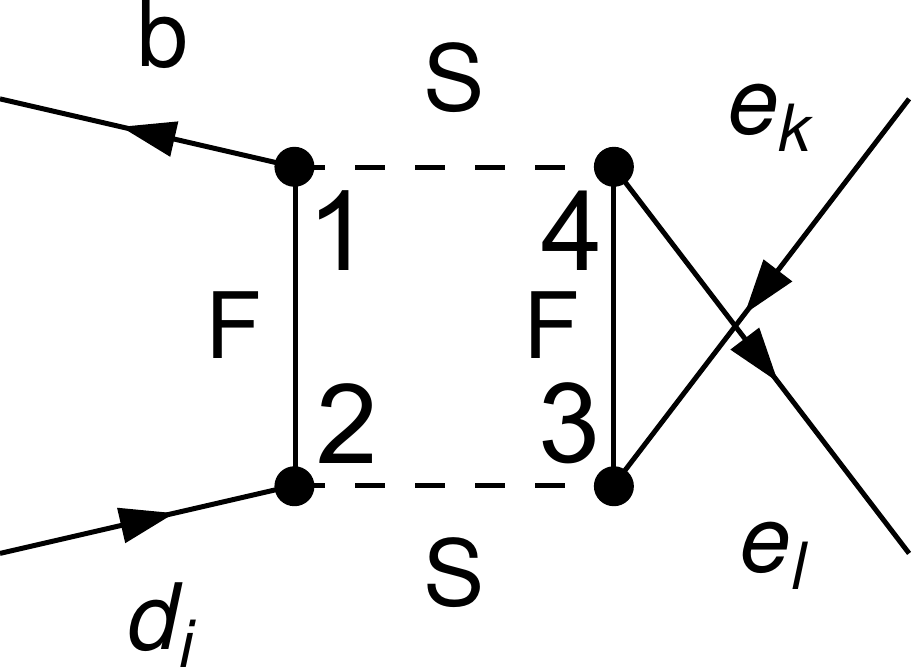} & \includegraphics[width=4cm]{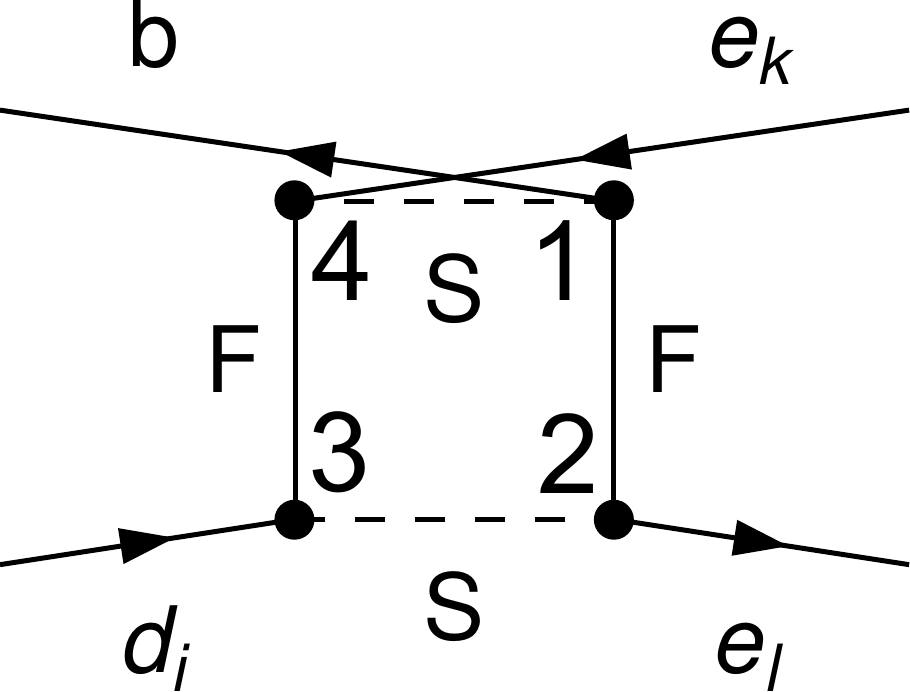} \\
  \end{tabular}
  \caption{Vertex number conventions for a set of representative box diagrams}
  \label{fig:boxvertex}
\end{figure}
\begin{figure}[hbt]
 \centering
 \begin{tabular}{ccc}
  \includegraphics[width=3cm]{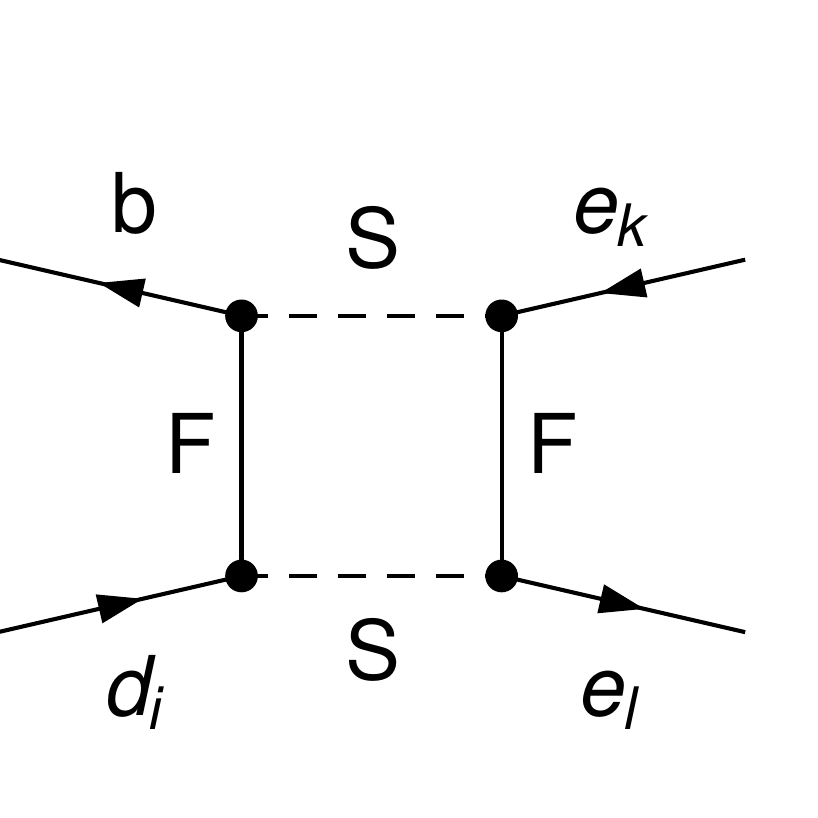} &   \includegraphics[width=3cm]{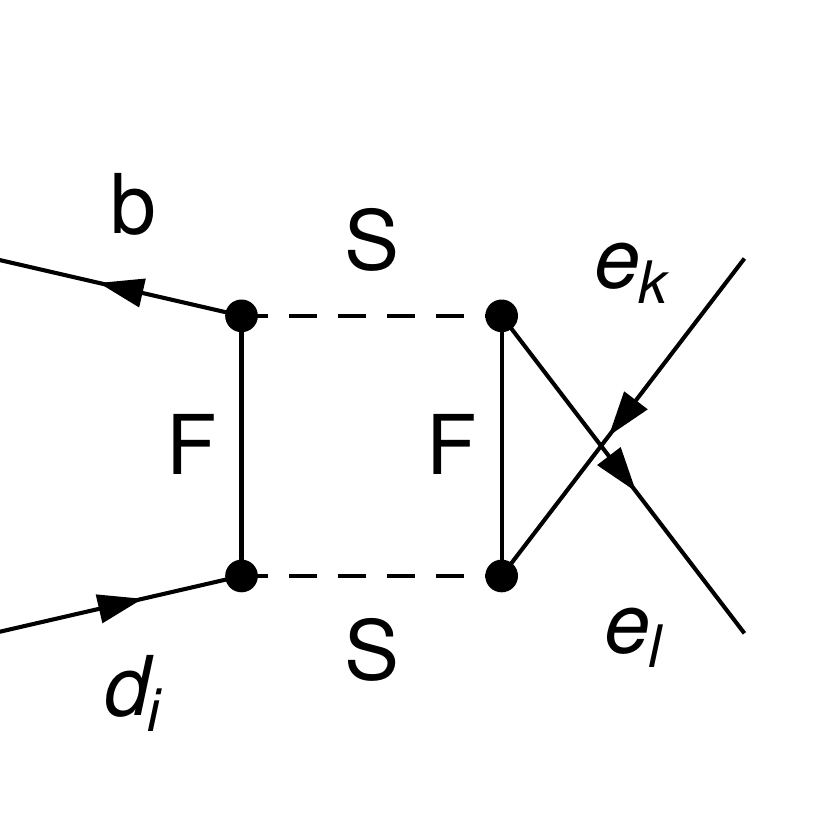} &   \includegraphics[width=3cm]{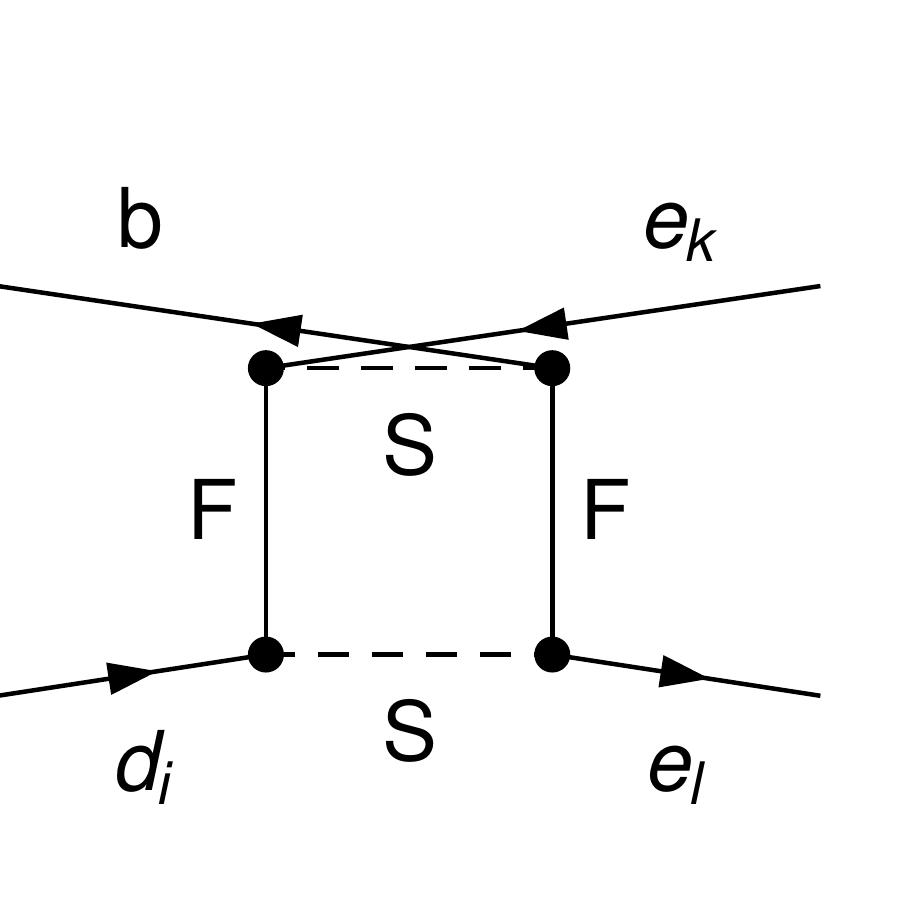} \\
  (a) & (b) & (c) \\
  \includegraphics[width=3cm]{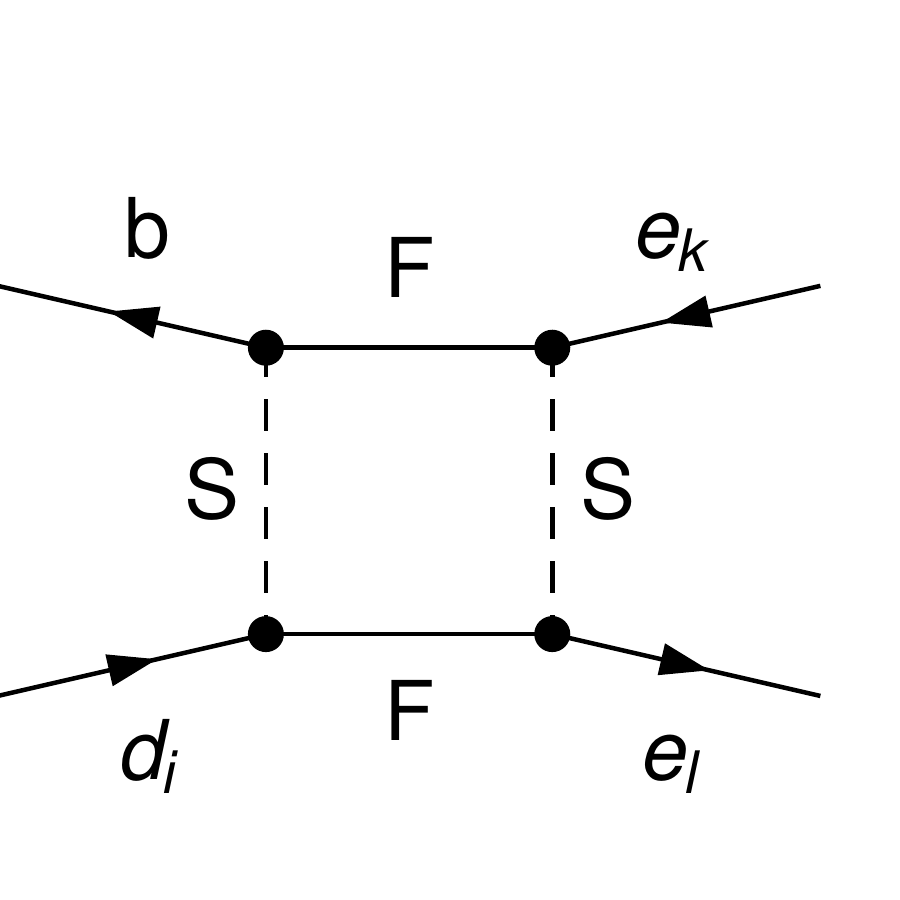} &   \includegraphics[width=3cm]{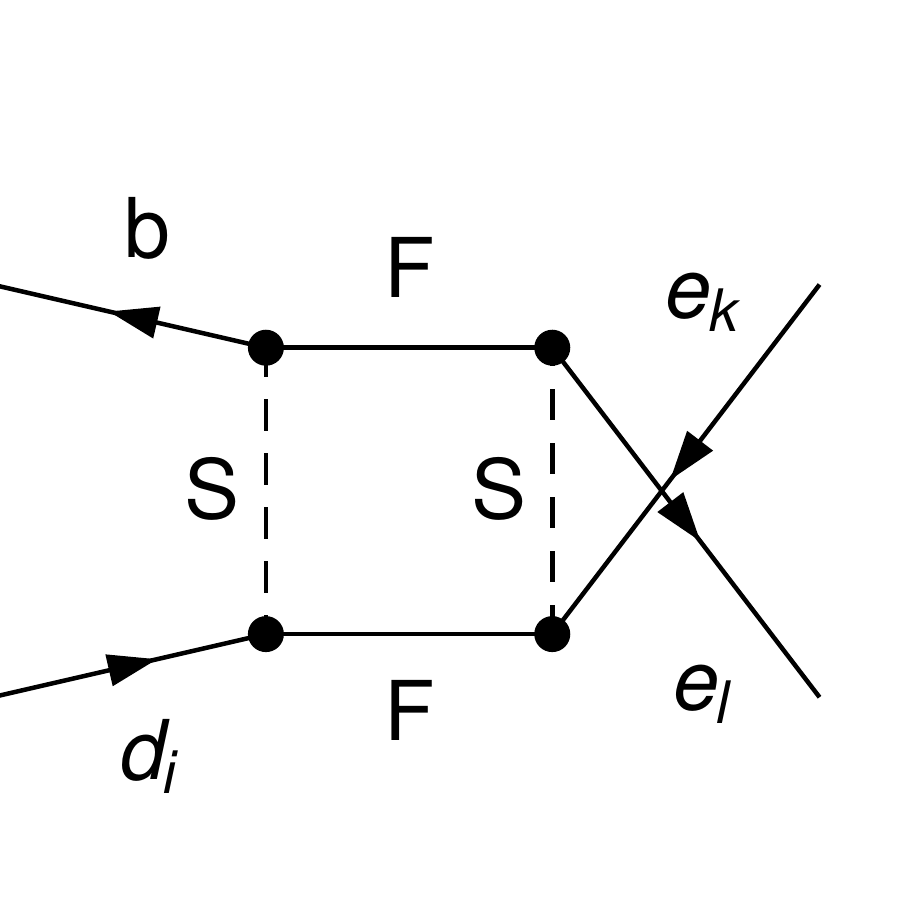} & 
  \includegraphics[width=3cm]{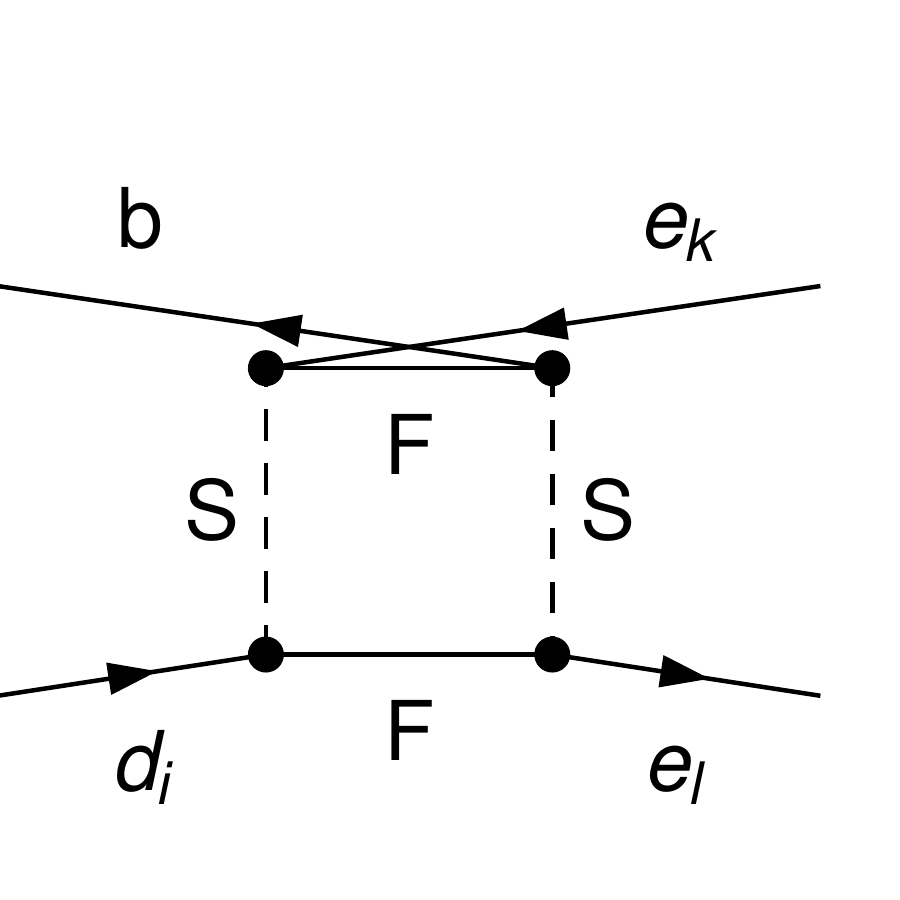} \\
  (d) & (e) & (f) \\
  \includegraphics[width=3cm]{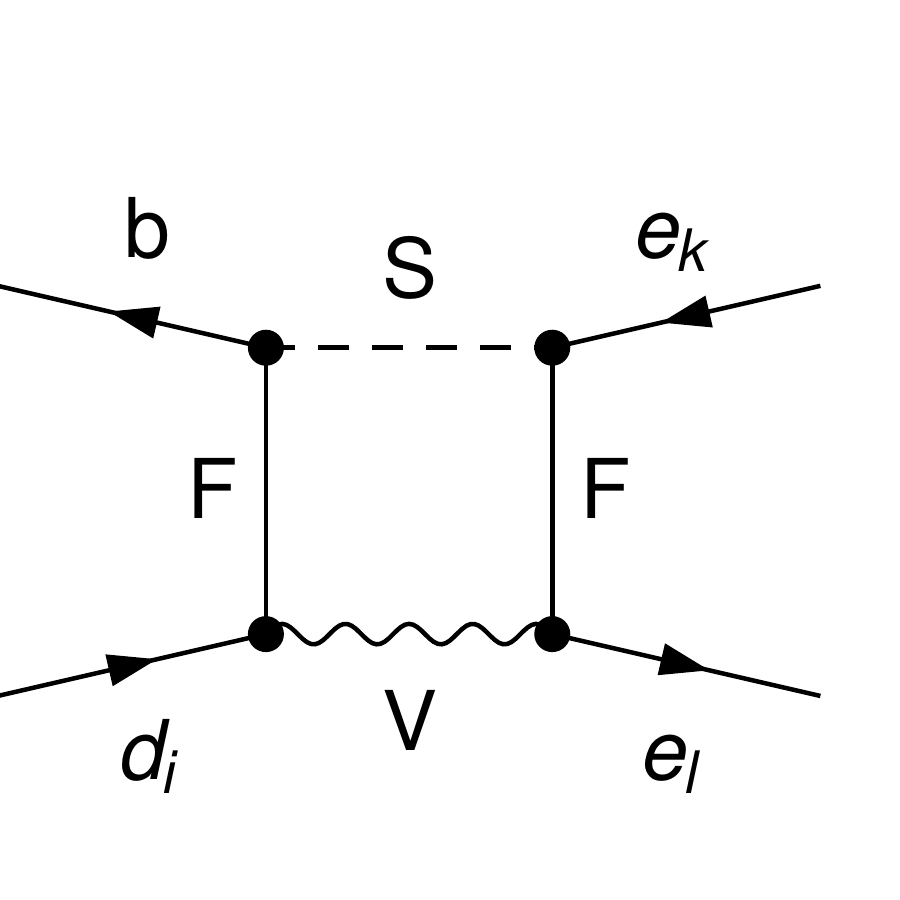} &   \includegraphics[width=3cm]{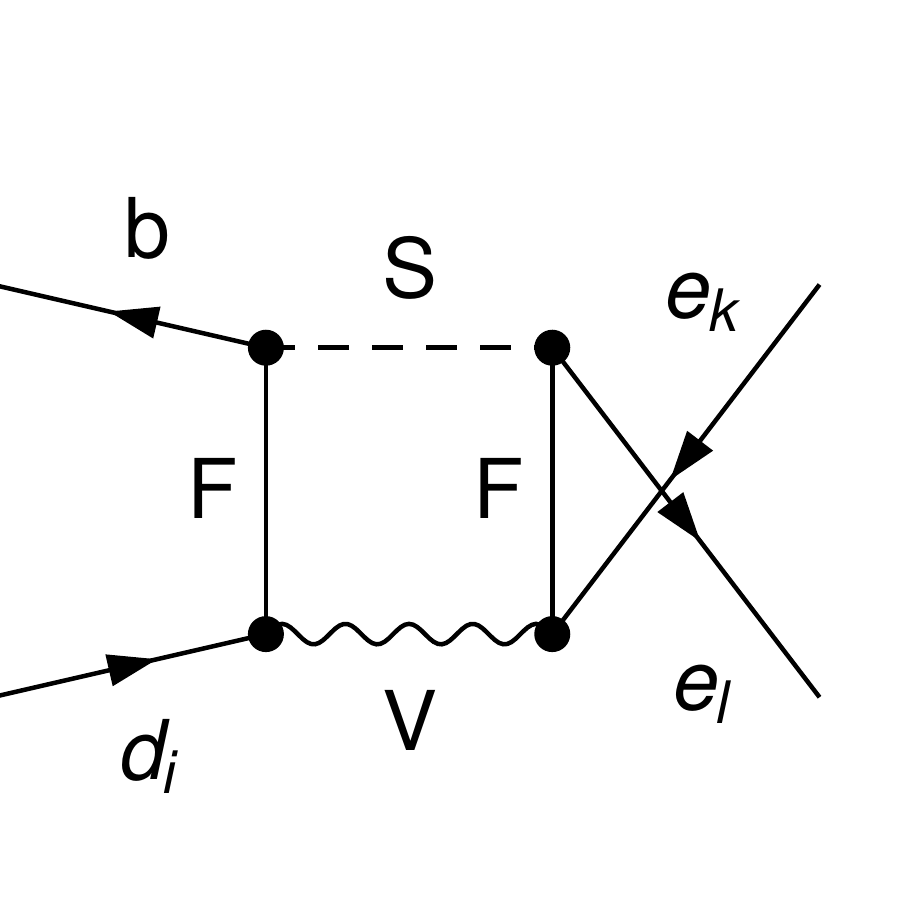} &
  \includegraphics[width=3cm]{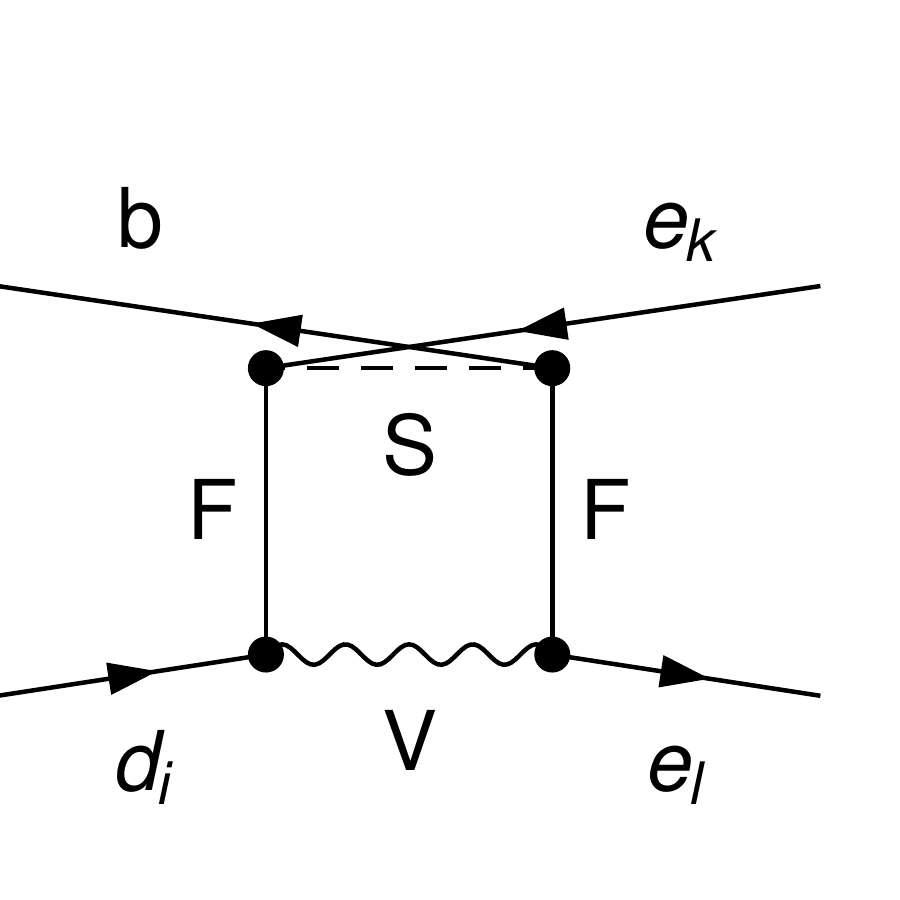} \\
  (g) & (h) & (i) \\
  \includegraphics[width=3cm]{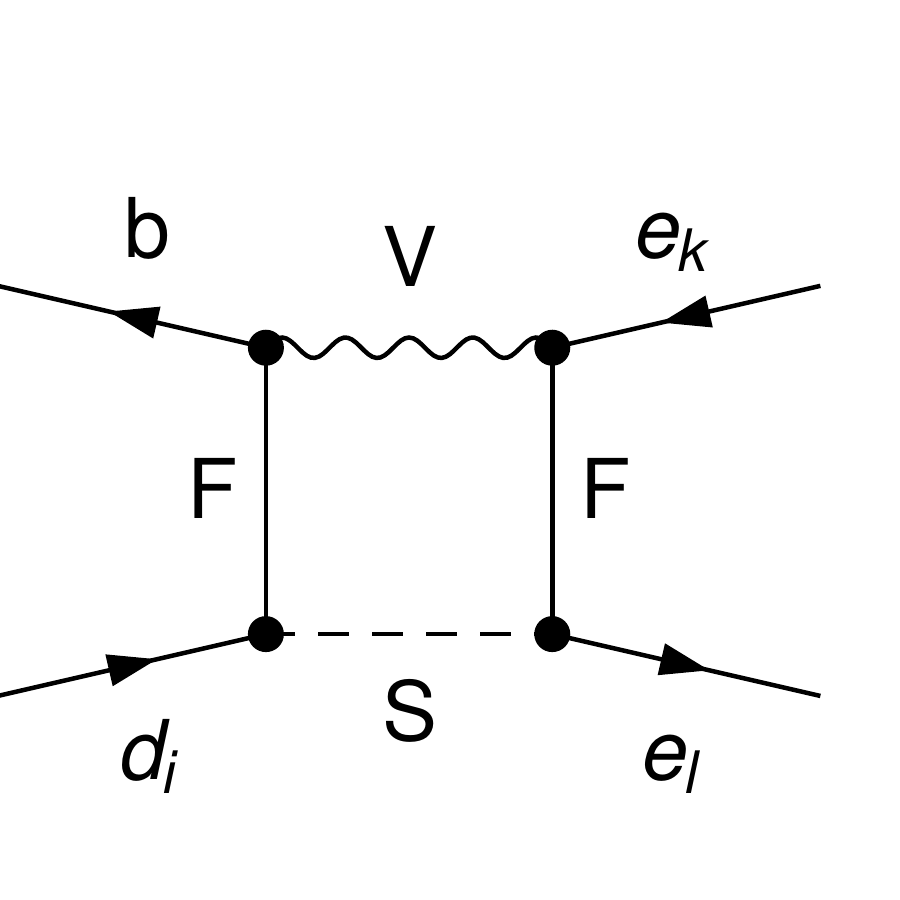} &   \includegraphics[width=3cm]{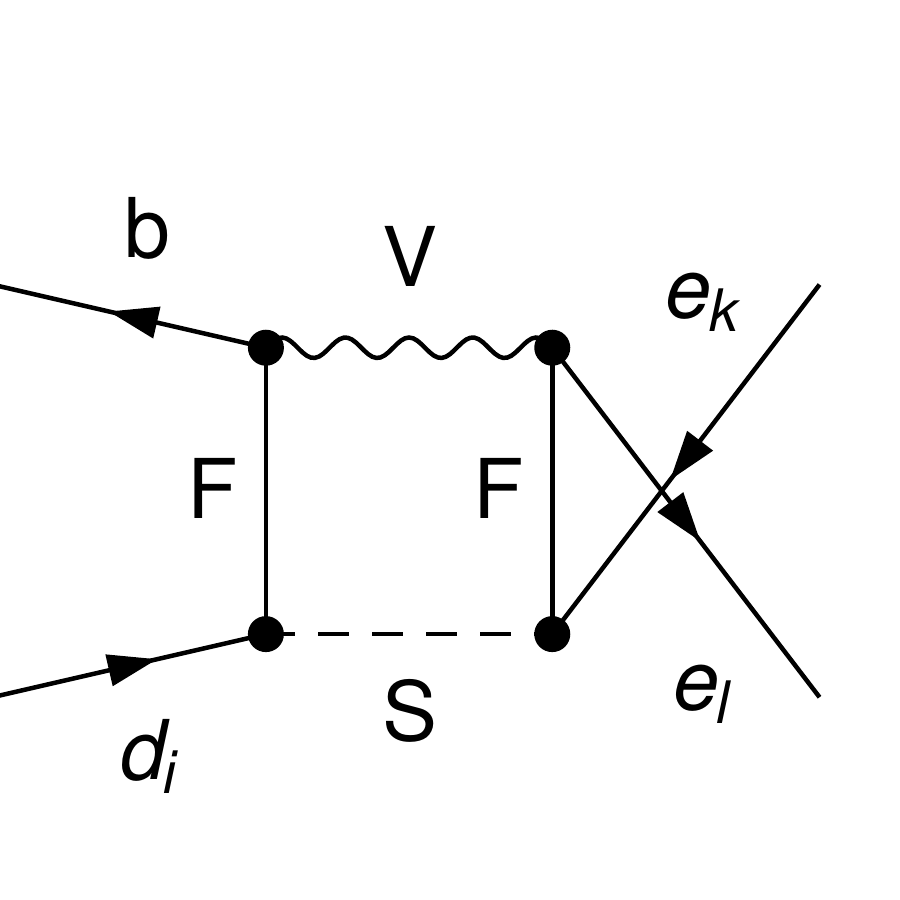} &
  \includegraphics[width=3cm]{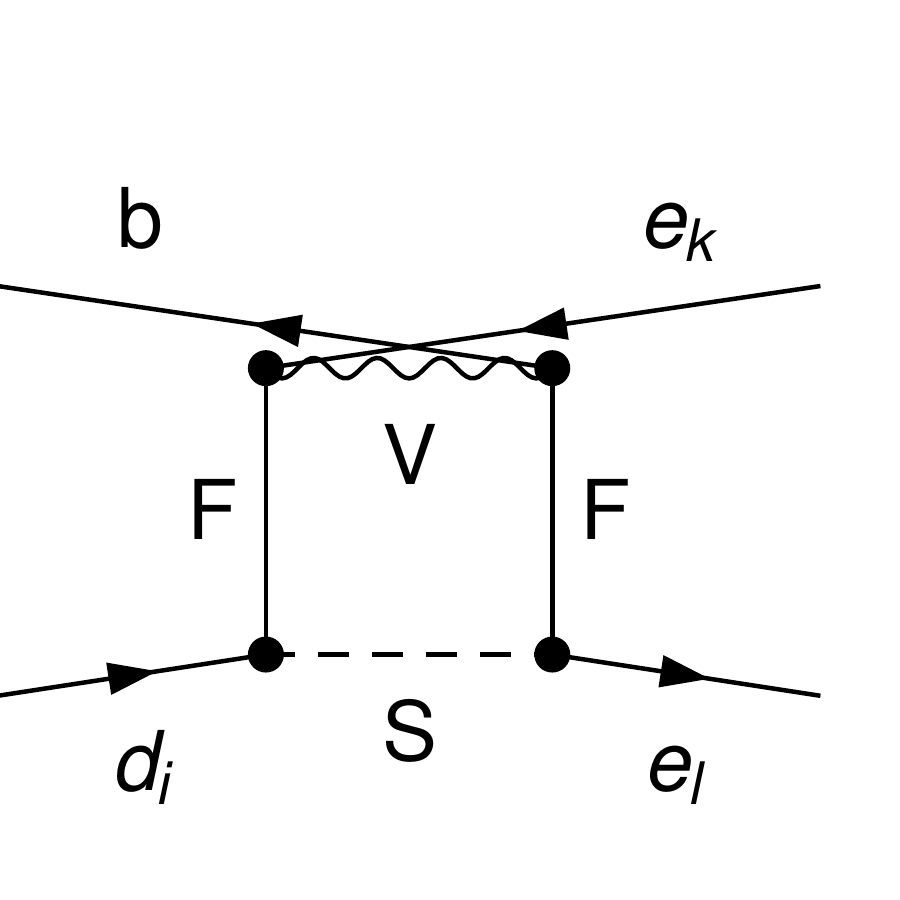} \\
  (j) & (k) & (l) 
 \end{tabular}
  \caption{Generic box diagrams I}
 \label{fig:box1}
\end{figure}
\begin{figure}[hbt]
 \centering
 \begin{tabular}{ccc}
  \includegraphics[width=3cm]{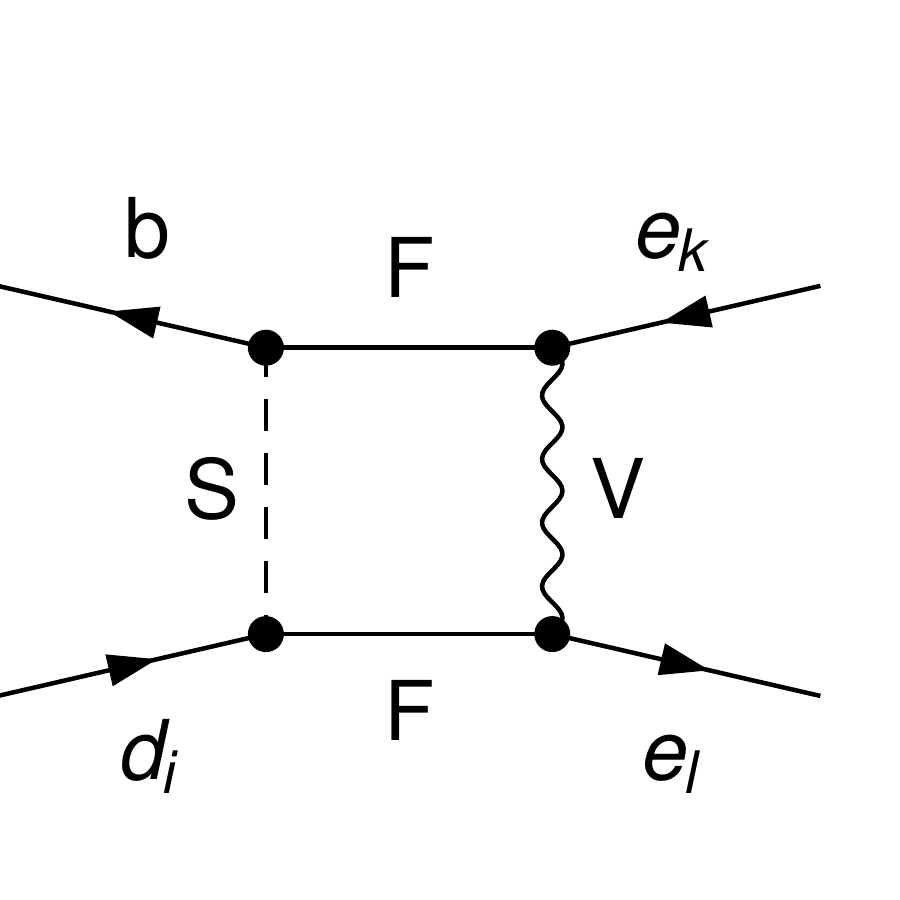} &   \includegraphics[width=3cm]{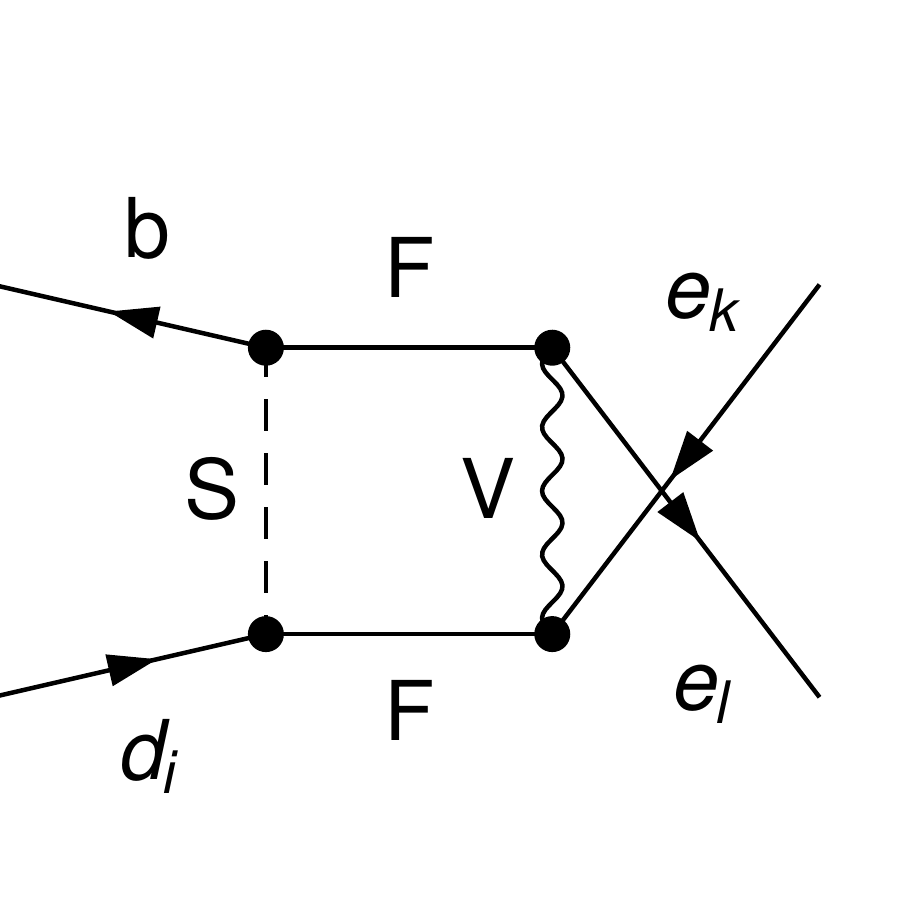} &
  \includegraphics[width=3cm]{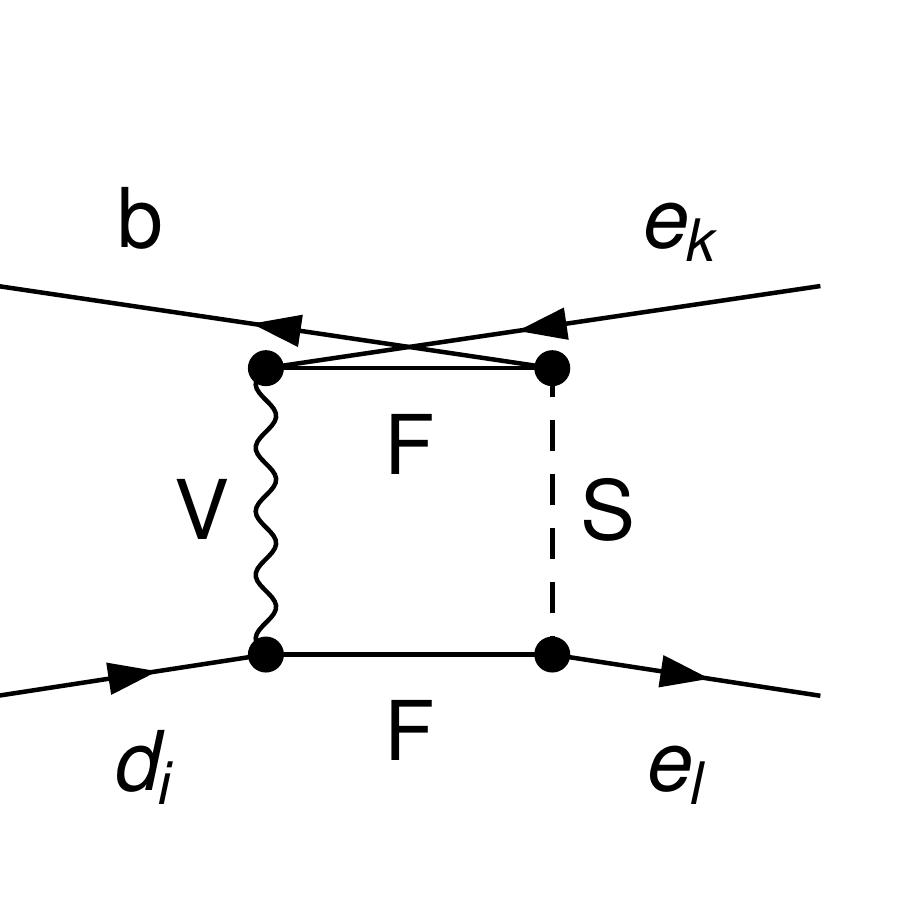} \\
  (m) & (n) & (o) \\
  \includegraphics[width=3cm]{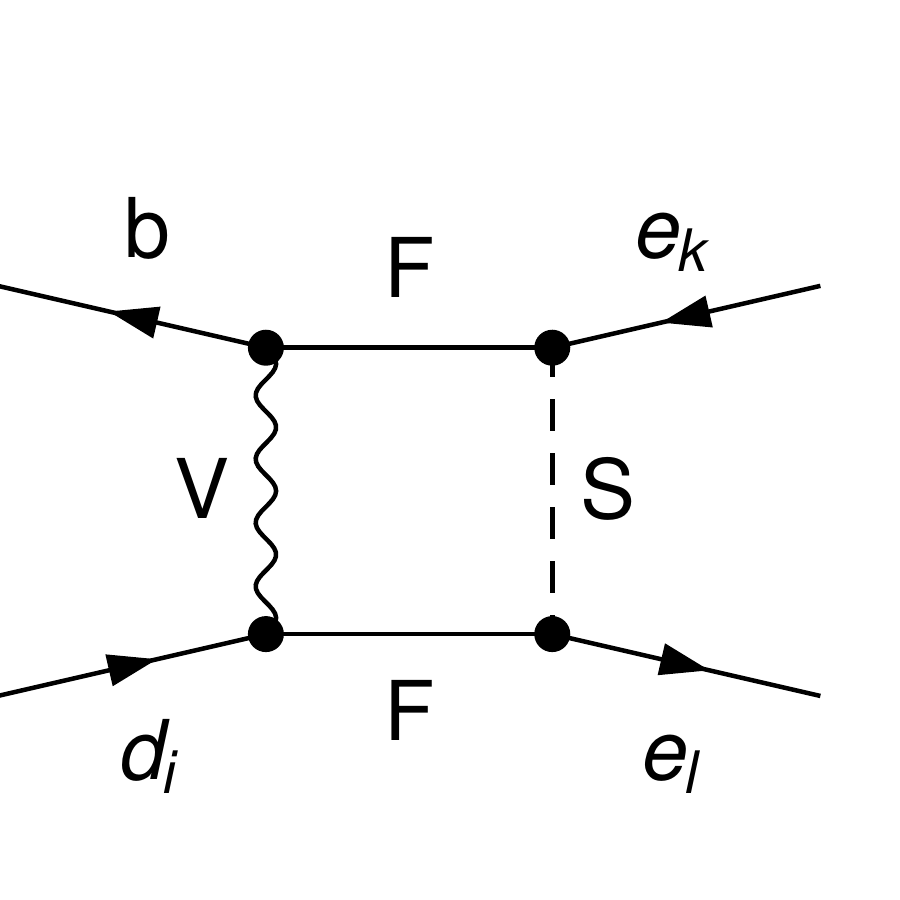} &   \includegraphics[width=3cm]{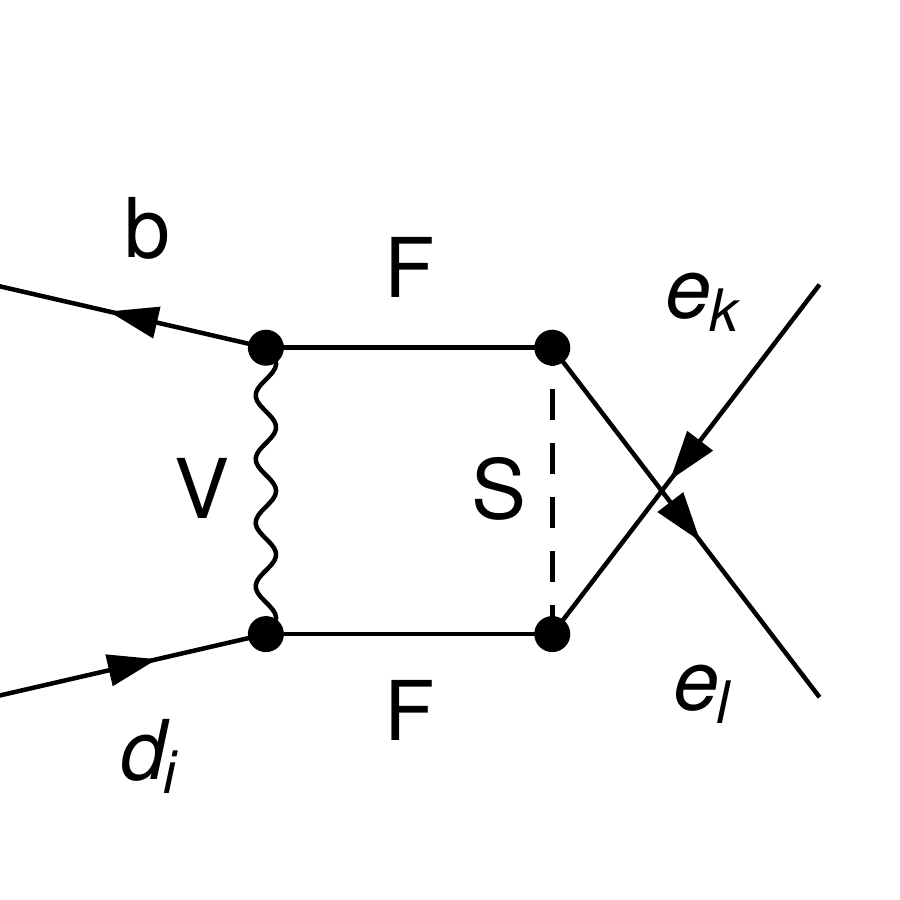} & 
  \includegraphics[width=3cm]{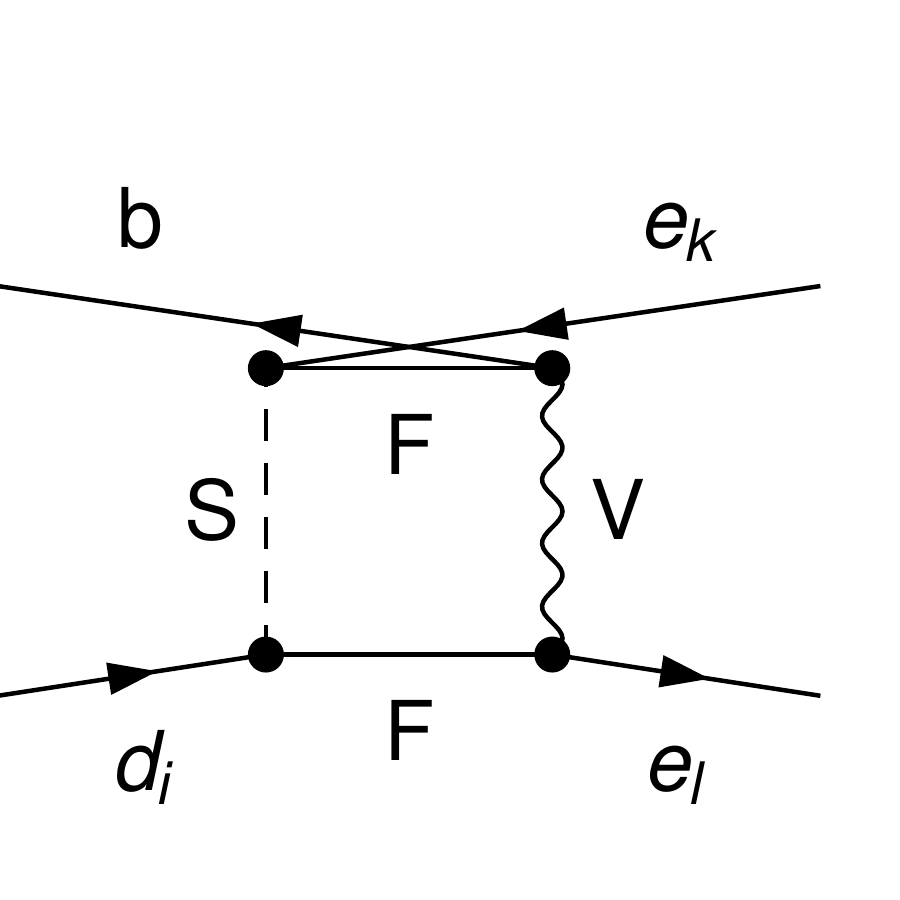} \\
 (p) & (q) & (r)  \\
  \includegraphics[width=3cm]{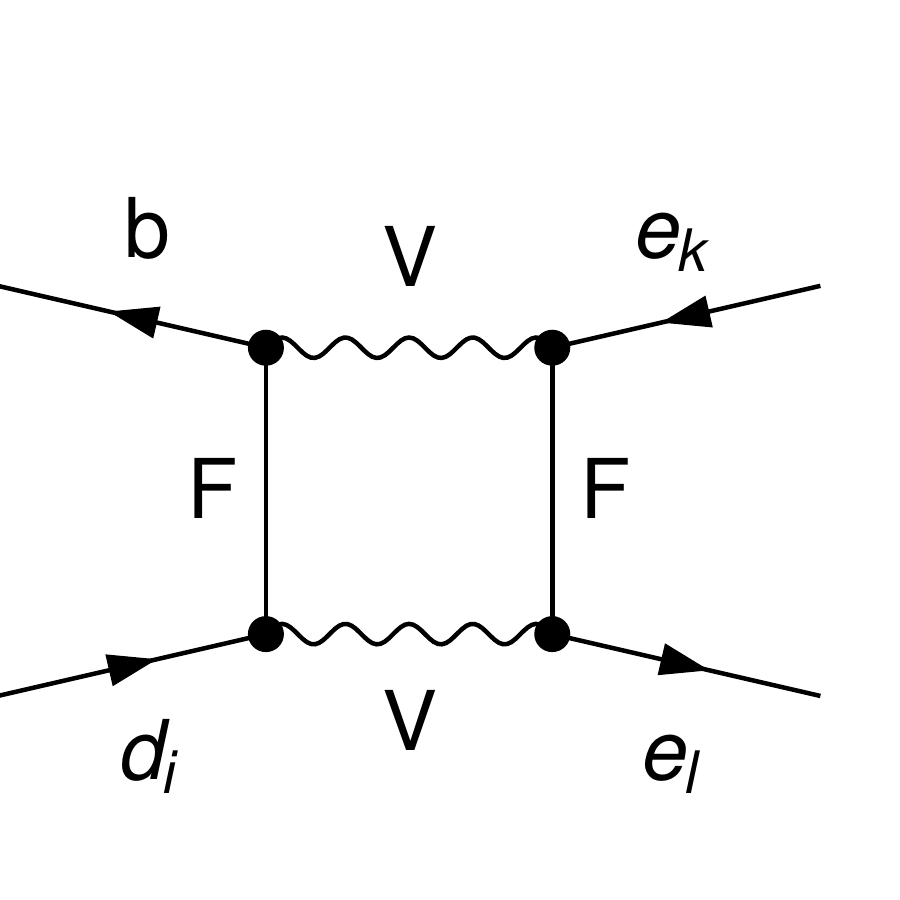} &   \includegraphics[width=3cm]{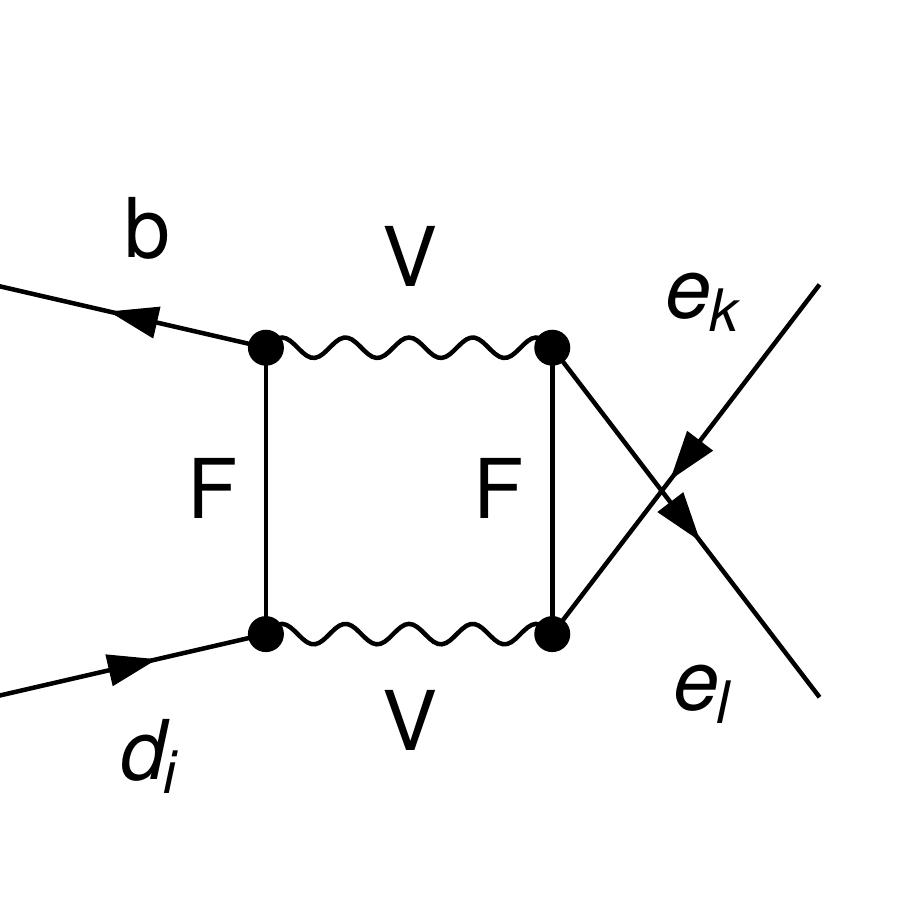} &
  \includegraphics[width=3cm]{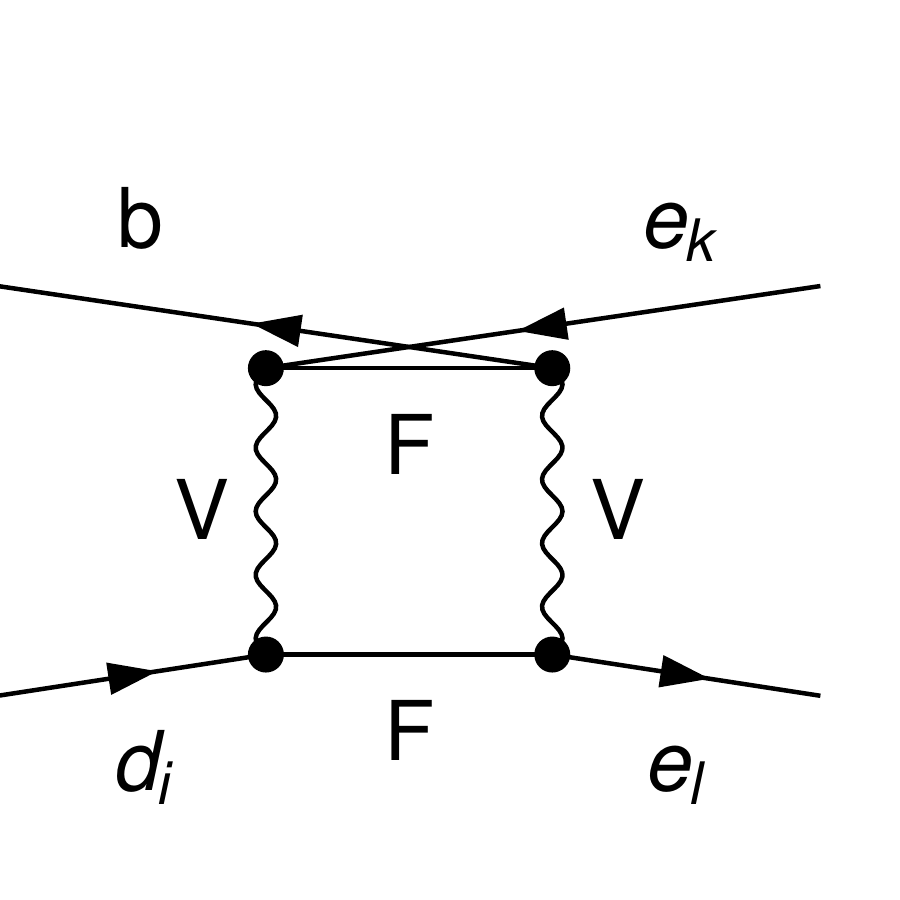} \\
  (s) & (t) & (u) \\
  \includegraphics[width=3cm]{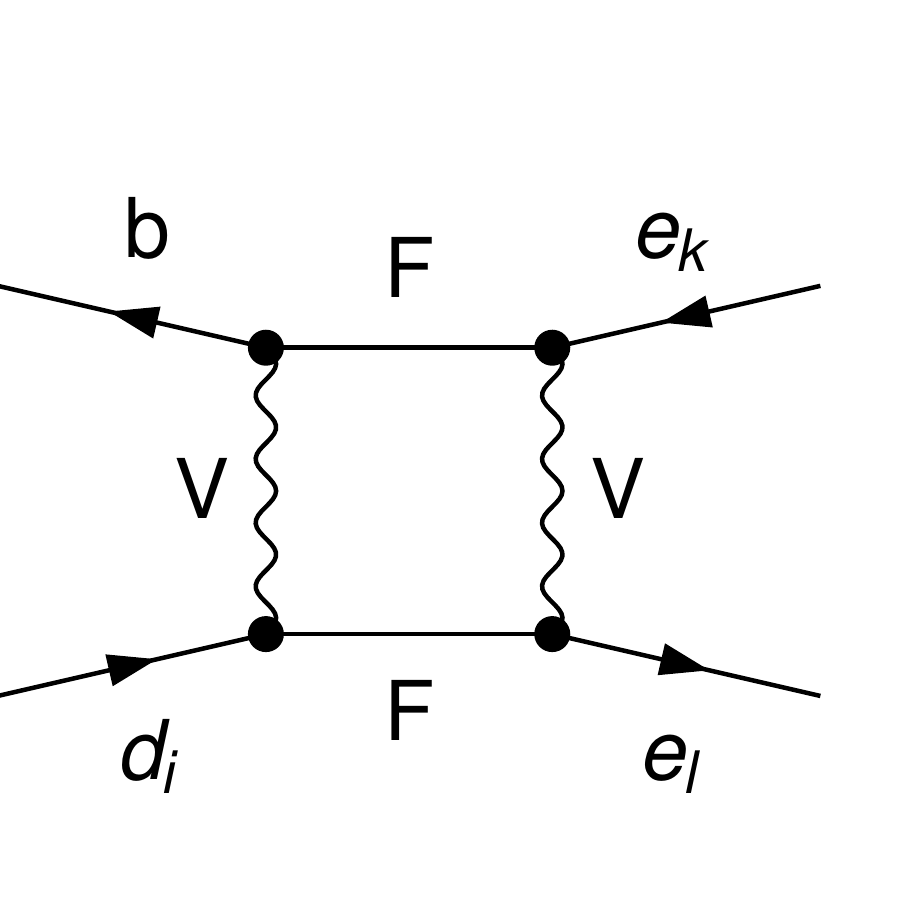} &   \includegraphics[width=3cm]{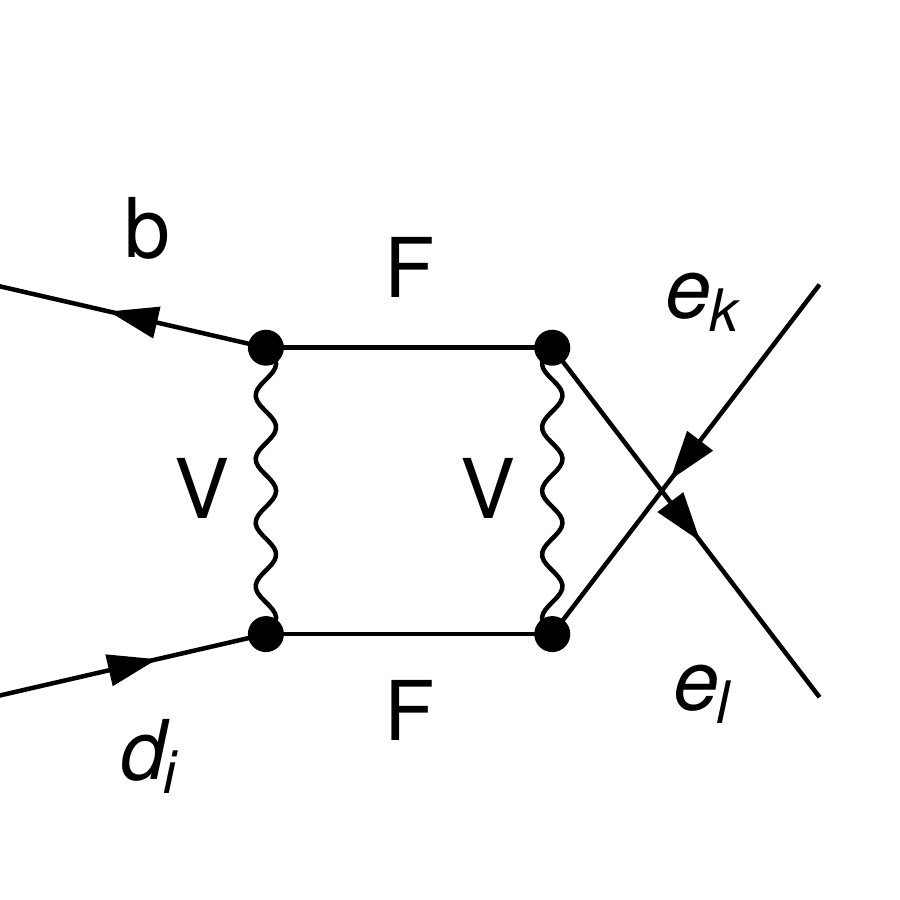} &
  \includegraphics[width=3cm]{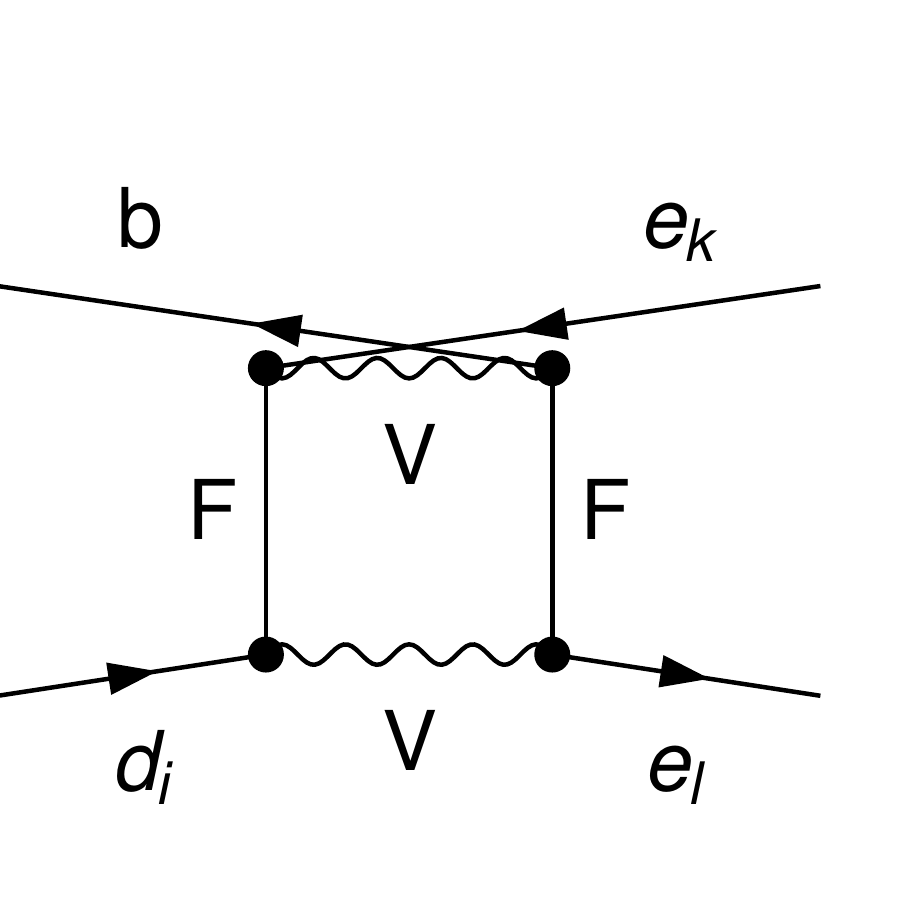} \\
  (v) & (w) & (x)
 \end{tabular}
 \caption{Generic box diagrams II}
 \label{fig:box2}
\end{figure}
The vertex number conventions for boxes are shown in figs.~\ref{fig:boxvertex}, while all possible generic diagrams are given in Figures~\ref{fig:box1} and \ref{fig:box2}. All vertices are chiral and they are parametrized as in eqs.~(\ref{eq:chiralvertices1})-(\ref{eq:chiralvertices2}) with $A=1,B=2$ for vertex 1, $A=3,B=4$ for vertex 2, $A=5,B=6$ for vertex 3 and $A=7,B=8$ for vertex 4. If there are two particles of equal type in a loop (say, two fermions), the one between vertices 1 and 2 (2 or 3) will be labelled $F1$ and the other one will be $F2$. The contributions to the different Wilson coefficients read 
\begin{align}
  C^{(a)}_{SLL} &= -G_1G_3G_5G_7M_{F1}M_{F2} \cdot D^{(a-c)}_0 \\
  C^{(a)}_{SLR} &= -G_1G_3G_ 6G_8M_{F1}M_{F2}\cdot D^{(a-c)}_0 \\
  C^{(a)}_{VLL}&=-G_2G_3G_6G_7 \cdot D^{(a-c)}_{00} \\
  C^{(a)}_{VLR}&=-G_2G_3G_5G_8 \cdot D^{(a-c)}_{00} \\
C^{(b)}_{SLL} &= -G_1G_3G_5 G_7 M_{F1}M_{F2} \cdot D^{(a-c)}_0 \\
C^{(b)}_{SLR} &= -G_1G_3G_ 6 G_8M_{F1}M_{F2}\cdot D^{(a-c)}_0 \\
C^{(b)}_{VLL}&=G_2G_3G_5 G_8  \cdot D^{(a-c)}_{00} \\
C^{(b)}_{VLR}&=G_2G_3G_6 G_7  \cdot D^{(a-c)}_{00} \\
C^{(c)}_{SLL} &= \frac 12 G_1G_3G_5G_7 M_{F1}M_{F2}D^{(a-c)}_0 \\
C^{(c)}_{SLR} &= -2 G_1G_4G_5G_8 D^{(a-c)}_{00} \\
C^{(c)}_{VLL} &= -\frac 12 G_2G_4G_5G_7 M_{F1}M_{F2} D^{(a-c)}_0 \\
C^{(c)}_{VLR} &= -G_2G_3G_5G_8 D^{(a-c)}_{00} \\
  C^{(d)}_{SLL} &= \frac 12 G_1G_3G_5G_7M_1M_2 \cdot D^{(d-f)}_0 \\
  C^{(d)}_{SLR} &= 2G_1G_3G_6G_8 \cdot D^{(d-f)}_{00} \\
  C^{(d)}_{VLL}&= -G_2G_3G_6G_7 \cdot D^{(d-f)}_{00} \\
  C^{(d)}_{VLR}&=\frac 12 G_2G_3G_5G_8M_1M_2 \cdot D^{(d-f)}_{0} \\
  C^{(e)}_{SLL} &= \frac 12 G_1G_3G_5 G_7 M_{F1}M_{F2} \cdot D^{(d-f)}_0 \\
 C^{(e)}_{SLR} &= 2G_1G_3G_6 G_8  \cdot D^{(d-f)}_{00} \\
 C^{(e)}_{VLL}&= -\frac 12G_2G_3G_5 G_8  M_{F1}M_{F2}\cdot D^{(d-f)}_{0} \\
 C^{(e)}_{VLR}&=  G_2G_3G_6 G_7 \cdot D^{(d-f)}_{00} \\
 C^{(f)}_{SLL} &= \frac 12 G_1G_3G_5G_7 M_{F1}M_{F2}D^{(d-f)}_0 \\
C^{(f)}_{SLR} &= -2G_1G_4G_5G_8 D^{(d-f)}_{00} \\
C^{(f)}_{VLL} &= G_2G_4G_5G_7 D^{(d-f)}_{00} \\
C^{(f)}_{VLR} &= \frac 12 G_2G_3G_5G_8 M_{F1}M_{F2}D^{(d-f)}_0 \\
  C^{(g)}_{SLL}&= 2 G_1G_3G_6G_7 \kl{C^{(g-i)}_0+M_{F1}^2D^{(g-i)}_0-2D^{(g-i)}_{00}}\\
  C^{(g)}_{SLR}&= 2 G_1G_3G_5G_8 \kl{C^{(g-i)}_0+M_{F1}^2D^{(g-i)}_0-2D^{(g-i)}_{00}}\\
  C^{(g)}_{VLL}&= G_2G_3G_5G_7 M_{F1}M_{F2} D^{(g-i)}_0 \\
  C^{(g)}_{VLR}&= G_2G_3G_6G_8 M_{F1}M_{F2} D^{(g-i)}_0 \\
C^{(h)}_{SLL}&= -4 G_1G_3G_5 G_7  D^{(g-i)}_{00}\\
  C^{(h)}_{SLR}&= -4 G_1G_3G_6 G_8  D^{(g-i)}_{00}\\
C^{(h)}_{VLL}&= G_2G_3G_5 G_8  M_{F1}M_{F2} D^{(g-i)}_0 \\
  C^{(g)}_{VLR}&= G_2G_3G_6 G_7  M_{F1}M_{F2} D^{(g-i)}_0 \\
C^{(i)}_{SLL}&= -G_1G_3G_5G_7\kl{C^{(g-i)}_0+M_S^2D^{(g-i)}_0-8D^{(g-i)}_{00}} \\
C^{(i)}_{SLR}&= 2G_1G_3G_5G_8 M_{F1}M_{F2}D^{(g-i)}_0 \\
C^{(i)}_{VLL}&= G_2G_3G_5G_7 \kl{C^{(g-i)}_0+M_S^2D^{(g-i)}_0-2D^{(g-i)}_{00}} \\
C^{(i)}_{VLR}&= G_2G_4G_5G_8 M_{F1}M_{F2} D^{(g-i)}_0 \\
C^{(j)}_{SLL}&= 2 G_2G_3G_5G_7 \kl{C^{(j-l)}_0+M_{F1}^2D^{(j-l)}_0-2D^{(j-l)}_{00}}\\
  C^{(j)}_{SLR}&= 2 G_2G_3G_6G_8 \kl{C^{(j-l)}_0+M_{F1}^2D^{(j-l)}_0-2D^{(j-l)}_{00}}\\
   C^{(j)}_{VLL}&= G_1G_3G_6G_7 M_{F1}M_{F2} D^{(j-l)}_0 \\
  C^{(j)}_{VLR}&= G_1G_3G_5G_8 M_{F1}M_{F2} D^{(j-l)}_0 \\
C^{(k)}_{SLL}&= -4 G_2G_3G_5 G_8  D^{(j-l)}_{00}\\
  C^{(k)}_{SLR}&= -4 G_2G_3G_6 G_7  D^{(j-l)}_{00}\\
C^{(k)}_{VLL}&= G_1G_3G_5 G_7  M_{F1}M_{F2} D^{(j-l)}_0 \\
  C^{(k)}_{VLR}&= G_1G_3G_6 G_8  M_{F1}M_{F2} D^{(j-l)}_0 \\
 C^{(l)}_{SLL}&= -G_1G_3G_5G_8(C^{(j-l)}_0+M_V^2D^{(j-l)}_0-8D^{(j-l)}_{00}) \\
  C^{(l)}_{SLR}&= 2G_1G_4G_5G_7 M_{F1}M_{F2}D^{(j-l)}_0 \\
  C^{(l)}_{VLL}&= G_2G_4G_5G_8 (C^{(j-l)}_0+M_V^2D^{(j-l)}_0-2D^{(j-l)}_{00}) \\
  C^{(l)}_{VLR}&= G_2G_3G_5G_7 M_{F1}M_{F2} D^{(j-l)}_0 \\
    C^{(m)}_{SLL}&= - G_1G_3G_6G_7 \left( C^{(m-o)}_0+M_S^2D^{(m-o)}_0 - \frac 14(13G_1G_3G_6G_7+3G_2G_4G_5G_8)D^{(m-o)}_{00}\right) \\
  C^{(m)}_{SLR}&=-2 G_1G_3G_5G_8 M_{F1}M_{F2} D^{(m-o)}_0\\
  C^{(m)}_{VLL}&=  G_2G_3G_5G_7 M_{F1}M_{F2} D^{(m-o)}_0\\
  C^{(m)}_{VLR}&= - G_2G_3G_6G_8 \kl{C^{(m-o)}_0+M_S^2D^{(m-o)}_0-2D^{(m-o)}_{00}}\\
C^{(n)}_{SLL}&= 8 G_1G_3G_6 G_8  D^{(m-o)}_{00}\\
C^{(n)}_{SLR}&= 2 G_1G_3G_5 G_7  M_{F1}M_{F2} D^{(m-o)}_0\\
C^{(n)}_{VLL}&=  -2 G_2G_3G_6 G_7  D^{(m-o)}_{00}\\
C^{(n)}_{VLR}&=  G_2G_3G_5 G_8 M_{F1}M_{F2}D^{(m-o)}_0\\
C^{(o)}_{SLL}&= -\frac 14 (13G_1G_3G_5G_7+3G_2G_4G_6G_8)D^{(m-o)}_{00} \\
  C^{(o)}_{SLR}&= -2G_1G_4G_5G_8 M_{F1}M_{F2}D^{(m-o)}_0 \\
C^{(o)}_{VLL}&= G_2G_4G_5G_7 M_{F1}M_{F2} D^{(m-o)}_0 \\
  C^{(o)}_{VLR}&= 2G_2G_3G_5G_8 D^{(m-o)}_{00} \\
  C^{(p)}_{SLL} &=- G_2G_3G_5G_7 \left( C^{(p-r)}_0+M_V^2D^{(p-r)}_0 -  \frac 14(13G_2G_3G_5G_7+3G_1G_4G_6G_8)D^{(p-r)}_{00} \right)\\ 
  C^{(p)}_{SLR}&= -2 G_2G_3G_6G_8 M_{F1}M_{F2} D^{(p-r)}_0\\
  C^{(p)}_{VLL}&=  G_1G_3G_6G_7 M_{F1}M_{F2} D^{(p-r)}_0\\
C^{(p)}_{VLR}&= - G_1G_3G_5G_8 \kl{C^{(p-r)}_0+M_V^2D^{(p-r)}_0-2D^{(p-r)}_{00}}\\
 C^{(r)}_{SLL}&= 8  G_1G_3G_5 G_7 D^{(p-r)}_{00}\\
  C^{(q)}_{SLR}&= 2  G_1G_3G_6 G_8  M_{F1}M_{F2} D^{(p-r)}_0\\
  C^{(q)}_{VLL}&= -2 G_2G_3G_5 G_8  D^{(p-r)}_{00}\\
  C^{(q)}_{VLR}&=    G_2G_3G_6 G_7  M_{F1}M_{F2} D^{(p-r)}_0\\
  C^{(r)}_{SLL}&= -\frac 14 (13G_2G_4G_5G_7+3G_1G_3G_6G_8)D^{(p-r)}_{00} \\
  C^{(r)}_{SLR}&= -2 G_2G_3G_5G_8 M_{F1}M_{F2}D^{(p-r)}_0+\frac 34 (G_2G_4G_5G_7-G_1G_3G_6G_8)D^{(p-r)}_{00} \\
  C^{(r)}_{VLL}&= G_1G_3G_5G_7M_{F1}M_{F2}D^{(p-r)}_0 \\
  C^{(r)}_{VLR}&= 2G_1G_4G_5G_8 D^{(p-r)}_{00} \\
  C^{(s)}_{SLL}&= -4 G_2G_3G_6G_7 M_{F1}M_{F2} D^{(s-u)}_{0}\\
  C^{(s)}_{SLR}&= -4 G_2G_3G_5G_8 M_{F1}M_{F2} D^{(s-u)}_0  \\
  C^{(s)}_{VLL}&= -4 G_1G_3G_5G_7 \kl{C^{(s-u)}_0+M_{F1}^2D^{(s-u)}_0-3D^{(s-u)}_{00}}\\
  C^{(s)}_{VLR}&= -4 G_1G_3G_6G_8 \kl{C^{(s-u)}_0+M_{F1}^2D^{(s-u)}_0}\\
  C^{(t)}_{SLL}&= -4 G_2G_3G_5 G_8  M_{F1}M_{F2} D^{(s-u)}_{0}\\
  C^{(t)}_{SLR}&= -4 G_2G_3G_6 G_7  M_{F1}M_{F2} D^{(s-u)}_0  \\
  C^{(t)}_{VLL}&= 16 G_1G_3G_5 G_7  D^{(s-u)}_{00}\\
  C^{(t)}_{VLR}&= 4 G_1G_3G_6 G_8  D^{(s-u)}_{00}\\
  C^{(u)}_{SLL}&= -4 G_2G_4G_5 G_7  M_{F1}M_{F2} D^{(s-u)}_{0}\\
  C^{(u)}_{SLR}&= -8 G_2G_3G_5 G_8  D^{(s-u)}_{00}  \\
  C^{(u)}_{VLL}&= 16 G_1G_3G_5 G_7  D^{(s-u)}_{00}\\
  C^{(u)}_{VLR}&= 2 G_1G_4G_5 G_8  M_{F1}M_{F2} D^{(s-u)}_{0}\\
  C^{(v)}_{SLL}&= 8 G_2G_3G_6G_7 M_{F1}M_{F2} D^{(v-x)}_{0}\\
  C^{(v)}_{SLR}&= 8 G_2G_3G_5G_8 \kl{C^{(v-x)}_0+M_{V1}^2D^{(v-x)}_0}  \\
  C^{(v)}_{VLL}&= -4 G_1G_3G_5G_7 \kl{C^{(v-x)}_0+M_{V1}^2D^{(v-x)}_0-3D^{(v-x)}_{00}}\\
  C^{(v)}_{VLR}&= 2 G_1G_3G_6G_8 M_{F1}M_{F2} D^{(v-x)}_{0}\\
  C^{(w)}_{SLL}&= 8 G_1G_3G_6 G_8  M_{F1}M_{F2} D^{(v-x)}_{0}\\
  C^{(w)}_{SLR}&= 32 G_1G_3G_5 G_7  D^{(v-x)}_{00}  \\
  C^{(w)}_{VLL}&= -2 G_2G_3G_6 G_7  M_{F1}M_{F2} D^{(v-x)}_0\\
  C^{(w)}_{VLR}&= 4 G_2G_3G_5 G_8  D^{(v-x)}_{00}\\
  C^{(x)}_{SLL}&= -4 G_1G_4G_5 G_8  M_{F1}M_{F2} D^{(v-x)}_{0}\\
  C^{(x)}_{SLR}&= -8 G_1G_3G_5 G_7 (C^{(v-x)}_0+M_{V1}^2D^{(v-x)}_0-3 D^{(v-x)}_{00})  \\
  C^{(x)}_{VLL}&= -2 G_2G_3G_5 G_8  M_{F1}M_{F2} D^{(v-x)}_0\\
  C^{(x)}_{VLR}&= -4 G_2G_4G_5 G_7(C^{(v-x)}_0+M_{V1}^2D^{(v-x)}_0)
  \end{align}
The arguments of the loop functions for the different amplitudes are
\begin{align}
D_X^{(a-c)}&= D_X(M_{F1}^2,M_{F2}^2,M_{S1}^2,M_{S2}^2) \\ 
D_X^{(d-f)}&= D_X(M_{F1}^2,M_{F2}^2,M_{S1}^2,M_{S2}^2) \\
C_X^{(g-i)}&= C_X(\vec 0_3, M_{F2}^2,M_V^2,M_{S}^2) \,\hspace{1cm}
D_X^{(g-i)}= D_X(M_{F1}^2,M_{F2}^2,M_{V}^2,M_{S}^2)\\
C_X^{(j-l)}&= C_X(\vec 0_3,M_{F2}^2,M_S^2,M_{V}^2) \,\hspace{1cm}
D_X^{(j-l)}= D_X(M_{F1}^2,M_{F2}^2,M_{S}^2,M_{V}^2)\\
C_X^{(m-o)}&= C_X(\vec 0_3, M_{F2}^2,M_{F1}^2,M_V^2) \,\hspace{1cm}
D_X^{(m-o)}= D_X(M_{F2}^2,M_{F1}^2,M_S^2,M_V^2)\\
C_X^{(p-r)}&= C_X(\vec 0_3, M_{F2}^2,M_{F1}^2,M_S^2) \,\hspace{1cm}
D_X^{(p-r)}= D_X(M_{F2}^2,M_{F1}^2,M_V^2,M_S^2) \\
C_X^{(s-u)}&= C_X(\vec 0_3, M_{F2}^2,M_{V1}^2,M_{V2}^2)\,\hspace{1cm}
D_X^{(s-u)}= D_X(M_{F1}^2,M_{F2}^2,M_{V1}^2,M_{V2}^2)  \\
C_X^{(v-x)}&= C_X(\vec 0_3, M_{F2}^2,M_{F1}^2,M_{V2}^2) \,\hspace{1cm}
D_X^{(v-x)}= D_X(M_{F2}^2,M_{F1}^2,M_{V1}^2,M_{V2}^2)  
\end{align}

\end{appendix}

\bibliographystyle{h-physrev}

\end{document}